\documentclass{aa}
\usepackage{graphicx}
\usepackage{txfonts}
\usepackage{lscape} 
\usepackage{morefloats}
\newcommand{\Prot}{P_{\rm rot}}
\newcommand{\Pcyc}{P_{\rm cyc}}

\newcommand{\Rvar}{R_{\rm var}}
\newcommand{\logRHK}{\log R'_{\rm HK}}

\begin{document}

\title{Transition from spot to faculae domination}
\subtitle{An alternate explanation for the dearth of intermediate \textit{Kepler} 
rotation periods}

\author{Timo Reinhold\inst{1,2}, 
	Keaton J.\ Bell\inst{1,2}, 
	James Kuszlewicz\inst{1,2}, 
	Saskia Hekker\inst{1,2},
	Alexander I.\ Shapiro\inst{1}
	}

\offprints{T. Reinhold,\email{reinhold@mps.mpg.de}}

\institute{
  \inst{1}
  Max-Planck-Institut f\"ur Sonnensystemforschung,  
  Justus-von-Liebig-Weg 3, 37077 G\"ottingen, Germany \\
  \inst{2}
  Stellar Astrophysics Centre, Department of Physics and Astronomy, 
  Aarhus University, 120 Ny Munkegade, Building 1520, DK-8000 Aarhus C, Denmark
  }
  
\date{Received day month year / Accepted day month year}

\abstract
{
The study of stellar activity cycles is crucial to understand the underlying 
dynamo and how it causes magnetic activity signatures such as dark spots and 
bright faculae. Having knowledge about the dominant source of surface activity 
might allow us to draw conclusions about the star's age and magnetic field 
topology, and to put the solar cycle in context.
}
{
We investigate the underlying process that causes magnetic activity by studying 
the appearance of activity signatures in contemporaneous photometric and 
chromospheric time series.
}
{
Lomb-Scargle periodograms are used to search for cycle periods present in the 
photometric and chromospheric time series. To emphasize the signature of the 
activity cycle we account for rotation-induced scatter in both data sets by 
fitting a quasi-periodic Gaussian process model to each observing season. After 
subtracting the rotational variability, cycle amplitudes and the phase difference 
between the two time series are obtained by fitting both time series simultaneously 
using the same cycle period.
}
{
We find cycle periods in 27 of the 30 stars in our sample. The phase difference 
between the two time series reveals that the variability in fast rotating active 
stars is usually in anti-phase, while the variability of slowly rotating inactive 
stars is in phase. The photometric cycle amplitudes are on average six times 
larger for the active stars. The phase and amplitude information demonstrates 
that active stars are dominated by dark spots, whereas less active stars are 
dominated by bright faculae. We find the transition from spot to faculae 
domination at the Vaughan-Preston gap, and around a Rossby number equal to one.
}
{
We conclude that faculae are the dominant ingredient of stellar activity cycles 
at ages $\gtrsim$2.55\,Gyr. The data further suggest that the Vaughan-Preston 
gap can not explain the previously detected dearth of \textit{Kepler} rotation 
periods between 15--25~days. Nevertheless, our results led us to propose an 
explanation for the rotation period dearth to be due to the non-detection of 
periodicity caused by the cancellation of dark spots and bright faculae at 
$\sim$800\,Myr.
}


\titlerunning{Transition from spot to faculae domination}
\authorrunning{Timo Reinhold et al.}
\maketitle

\section{Introduction}
In the mid Sixties, O.C. Wilson and collaborators started flux measurements at 
the centers of the \ion{Ca}{ii} H+K lines ``[...] for the purpose of initiating 
a search for stellar analogues of the solar cycle.'' \citep{Wilson1968}. Ten 
years later, \citet{Wilson1978} presented long-term \ion{Ca}{ii} H+K flux 
measurements of 91 main-sequence stars, providing the first evidence for cyclic 
activity in some stars. Over the years an empirical activity index describing 
the emission in the H and K line cores was established \citep{Vaughan1978}, 
now referred to as the Mount Wilson S-index. This index can be converted 
into the chromospheric H and K line surface flux \citep{Middelkoop1982, Rutten1984}.
Subtracting the photospheric flux from this quantity, one arrives at the currently
most commonly used activity indicator $\logRHK$, where $R'_{\rm HK}$ denotes the 
chromospheric flux in the \ion{Ca}{ii} H+K lines normalized by the bolometric 
flux \citep{Linsky1979, Noyes1984}.

Most stellar activity cycles known these days rely on periodicities measured in 
long-term observations of the S-index \citep{Baliunas1995, SB1999, BV2007}. 
Constraining the basic physical parameters that determine the nature of the 
activity cycle, especially its period, is a matter of current research. Various 
authors \citep{SB1999, BV2007, Lehtinen2016} identified a strong dependence of 
the cycle period on the rotation period, and, as a consequence thereof, the 
existence of certain sequences (\textit{active} and \textit{inactive} sequence) 
in the $\Prot-\Pcyc$ plane. \citet{BV2007} suggested that the two distinct 
sequences might indicate the existence of two dynamos operating at different 
depths in the star. In contrast, the work by \citet{Reinhold2017} and 
\citet{Saikia2018} strongly question the existence of the two sequences, 
especially for the \textit{active} sequence. Moreover, \citet{Metcalfe2016} and
\citet{Metcalfe2017} proposed a secular increase of the cycle period once the 
star reaches a critical Rossby number $\rm{Ro} \sim 2$.

The periodic changes of the \ion{Ca}{ii} H+K emission originate from a change in 
the magnetic field generated by the stellar dynamo. Concurrently, the emergence 
of active regions in the photosphere (such as dark spots and bright faculae) is 
expected during activity maximum, as is observed in the Sun. Over the course 
of the solar cycle, variability is seen both in the total solar irradiance (TSI) 
data (e.g. \citealt{Shapiro2016}), as well as in the S-index (e.g. 
\citealt{Egeland2017}). In the Sun the photometric (TSI) and chromospheric
time series (disk-integrated \ion{Ca}{ii} K-line) are in phase \citep{Preminger2011}. 
At activity maximum the Sun is brightest due to the presence of bright faculae 
regions, slightly overcompensating the contributions of dark spots. 

Active stars are known to show photometric variability caused by dark spots and
bright faculae, as well as enhanced chromospheric emission in the \ion{Ca}{ii} 
H+K line cores. Whether all active stars exhibit an underlying activity cycle, 
i.e., if the surface activity changes periodically or randomly over time, is 
hitherto hardly known. We aim to understand the physical mechanism that drives 
long-term photometric and chromospheric brightness variations. To reach this goal, 
we search for cycle periods in contemporaneous photometric and chromospheric time 
series. Assuming a common origin of the periodicity, we fit the same cycle period 
to both time series, which allows us to measure photometric and chromospheric 
amplitudes, and the phase difference between the variations in both time series. 
We show that the phase difference reveals the dominant type of stellar activity, 
i.e., dark spots or bright faculae. Moreover, the phase difference shows 
a strong dependence on the stellar activity level, the rotation period, and the 
photometric amplitude. These correlations may help to understand stellar dynamos
as a function of activity level, rotation period, and age. 
In the following we study a sample of 30 stars, for which chromospheric activity 
cycles have been reported for 18 stars (17 in \citealt{Baliunas1995}, 14 in 
\citealt{Saikia2018}, and 6 in \citealt{Olspert2017_II_arxiv}), and 
photometric activity cycles for 6 stars \citep{Messina2002}. Taking advantage of 
analyzing both time series simultaneously, we detect activity cycles in 27 stars 
in total. In 8 stars thereof, activity cycles have not been measured before.

\section{Data}\label{data}
In the current study, data from three different sources have been used. 
Long-term V~band and Str\"omgren b and y photometric time series have 
successfully been requested from \citet{Messina2002} and \citet{Lockwood2007}, 
respectively. For all stars, chromospheric emission data from the Mount Wilson 
survey are publicly available\footnotemark\footnotetext{The data can be found at 
\url{ftp://solis.nso.edu/MountWilson_HK/} and a brief manual at \url{ftp://solis.nso.edu/MountWilson_HK/HK_data_ReadMe.docx}.}. In total our 
sample contains 30 stars. The Julian date range, number of data points, 
mean time stamp, brightness, and standard deviation of each observing season is 
listed in appendices \ref{app_A} and \ref{app_B} for the photometric and 
chromospheric time series, respectively. Their basic stellar parameters are 
listed in Table~\ref{params}. Effective temperatures, surface gravities and 
metallicities are taken from the references given. The spectral types have been 
adopted from \citet{Messina2002} and \citet{Lockwood2007}. The different data 
sets are briefly described in the following.
\begin{table*}
  \centering
  \begin{tabular}{cccccccc}
\hline\hline
HD & Name & Object type & SpT & $T_{\rm eff}$ (K) & log g (dex) & Fe/H (dex) & Reference \\
\hline
1835 & 9 Cet & BY* & G2.5 V & $5723\substack{+148 \\ -41}$ & $4.48\substack{+0.14 \\ -0.09}$ & $0.19\substack{+0.04 \\ -0.01}$ & \citet{Boeche2016} \\
10476 & 107 Psc & PM* & K1 V & $5125\substack{+30 \\ -49}$ & $4.47\substack{+0.03 \\ -0.04}$ & $-0.04\substack{+0.04 \\ -0.04}$ & \citet{Boeche2016} \\
13421 & 64 Cet & PM* & G0 IV & $6066\substack{+42 \\ -42}$ & $3.81\substack{+0.09 \\ -0.09}$ & $0.14\substack{+0.04 \\ -0.04}$ & \citet{Niedzielski2016} \\
18256 & $\rho$ Ari & SB* & F6 V & $6380\substack{+80 \\ -80}$ & $4.17\substack{+0.20 \\ -0.20}$ & $-0.23\substack{+0.10 \\ -0.10}$ & \citet{Balachandran1990} \\
20630 & $\kappa^1$ Cet & BY* & G5 V & $5605\substack{+55 \\ -15}$ & $4.40\substack{+0.07 \\ -0.02}$ & $0.04\substack{+0.03 \\ -0.01}$ & \citet{Boeche2016} \\
25998 & 50 Per & RS* & F7 V & $6147\substack{+70 \\ -70}$ & $4.35\substack{+0.10 \\ -0.10}$ & $-0.11\substack{+0.10 \\ -0.10}$ & \citet{Chen2000} \\
35296 & 111 Tau & BY* & F8 V & $6171\substack{+63 \\ -63}$ & $4.31\substack{+0.09 \\ -0.09}$ & $0.01\substack{+0.05 \\ -0.05}$ & \citet{Prugniel2011} \\
39587 & $\chi^1$ Ori & RS* & G0-V & $5935\substack{+89 \\ -16}$ & $4.52\substack{+0.06 \\ -0.06}$ & $0.00\substack{+0.04 \\ -0.01}$ & \citet{Boeche2016} \\
72905 & $\pi^1$ UMa & BY* & G1.5 V & $5814\substack{+67 \\ -19}$ & $4.55\substack{+0.08 \\ -0.09}$ & $-0.06\substack{+0.03 \\ -0.01}$ & \citet{Boeche2016} \\
75332 & - & PM* & F7 Vn & $6130\substack{+70 \\ -70}$ & $4.32\substack{+0.10 \\ -0.10}$ & $0.00\substack{+0.10 \\ -0.10}$ & \citet{Chen2000} \\
81809 & - & SB* & G2 V & $5667\substack{+18 \\ -102}$ & $3.85\substack{+0.03 \\ -0.14}$ & $-0.37\substack{+0.02 \\ -0.03}$ & \citet{Boeche2016} \\
82443 & DX Leo & BY* & K0 V & $5334\substack{+80 \\ -80}$ & $4.40\substack{+0.17 \\ -0.17}$ & $-0.03\substack{+0.08 \\ -0.08}$ & \citet{Mishenina2013} \\
82885 & 11 LMi & RS* & G8 IV-V & $5438\substack{+47 \\ -58}$ & $4.34\substack{+0.14 \\ -0.03}$ & $0.33\substack{+0.02 \\ -0.02}$ & \citet{Boeche2016} \\
103095 & - & PM* & G8 V & $4947\substack{+75 \\ -80}$ & $4.56\substack{+0.07 \\ -0.28}$ & $-1.41\substack{+0.04 \\ -0.03}$ & \citet{Boeche2016} \\
115383 & $\eta$ Vir & PM* & G0 Vs & $6087\substack{+100 \\ -12}$ & $4.42\substack{+0.08 \\ -0.02}$ & $0.13\substack{+0.03 \\ -0.01}$ & \citet{Boeche2016} \\
115404 & - & ** & K1 V & $4901\substack{+47 \\ -25}$ & $4.43\substack{+0.07 \\ -0.02}$ & $-0.17\substack{+0.01 \\ -0.02}$ & \citet{Boeche2016} \\
120136 & $\tau$ Boo & ** & F6 IV & $6462\substack{+78 \\ -16}$ & $4.37\substack{+0.13 \\ -0.02}$ & $0.15\substack{+0.03 \\ -0.01}$ & \citet{Boeche2016} \\
124570 & 14 Boo & ** & F6 IV & $6109\substack{+100 \\ -100}$ & $3.85\substack{+0.10 \\ -0.10}$ & $0.07\substack{+0.10 \\ -0.10}$ & \citet{Takeda2007} \\
129333 & EK Dra & BY* & G0 V & $5700\substack{+70 \\ -70}$ & $4.37\substack{+0.10 \\ -0.10}$ & $-0.16\substack{+0.07 \\ -0.07}$ & \citet{Koenig2005} \\
131156A & $\xi$ Boo A & PM* & G8 V & $5410\substack{+0 \\ -0}$ & $4.48\substack{+0.00 \\ -0.00}$ & $-0.05\substack{+0.00 \\ -0.00}$ & \citet{Boeche2016} \\
143761 & $\rho$ CrB & PM* & G0+ Va & $5745\substack{+14 \\ -89}$ & $3.99\substack{+0.06 \\ -0.09}$ & $-0.30\substack{+0.02 \\ -0.02}$ & \citet{Boeche2016} \\
149661 & 12 Oph & BY* & K2 V & $5156\substack{+30 \\ -14}$ & $4.47\substack{+0.03 \\ -0.05}$ & $0.05\substack{+0.01 \\ -0.01}$ & \citet{Boeche2016} \\
158614 & - & SB* & G9 IV-V & $5461\substack{+35 \\ -24}$ & $4.20\substack{+0.08 \\ -0.02}$ & $-0.02\substack{+0.02 \\ -0.01}$ & \citet{Boeche2016} \\
161239 & 84 Her & PM* & G2 IIIb & $5742\substack{+15 \\ -15}$ & $3.68\substack{+0.04 \\ -0.04}$ & $0.21\substack{+0.02 \\ -0.02}$ & \citet{Niedzielski2016} \\
182572 & 31 Aql & PM* & G7 IV & $5561\substack{+60 \\ -43}$ & $4.12\substack{+0.14 \\ -0.02}$ & $0.34\substack{+0.03 \\ -0.02}$ & \citet{Boeche2016} \\
185144 & $\sigma$ Dra & PM* & K0 V & $5132\substack{+53 \\ -28}$ & $4.35\substack{+0.08 \\ -0.03}$ & $-0.26\substack{+0.02 \\ -0.01}$ & \citet{Boeche2016} \\
190007 & - & BY* & K4 V & $4541\substack{+80 \\ -43}$ & $4.26\substack{+0.24 \\ -0.02}$ & $0.11\substack{+0.02 \\ -0.05}$ & \citet{Boeche2016} \\
201091 & 61 Cyg A & BY* & K5 V & $4310\substack{+61 \\ -7}$ & $4.14\substack{+0.16 \\ -0.00}$ & $-0.36\substack{+0.01 \\ -0.02}$ & \citet{Boeche2016} \\
201092 & 61 Cyg B & Fl* & K7 V & $3945\substack{+107 \\ -9}$ & $4.03\substack{+0.23 \\ -0.10}$ & $-0.63\substack{+0.07 \\ -0.09}$ & \citet{Boeche2016} \\
206860 & HN Peg & BY* & G0 V & $5961\substack{+85 \\ -85}$ & $4.45\substack{+0.03 \\ -0.03}$ & $-0.06\substack{+0.07 \\ -0.07}$ & \citet{Ramirez2013} \\
\hline
\end{tabular}

  \caption{Basic stellar parameters of the sample taken from Simbad. Spectral 
  types have been adopted from \citet{Messina2002} and \citet{Lockwood2007}. The 
  object type reads as follows: BY* = Variable of BY~Dra~type, RS* = Variable of 
  RS~CVn~type, PM* = High proper-motion star, SB* = Spectroscopic binary, 
  ** = Double or multiple star, Fl* = Flare star.}
  \label{params}
\end{table*}

\subsection{Photometric data}
\citet{Messina2002, Messina2003} studied the long-term magnetic activity of five 
young solar analogues (HD\,1835, HD\,20630, HD\,72905, HD\,129333, HD\,206860), 
and a young K~dwarf (HD\,82443). These authors compiled photometric observations 
available from the literature (with different band passes) and new observations 
taken in Johnson V~band into one time series covering up to $\sim$16~years. 
The authors state that the typical standard deviation of the V~band magnitudes is 
equal to $0.007$~mag, which we adopt as photometric uncertainty of the measurements. 
For all stars in their sample, \citet{Messina2002, Messina2003} detected a photometric 
activity cycle. For details, we refer the reader to the above publications.

\citet{Lockwood2007} presented up to 20 years of differential Str\"omgren b and 
y photometry of 32 stars with contemporaneous observations of the Mount Wilson 
S-index (see also \citealt{Lockwood1997, Radick1998, Radick2018}) with the 
purpose of studying long-term photometric and chromospheric variability. We 
received data for 26 out of the 32 stars in their sample. Photometric 
uncertainties $\sigma_{by}$ have been adopted from Table~2 in \citet{Lockwood2007}.
In total, our sample consists of 30 stars as two stars (HD\,1835 and HD\,129333) 
are present in both photometric samples.

\subsection{Chromospheric emission data}
The Mount Wilson survey has taken chromospheric emission data from 1966-1995 for 
the majority of stars in their sample, and for 35 stars until the year 2001. 
Thereof, 24 stars belong to our sample and have been observed for more than 33 
years. The remaining six stars have shorter observing time spans between 9 to 28 
years. Recently, this unprecedented data set became publicly 
available\footnotemark[\value{footnote}]. The Mount Wilson S-index is the main 
quantity of interest describing the amount of chromospheric emission in the 
centers of the \ion{Ca}{ii} H+K lines. A definition of this quantity can be 
found in \citet{Vaughan1978}. From the year 1977 on, an observational weight, 
\textit{WT}, accounting for the photon noise is supplied. This weight can be 
converted into a relative S-index uncertainty $\sigma_S/S = WT^{-1/2}$ 
\citep{Duncan1991}. These uncertainties were used in our computations whenever 
available.

\section{Detection of activity cycle period and phase}\label{method}
\begin{figure*}
  \centering
  \includegraphics[width=17cm]{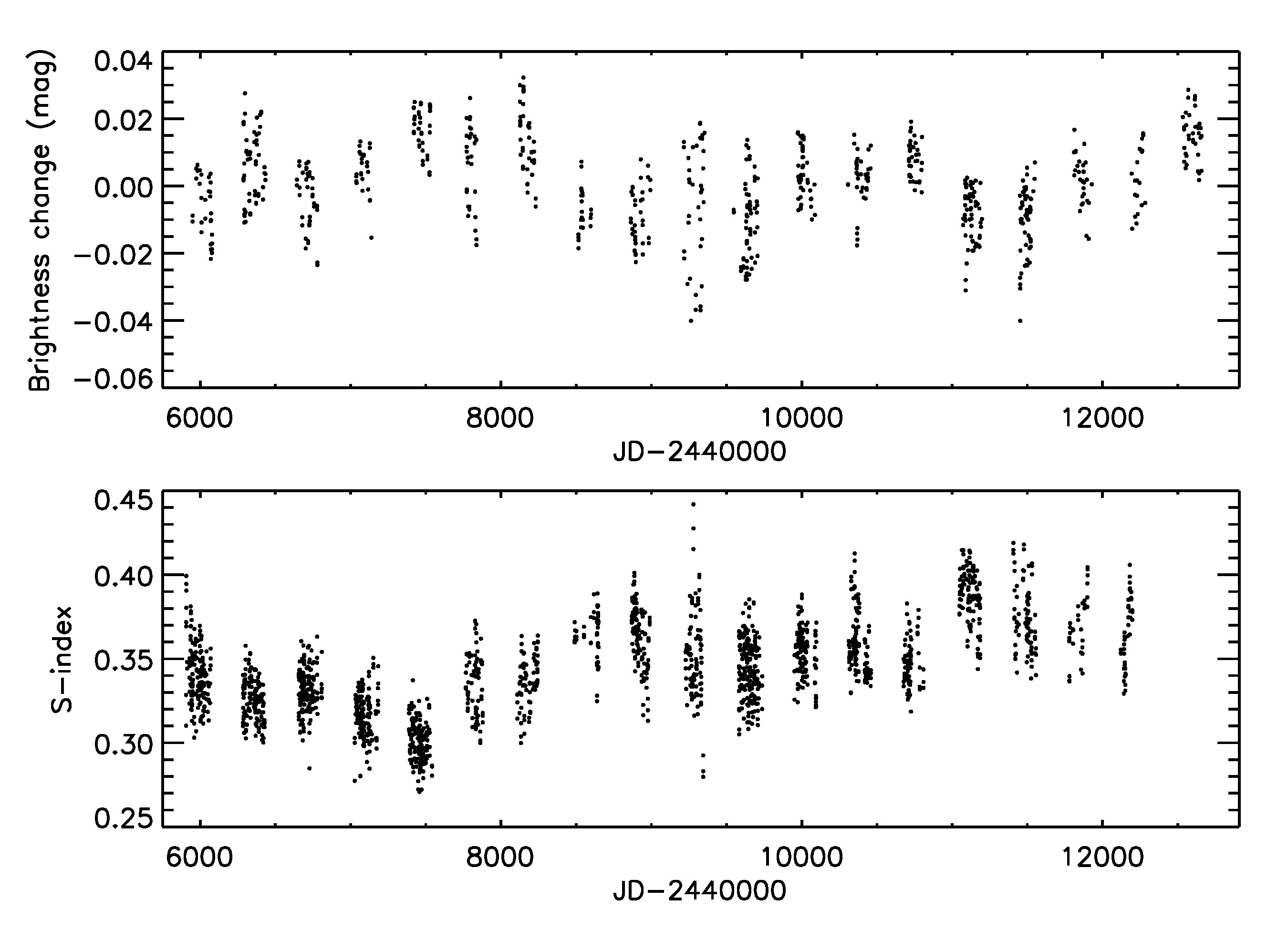}
  \caption{Photometric (top panel) and chromospheric (bottom panel) time series 
  of the star HD\,1835. The red box indicates the observing season shown in 
  Fig.~\ref{S28}.}
  \label{HD1835}
\end{figure*}

In Fig.~\ref{HD1835} we show an example of contemporaneous 
photometric (top panel) and chromospheric (bottom panel) observations. Both time 
series show short-term (seasonal) variability, partly caused by active regions 
on the stellar surface rotating in and out of view, and long-term brightness 
changes over several seasons, possibly caused by an underlying activity cycle. 
However, detecting a cycle period (and in a next step, the phase difference 
between the two time series) is challenging because the scatter in the individual 
observing seasons hampers the detection of long-term brightness changes. In the 
following we present a model capable of detecting stellar rotation signals. This 
model is applied to each observing season and subtracted from the data in order 
to enhance the visibility of the signal from the activity cycle. Afterwards we 
search for a cycle period present in both time series in order to compute the 
phase difference between the two time series.

\subsection{The Gaussian process model}
To account for the variation due to stellar rotation in each observing season, 
we apply a quasi-periodic Gaussian process\footnote{An introduction to Gaussian 
process regression can be found in \citet{Rasmussen2005}.} (GP) model that 
accounts for periodic and long-term trends in the data. 
The GP was implemented using the python package George \citep{Ambikasaran2015}.
A quasi-periodic GP model has successfully been applied to \textit{Kepler} data 
to infer stellar rotation periods \citep{Angus2018}. The underlying kernel 
function of this GP model reads
\begin{equation}
 \begin{aligned}
  \label{kernel}
  k(x_i,x_j) &= h^2 \underbrace{\exp\left(-\frac{(x_i-x_j)^2}{2l^2}\right)}_{\rm SE} 
  \cdot \underbrace{\exp\left(-\Gamma\sin^{2}\left[\frac{\pi}{P}(x_{i}-x_{j})
  \right]\right)}_{\rm ESS} \\
  &+ \sigma^2\,\delta_{ij}. 
 \end{aligned}
\end{equation}
In general, a kernel function $k(x_i,x_j)$ describes the covariance between 
arbitrary data points $x_i$ and $x_j$. The quasi-periodic kernel is composed of 
two well-established kernels, namely the squared exponential (SE) and the 
exponential sine squared (ESS) kernel. The SE kernel contains the length scale 
parameter $l$, which accounts for the covariance between observations taken at 
times $x_i$ and $x_j$. The ESS kernel contains two parameters: the period $P$ to 
account for periodicities in the data, and the parameter $\Gamma$ to control the 
harmonic behavior of the periodicity. The parameter $h$ is sometimes called the 
``amplitude'' of the covariance because it is linked to the 
variance in the time series. The second term, $\sigma^2\,\delta_{ij}$, accounts 
for the white noise in the time series, with $\delta_{ij}$ being the Kronecker 
delta. For all stars in our sample, the rotation periods have been measured 
\citep{Baliunas1996}, and we use them as initial period guesses $P$ in the GP 
prediction. Finally the full set of so-called hyper-parameters 
$\{\sigma,h,l,\Gamma,P\}$ is estimated by fitting the above model to the data 
assuming Gaussian-distributed uncertainties. 

We applied uniform priors to all parameters, with the period being 
limited to $\pm$30\% of the initial value. It is worth noting that these 
limits do not exclude possible harmonics of this period needed to fit a double-dip 
shape in the time series arising from spots on opposite sides of the star. The 
parameter $\Gamma$ in the GP accounts for the harmonic behavior of the fit, given 
an initial period. Additionally, we checked the initial values reported by 
\citet{Baliunas1996} and \citet{Messina2002}, and can confirm all rotation 
periods within $\pm$20\%.

The parameter estimation is performed using the affine-invariant ensemble 
sampler \texttt{emcee} \citep{Foreman-Mackey2013, Goodman2010}. In total we used 
32 walkers and 2000 iterations when running \texttt{emcee}. The walkers were 
initialized around the solution found by a previous Maximum Likelihood Estimation 
(MLE) step and in almost all cases this provided a starting point very close to 
the final parameters found after the MCMC. For each run we 
calculated the integrated auto-correlation time and thinned the chains accordingly. 
In addition to computing the effective sample size, we also visually inspected 
the chains and the marginal posterior distributions to ensure that they looked 
reasonable. This allows us to estimate the marginalized posterior probability 
density of the period given the other hyper-parameters. Additionally, the mean 
flux value of each observing season is predicted by the MCMC sampler.

\begin{figure}
    \resizebox{\hsize}{!}{\includegraphics{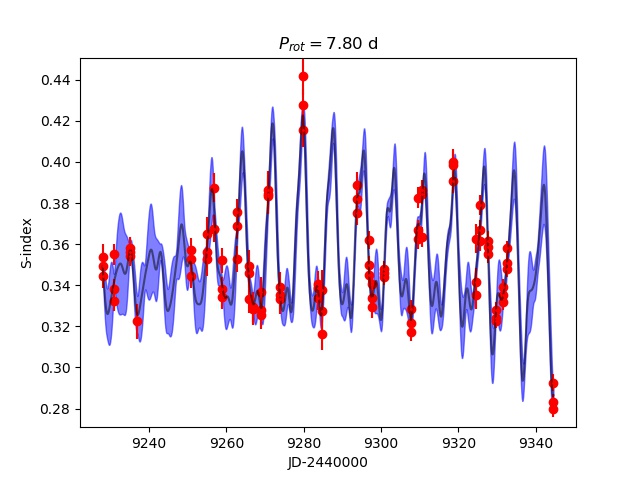}}
    \caption{Selected observing season of Mount Wilson S-index time series of 
    the star HD\,1835 (indicated by the red box in Fig.~\ref{HD1835}). The GP 
    prediction is shown as solid line with the $1\sigma$ uncertainties of the 
    prediction in blue. The GP model predicts a rotation period of 
    $\Prot=7.80$\,d.}
  \label{S28}
\end{figure}

In Fig.~\ref{S28} we present data from the observing season indicated by the red 
box in Fig.~\ref{HD1835} and the optimized GP model including $1\sigma$ 
uncertainties. The GP model predicts a rotation period of $\Prot=7.80$\,d, close 
to the literature value of $\Prot=7.756$\,d \citep{Messina2002}. The length scale 
parameter $l$ accounts for long-term trends over the course of the observing season, 
likely caused by the active regions' lifetimes.

The main purpose of using the GP model is to remove the rotational short-term 
activity rather than measuring the rotation period (a Lomb-Scargle periodogram 
or a Fourier transform would be sufficient for the latter). Whenever it was not 
possible to detect a periodic signal (either due to the absence of active 
regions or their short life time compared to the stellar rotation period), the 
GP model predicts a flat line being a fair model in quiet epochs. The GP model 
presented in Eq.~\ref{kernel} is applied to each observing season in each data 
set.

\subsection{The reduced time series}\label{reduced}
\begin{figure}
  \resizebox{\hsize}{!}{
  \includegraphics{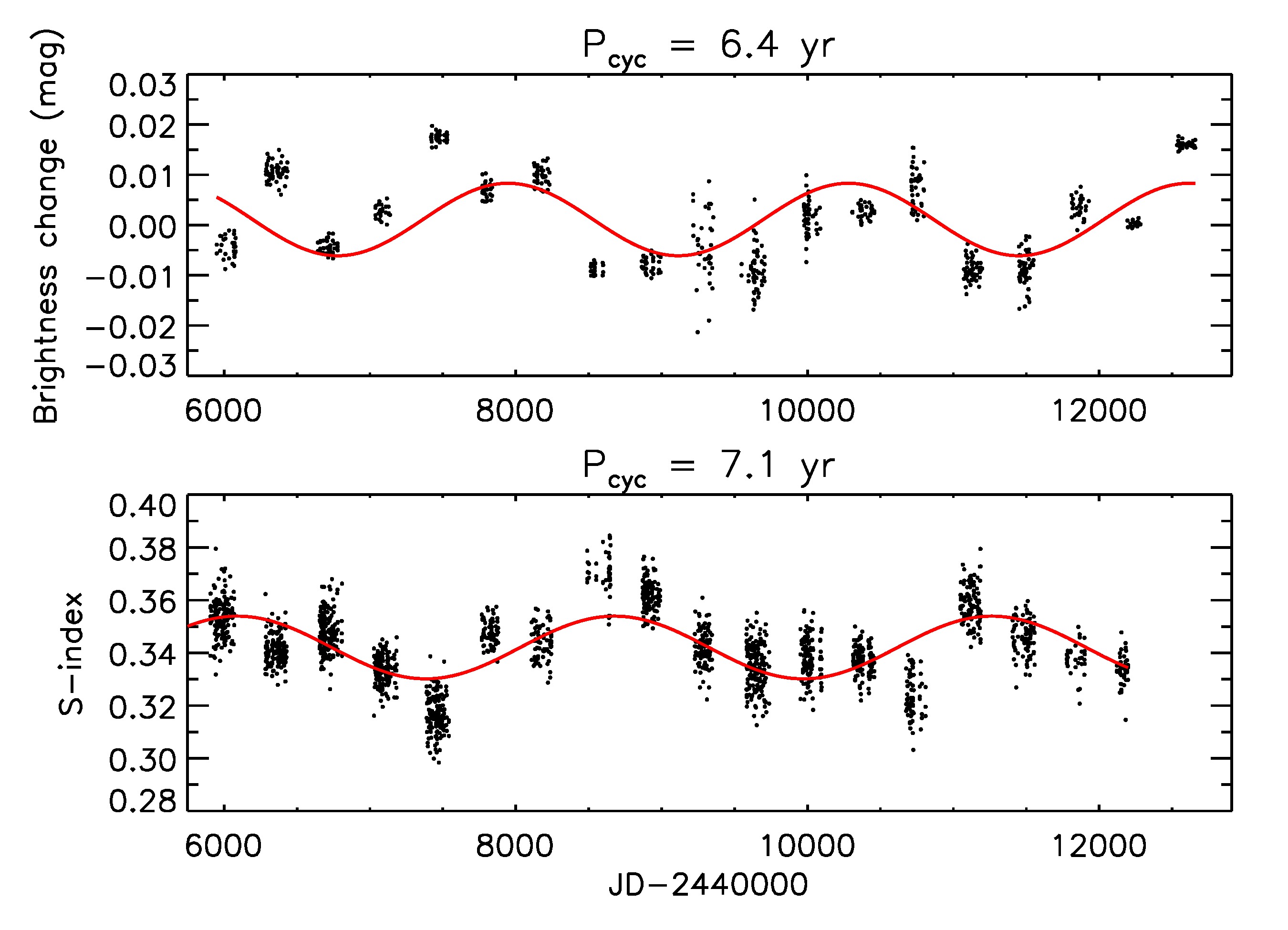}}
  \caption{Reduced photometric (top panel) and chromospheric (bottom panel) time 
  series of Fig.~\ref{HD1835}. Sine fits to the data are shown in red with their 
  cycle periods given at the top of each panel. In the lower panel a sine wave
  with a long period of $28.1$\,yr has been subtracted (see Table~\ref{Pcyc_table}).}
  \label{reduced_time_series}
\end{figure}

To remove the rotation signature from each observing season, we subtract the GP 
model from the data and add the mean flux value of the season as predicted by the 
MCMC sampler. This preserves the seasonally-averaged mean flux values and pronounces 
the signature of the activity cycle. The reduced time series of the star HD\,1835 
is shown in Fig.~\ref{reduced_time_series}. Comparing the reduced time series to 
the original data in Fig.~\ref{HD1835} shows that the scatter in each season is 
reduced significantly.
Alternatively to using all data points, the time series can be simplified 
by compressing each season to the predicted mean value, and using the standard 
deviation of the residuals (i.e. data minus GP model) as uncertainties. Since 
these uncertainties reflect the residual activity of each season, and not the 
uncertainty of the predicted mean value, we decided to use the full time series 
in the following. Note that we also searched 
for cyclic behavior of the rotation periods inferred from the GP analysis as 
done by \citet{Messina2003}. These authors interpreted (cyclic) changes of the 
rotation period in terms of a possible butterfly diagram. However, our analysis 
of the ``rotation period time series'' does not show a significant cycle period.

The reduced time series are used to search for cycle periods in the following 
way. Using Lomb-Scargle periodograms \citep{Zechmeister2009} we first search for 
periodicity in the photometric and chromospheric time series individually. In a 
second step we search for a single cycle period fitting both data sets at the same 
time (see Sect.~\ref{joint}). However, finding the correct cycle period requires 
visual inspection of each time series. The data often show long trends on the order of 
the observing time span, appearing as the highest peak in the periodogram. Whenever 
periods longer than 75\% of the observing time span are detected, we subtract a 
long sine wave from the data and compute the periodogram again. Since it cannot 
be decided whether such a trend is caused by a very long activity cycle or if it 
is an instrumental artifact, we report these long periods in Table~\ref{Pcyc_table} 
as potential long activity cycles.

Another problem arises for the Mount Wilson data before the year 1980. At these 
early epochs the data sampling is much sparser due to a different observing 
strategy. It seems that including these early observations generates a long 
trend over the full observing time span, although no systematic offset or a 
common trend could be identified for all stars. Thus, for 18 stars in our sample, 
early Mount Wilson data have been discarded in the further analysis. 
All Mount Wilson observing seasons used in the following can be found in 
appendix \ref{app_B}.

\subsection{Detection of a single cycle period and phase difference}\label{joint}
\begin{figure}
  \resizebox{\hsize}{!}{
  \includegraphics{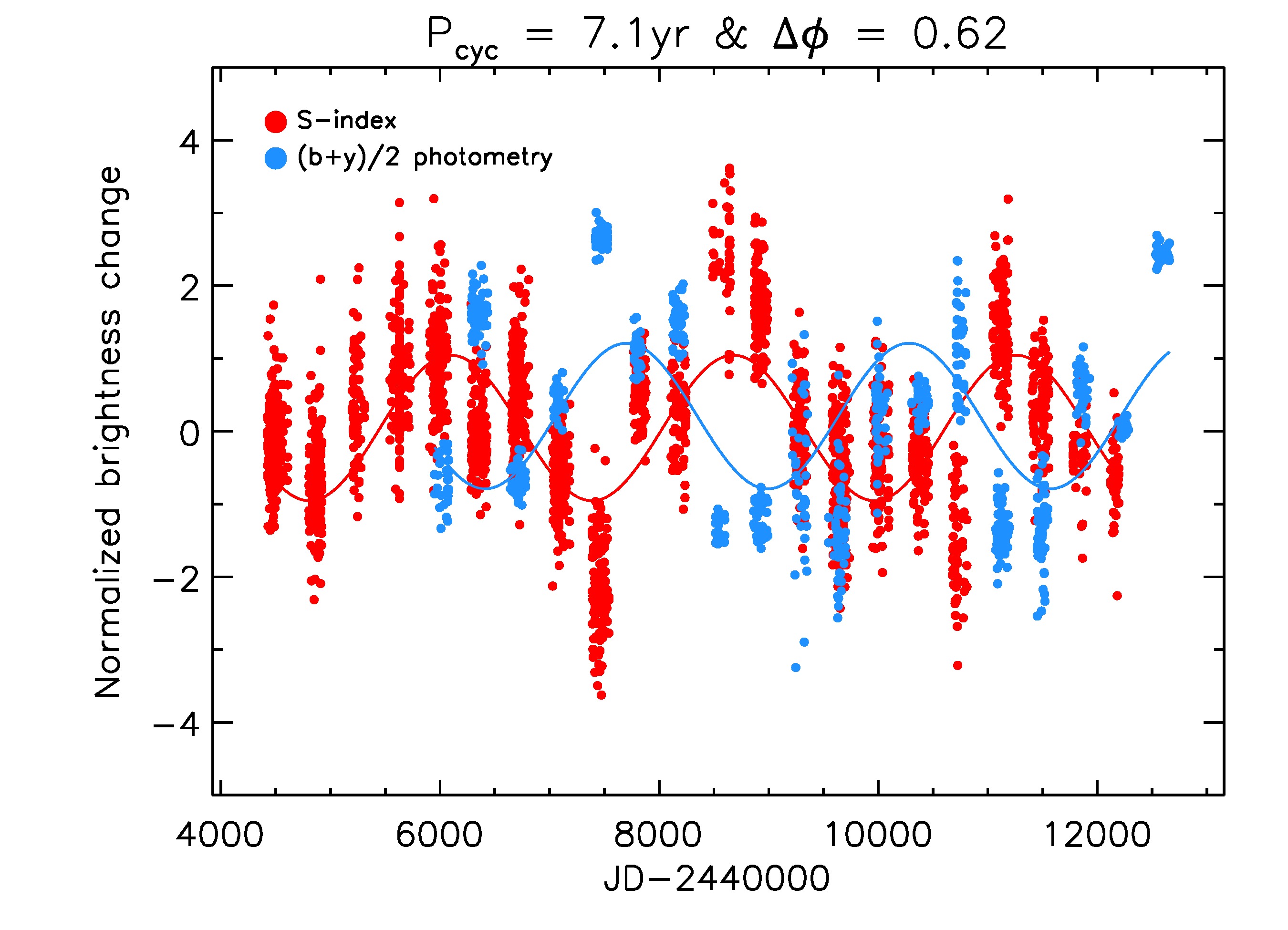}}
  \caption{Photometric (blue) and chromospheric (red) time series of the star 
  HD\,1835. The simultaneous sine fit to the data (same colors, respectively) 
  yields a cycle period of $\Pcyc=7.07\pm0.02$\,yr. Both data sets were
  normalized by their respective sine amplitudes.}
  \label{joint_period}
\end{figure}

After the subtraction of potential long activity cycles we search for periods
present in both the photometric and chromospheric time series. Computing the 
Lomb-Scargle periodogram and selecting the highest peak usually yields a good 
estimate of the true cycle period. In  Fig.~\ref{reduced_time_series} the 
Lomb-Scargle periodogram returns a cycle period of $\Pcyc=6.4$\,yr for the 
photometric time series, and a cycle period of $\Pcyc=7.1$\,yr for the 
chromospheric time series. To compute the phase difference between the two time 
series one has to fit the same period in both data sets simultaneously.
Thus, a sine fit was applied to fit both time series at the same time using a 
single period, with individual amplitudes, phases and offsets for the two data 
sets. As an initial guess of the true cycle period, we use the mean of the cycle 
periods inferred from the individual time series. The result of the fit to both 
data sets of HD\,1835 simultaneously is shown in Fig.~\ref{joint_period}. To 
combine the two different types of data into a dimensionless quantity the time 
series have been normalized by their sine amplitudes. The simultaneous sine fit 
returns a single cycle period of $\Pcyc=7.07\pm0.02$\,yr. From this fit we can 
compute the phase difference between the two time series to check for possible 
phase shifts. These results are presented in Sect.~\ref{phase_diff}. The sine fit 
was performed using the Levenberg-Marquardt least-squares minimization method. 
We also tried fitting the sine wave using an MCMC approach to retrieve more 
realistic uncertainties. This approach did not converge unless we restricted the 
priors of the parameters to a very small region around the best values returned 
by the least-squares fit. Given the nature of the model, where interpretability 
in terms of phase differences is preferred over being optimal in terms of model 
structure (i.e., more activity cycle-based), the application of MCMC was not 
ideal and a simpler least-squares approach was favored. Thus, we report the 
formal uncertainties of the fit parameters, which should be considered as lower 
limits. We did also try modeling the 
cycle behavior with the same GP model that was used for the rotation, and use 
cross correlation to determine the relative phases; however, this was found to 
be unreliable. A reason that we did not try to add a phase difference into the 
kernel of the GP is that understanding the behavior of the resulting covariance 
matrix is not straight forward.

\section{Results}\label{results}
\subsection{Cycle periods}
Cycle periods could be detected for 27 stars in total. These periods are given 
in Table~\ref{Pcyc_table}. Following the cycle classification as introduced by 
\citet{Baliunas1995}, we provide a quality flag for each cycle period that has 
been assigned by visual inspection of both the photometric and the chromospheric 
time series individually. A higher number indicates stronger periodicity of the 
data, with integer values between 1 (weak periodicity), and 3 (excellent periodicity).
Two example light curves are shown in appendix \ref{app_C}.
We identified a subset of eight stars exhibiting an excellent chromospheric activity 
cycle. We return to these stars in Sect.~\ref{phase_diff} and Sect.~\ref{discussion}.
To quantify the goodness of the periodicity we computed the reduced chi-square 
values of the fit to each data set individually ($\chi^2_{\rm phot}$ and 
$\chi^2_{\rm chrom}$), and also to both data sets simultaneously ($\chi^2_{\rm sim}$).
The values are reported in the last three columns of Table~\ref{Pcyc_table}.
However, we found that these values do not correlate with the periodicity of the 
data. This is owing to the fact that both the period and the amplitude of the 
activity cycle can vary from cycle to cycle. For instance, the varying cycle 
amplitude of the star HD\,201091 in the Mount Wilson time series creates quite 
a high value for $\chi^2_{\rm chrom}$, although the periodicity is evident in 
this star ($\rm flag_{chrom} = 3$). Moreover, the shape of the activity cycle 
does not need to be sinusoidal (as assumed by the sine fit), even when the 
period and amplitude are constant over time.
\longtab{
  \begin{landscape}
  \begin{longtable}{ccccccccccccccc}
\hline\hline
HD & $B-V$ & $\log R'_{\rm HK}$ & $P_{\rm rot}$ & $P_{\rm cyc}$ & $P_{\rm cyc,\,phot}^{\rm long}$ & $P_{\rm cyc,\,chrom}^{\rm long}$ & $\rm amp_{phot}$ & $\rm amp_{chrom}$ & $\Delta \phi$ & $\rm flag_{phot}$ & $\rm flag_{chrom}$ & $\chi^2_{\rm phot}$ & $\chi^2_{\rm chrom}$ & $\chi^2_{\rm sim}$ \\
 &  &  &  (days) & (yr) & (yr) & (yr) & $(*10^{-3}\,\rm mag)$ & $(*10^{-3})$ &  &  &  &  &  &  \\
\hline
1835 & 0.66 & -4.445 & 7.76 & $7.07 \pm 0.02$ & - & 28.1 & $6.56 \pm 0.45$ & $11.90 \pm 0.15$ & $0.62 \pm 0.01$ & 2 & 2 & 0.72 & 3.71 & 3.06 \\
10476 & 0.84 & -4.938 & 35.00 & $10.33 \pm 0.02$ & - & - & $1.23 \pm 0.07$ & $13.98 \pm 0.08$ & $-0.04 \pm 0.01$ & 2 & 3 & 1.01 & 9.82 & 7.27 \\
13421 & 0.56 & -5.217 & 17.00 & $7.71 \pm 0.05$ & - & - & $0.66 \pm 0.04$ & $1.20 \pm 0.04$ & $0.17 \pm 0.01$ & 1 & 1 & 4.26 & 2.12 & 2.48 \\
18256 & 0.43 & -4.758 & 3.00 & $7.65 \pm 0.02$ & - & - & $0.96 \pm 0.08$ & $7.39 \pm 0.07$ & $0.04 \pm 0.01$ & 1 & 1-2 & 1.31 & 10.44 & 7.82 \\
20630 & 0.68 & -4.420 & 9.21 & $6.03 \pm 0.01$ & - & - & $14.12 \pm 0.50$ & $14.50 \pm 0.12$ & $0.54 \pm 0.01$ & 3 & 2-3 & 0.35 & 6.09 & 5.17 \\
25998 & 0.46 & -4.489 & 2.00 & $4.51 \pm 0.01$ & - & 65.5 & $2.84 \pm 0.29$ & $4.54 \pm 0.11$ & $0.55 \pm 0.01$ & 3 & 2 & 0.97 & 4.12 & 3.80 \\
35296 & 0.53 & -4.438 & 4.00 & $4.16 \pm 0.01$ & - & 26.9 & $1.87 \pm 0.32$ & $4.54 \pm 0.06$ & $0.51 \pm 0.03$ & 3 & 1-2 & 0.30 & 6.60 & 5.76 \\
39587 & 0.59 & -4.460 & 5.00 & $6.24 \pm 0.01$ & - & 16.7 & $2.52 \pm 0.47$ & $5.28 \pm 0.05$ & $0.54 \pm 0.03$ & 2 & 1-2 & 0.55 & 9.02 & 8.27 \\
72905 & 0.62 & -4.375 & 4.91 & $4.22 \pm 0.03$ & - & - & $4.25 \pm 0.33$ & $6.94 \pm 0.24$ & $0.71 \pm 0.02$ & 2-3 & 2 & 0.62 & 2.46 & 1.48 \\
75332 & 0.49 & -4.474 & 4.00 & $10.91 \pm 0.13$ & - & 721.7 & $5.41 \pm 0.29$ & $2.65 \pm 0.12$ & $0.48 \pm 0.01$ & 1 & 1 & 0.35 & 5.02 & 3.82 \\
81809 & 0.64 & -4.927 & 41.00 & $8.19 \pm 0.01$ & - & - & $0.63 \pm 0.06$ & $10.07 \pm 0.08$ & $0.08 \pm 0.01$ & 1 & 3 & 2.65 & 2.79 & 2.75 \\
82443 & 0.77 & -4.211 & 5.42 & $2.69 \pm 0.01$ & - & 11.3 & $6.49 \pm 0.27$ & $35.83 \pm 0.46$ & $0.57 \pm 0.01$ & 3 & 2 & 0.70 & 3.90 & 2.52 \\
82885 & 0.77 & -4.674 & 18.00 & $9.29 \pm 0.01$ & - & 30.5 & $2.42 \pm 0.21$ & $15.93 \pm 0.12$ & $0.40 \pm 0.01$ & 1 & 1-2 & 0.61 & 11.20 & 7.19 \\
103095 & 0.75 & -4.899 & 31.00 & $7.20 \pm 0.01$ & - & - & $0.69 \pm 0.05$ & $12.68 \pm 0.11$ & $-0.10 \pm 0.01$ & 1 & 3 & 1.35 & 2.38 & 2.15 \\
115383 & 0.59 & -4.486 & 3.00 & $6.51 \pm 0.03$ & - & 29.4 & $4.15 \pm 0.23$ & $3.32 \pm 0.13$ & $0.34 \pm 0.01$ & 2 & 2-3 & 0.49 & 6.77 & 4.20 \\
115404 & 0.94 & -4.529 & 18.00 & $10.31 \pm 0.03$ & 70.2 & - & $2.37 \pm 0.33$ & $32.09 \pm 0.27$ & $0.34 \pm 0.02$ & 2 & 1-2 & 0.13 & 5.83 & 4.08 \\
120136 & 0.48 & -4.742 & 4.00 & $5.60 \pm 0.01$ & 15.9 & - & $1.30 \pm 0.13$ & $2.46 \pm 0.03$ & $0.03 \pm 0.02$ & 1-2 & 1-2 & 0.29 & 4.77 & 3.74 \\
124570 & 0.54 & -5.156 & 26.00 & - & - & - & - & - & - & - & - & - & - & - \\
129333 & 0.61 & -4.148 & 2.68 & $10.22 \pm 0.38$ & 547.0 & 15.9 & $14.89 \pm 1.78$ & $1.33 \pm 0.49$ & $0.33 \pm 0.05$ & 3 & 2 & 0.15 & 1.76 & 1.11 \\
131156A & 0.76 & -4.387 & 6.00 & $3.54 \pm 0.00$ & - & - & $6.87 \pm 0.70$ & $16.93 \pm 0.08$ & $0.53 \pm 0.02$ & 1-2 & 1 & 0.47 & 40.04 & 36.67 \\
143761 & 0.60 & -5.046 & 17.00 & - & - & - & - & - & - & - & - & - & - & - \\
149661 & 0.82 & -4.613 & 21.00 & $11.60 \pm 0.02$ & 40.2 & - & $3.26 \pm 0.36$ & $26.83 \pm 0.11$ & $0.47 \pm 0.02$ & 1-2 & 2 & 0.54 & 16.15 & 13.82 \\
158614 & 0.72 & -5.023 & 34.00 & $11.00 \pm 0.10$ & 253.6 & - & $1.07 \pm 0.15$ & $2.74 \pm 0.10$ & $0.23 \pm 0.02$ & 2 & 3 & 1.08 & 2.16 & 1.99 \\
161239 & 0.65 & -5.180 & 29.00 & $5.03 \pm 0.01$ & - & - & $1.33 \pm 0.11$ & $4.23 \pm 0.06$ & $-0.04 \pm 0.01$ & 2 & 3 & 2.30 & 2.53 & 2.51 \\
182572 & 0.77 & -5.093 & 41.00 & $12.51 \pm 0.07$ & - & - & $0.86 \pm 0.05$ & $3.35 \pm 0.07$ & $0.17 \pm 0.01$ & 2-3 & 3 & 1.97 & 8.51 & 6.73 \\
185144 & 0.79 & -4.823 & 27.00 & $6.55 \pm 0.01$ & 19.1 & - & $0.63 \pm 0.11$ & $21.43 \pm 0.09$ & $-0.06 \pm 0.03$ & 1 & 3 & 0.80 & 21.44 & 15.58 \\
190007 & 1.14 & -4.711 & 29.00 & $8.98 \pm 0.03$ & - & 38.0 & $4.36 \pm 0.28$ & $39.17 \pm 0.53$ & $0.48 \pm 0.01$ & 3 & 2 & 0.60 & 6.05 & 4.61 \\
201091 & 1.18 & -4.765 & 35.00 & $7.13 \pm 0.00$ & - & - & $0.58 \pm 0.13$ & $72.88 \pm 0.18$ & $0.00 \pm 0.04$ & 2-3 & 3 & 0.78 & 20.92 & 17.78 \\
201092 & 1.37 & -4.910 & 38.00 & - & - & - & - & - & - & - & - & - & - & - \\
206860 & 0.59 & -4.416 & 4.84 & $5.36 \pm 0.02$ & - & - & $4.44 \pm 0.49$ & $6.37 \pm 0.11$ & $0.41 \pm 0.02$ & 2 & 2 & 0.46 & 5.18 & 4.66 \\
\hline
\caption{\label{Pcyc_table} Physical parameters of the sample. $B-V$ colors and 
activity levels $\logRHK$ have been adopted from \citet{Lockwood2007} and 
\citet{Baliunas1996}. Rotation periods have been taken from \citet{Baliunas1996} 
and \citet{Messina2002}. Cycle periods $\Pcyc$ are the best sine fit periods to 
both data sets simultaneously. Possible long periods subtracted from the photometric 
and chromospheric time series are given by $P_{\rm cyc,\,phot}^{\rm long}$ and 
$P_{\rm cyc,\,chrom}^{\rm long}$, respectively. The amplitudes to the photometric 
and the chromospheric time series are given by $\rm amp_{phot}$ and $\rm amp_{chrom}$,
respectively. The phase difference between the two time series is indicated with 
$\Delta\phi$. The values for $\rm flag_{phot}$ and $\rm flag_{chrom}$ provide 
quality flags for the goodness of the periodicity in the respective time series. 
The values are in place of 1 = weak periodicity, 2 = moderate to good periodicity, 
3 = excellent periodicity. The last three columns contain the reduced chi-square values 
of the fit to the photometric, chromospheric, and combined time series, respectively.}
\end{longtable}

  \end{landscape}
}

\begin{figure*}
  \centering
  \includegraphics[width=17cm]{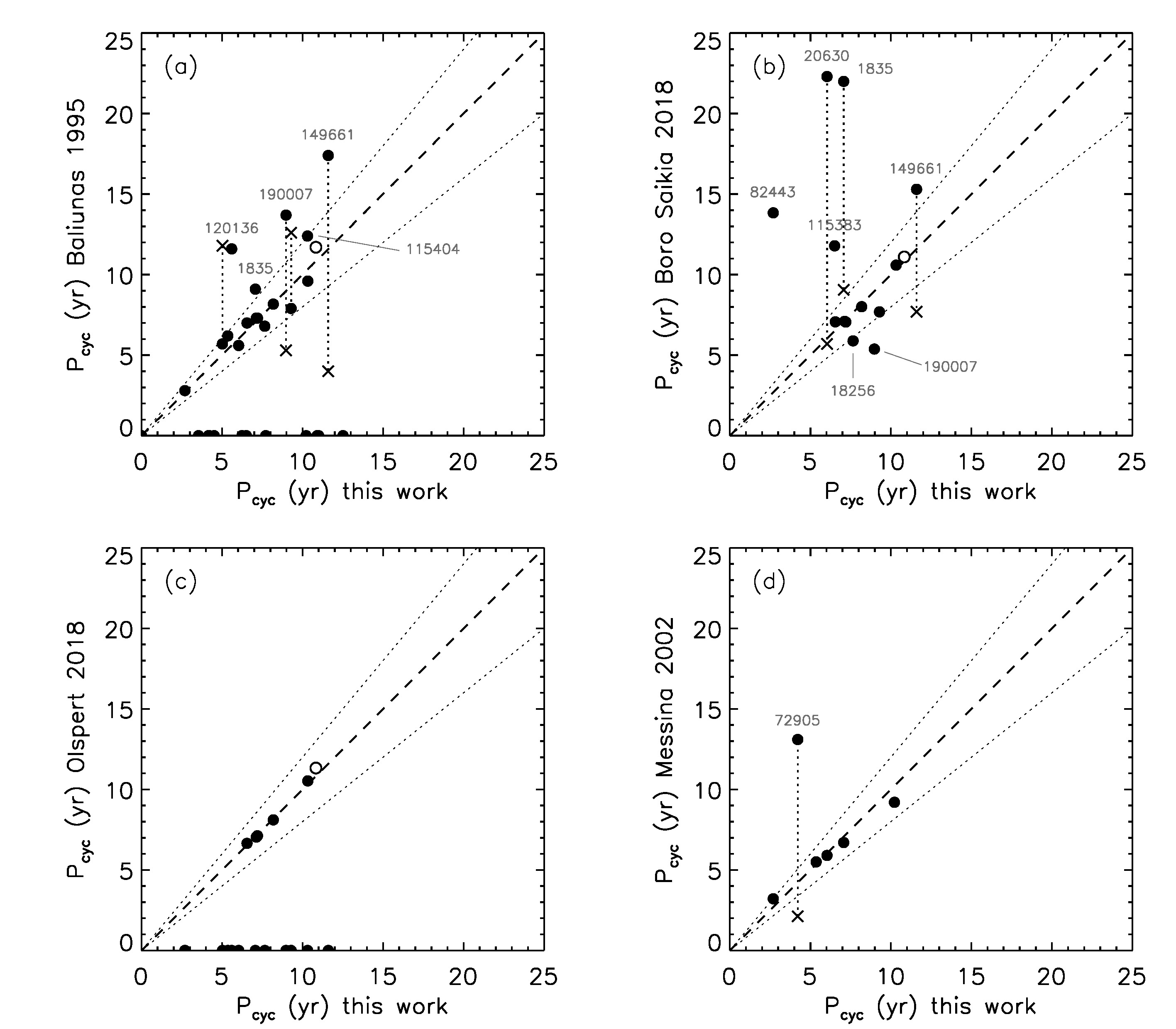}
  \caption{Comparison of our cycle period measurements to results from (a) 
  \citet{Baliunas1995}, (b) \citet{Saikia2018}, (c) \citet{Olspert2017_II_arxiv}, 
  and (d) \citet{Messina2002}. The open circle indicates the star HD\,201092 for 
  which only a chromospheric cycle period was measured. Crosses indicate secondary 
  periods connected by vertical dotted lines to the primary periods. The dashed 
  line shows the 1:1 identity, and the dotted lines to either side mark the 
  $\pm20$\% range. We set the cycle period to zero for stars in common for which 
  no cycle period was reported.}
  \label{compare}
\end{figure*}

We compare our cycle periods to measurements from the literature in Fig.~\ref{compare}. 
Panel (a) shows the measurements by \citet{Baliunas1995}, that contain all the 
30 stars from our sample. These authors found periodicity in 17 out of the 30 stars. 
We also detect periodicity in these 17 stars, although no periodicity could be 
detected in the photometric time series of the star HD\,201092. In panel (a)-(c) 
this star is shown as open circle to indicate that the periodicity has only been 
measured in the chromospheric time series. Taking advantage of up to ten years of 
additional Mount Wilson data, we detect periodicity in 11 stars where \citet{Baliunas1995} 
did not report a cycle. Comparing the period measurements to one another generally 
shows good agreement within $\pm20$\% (dotted lines), although some outliers are apparent. 
In HD\,120136, \citet{Baliunas1995} detected a ``poor'' cycle of $\sim$12\,yr, which 
can be disproved by considering the additional data. In HD\,149661, these authors 
detected a ``good'' cycle of $\sim$17.4\,yr. Again we cannot confirm this long 
period, which might be owed to the sparse data of the early observations. For the 
star HD\,190007, \citet{Baliunas1995} detected a ``fair'' $\sim$13.7\,yr cycle. 
This period might be a modulation of the potentially very long cycle of $\sim$38\,yr 
that has been subtracted in our analysis. The two stars HD\,1835 and HD\,115404 
lie close to the upper 20\% limit. The periods determined by \citet{Baliunas1995} 
are likely affected by trends in the early observations.

At first glance, the measurements by \citet{Saikia2018} in panel (b) of 
the 14 stars that we have in common seem to show stronger deviations to our 
results. The overall bigger deviations might arise from the fact that these 
authors also considered the much sparser data of the early seasons in their 
analysis. For HD\,1835, HD\,82443, and HD\,115383, long periods have been 
subtracted in our analysis, which explains the large deviation. HD\,20630 
shows a long trend in the early data, which is not visible anymore in the later 
seasons. Our cycle periods for HD\,1835 and HD\,20630 have been identified as 
secondary cycles in \citet{Saikia2018}. For HD\,149661, \citet{Saikia2018}
claimed a primary cycle of 15.3\,yr and a secondary cycle of 7.7\,yr, respectively. 
Our analysis does not confirm any of these periods, not even when considering 
all available ``raw'' data. Furthermore, we cannot reproduce the period claimed 
for HD\,18256, although the deviation to our results is smaller in this case.
For HD\,190007, \citet{Saikia2018} claimed a cycle period of 5.4\,yr. This 
period is visible in the computed periodograms with a smaller peak height. 
Thus, we expect this period to be a modulation of the longer one. Although most 
deviations can be explained by a different data analysis, some period detections 
remain mysterious.

A comparison of our results to those by \citet{Olspert2017_II_arxiv} in panel (c) 
shows remarkably good agreement. In total, 17 stars are contained in both samples. 
All six stars for which both authors detected periodicity almost lie on the 1:1 
line. As mentioned above, for the star HD\,201092 we do not detect a photometric 
cycle period. All cycle period detections in panel (a)-(c) are based solely on 
the Mount Wilson data set. In contrast to \citet{Baliunas1995} and 
\citet{Saikia2018}, \citet{Olspert2017_II_arxiv} use a similar approach 
as described here based on GP models. Additionally, these authors account for 
long-timescale trends in the time series, which explains the overall good agreement 
to our results.

Finally, we compare our measurements to the cycle periods detected by 
\citet{Messina2002} in panel (d). In contrast to the measurements in panel (a)-(c), 
these periods are solely based on photometric data. We find good agreement to our 
results for five of the six stars within the $\pm20$\% range. For HD\,72905, 
two different photometric data sets have been combined in the year 1993 (compare 
Table~5 and Fig.~6 in \citealt{Messina2002}). Comparing the photometric and the 
chromospheric time series of this star raises the question whether this combination 
may have introduced an offset between the different photometric data sets because 
no such offset is seen in the chromospheric time series. We calculated the median 
brightness of the data before and after the year 1993, and added the difference 
to the data taken after the year 1993. This removes the long period found by 
\citet{Messina2002}, and yields a cycle period of $\sim$4.2\,yr, which almost 
equals twice the period reported by these authors.

\subsection{Activity cycle phase difference}\label{phase_diff}
\begin{figure}
  \resizebox{\hsize}{!}{\includegraphics{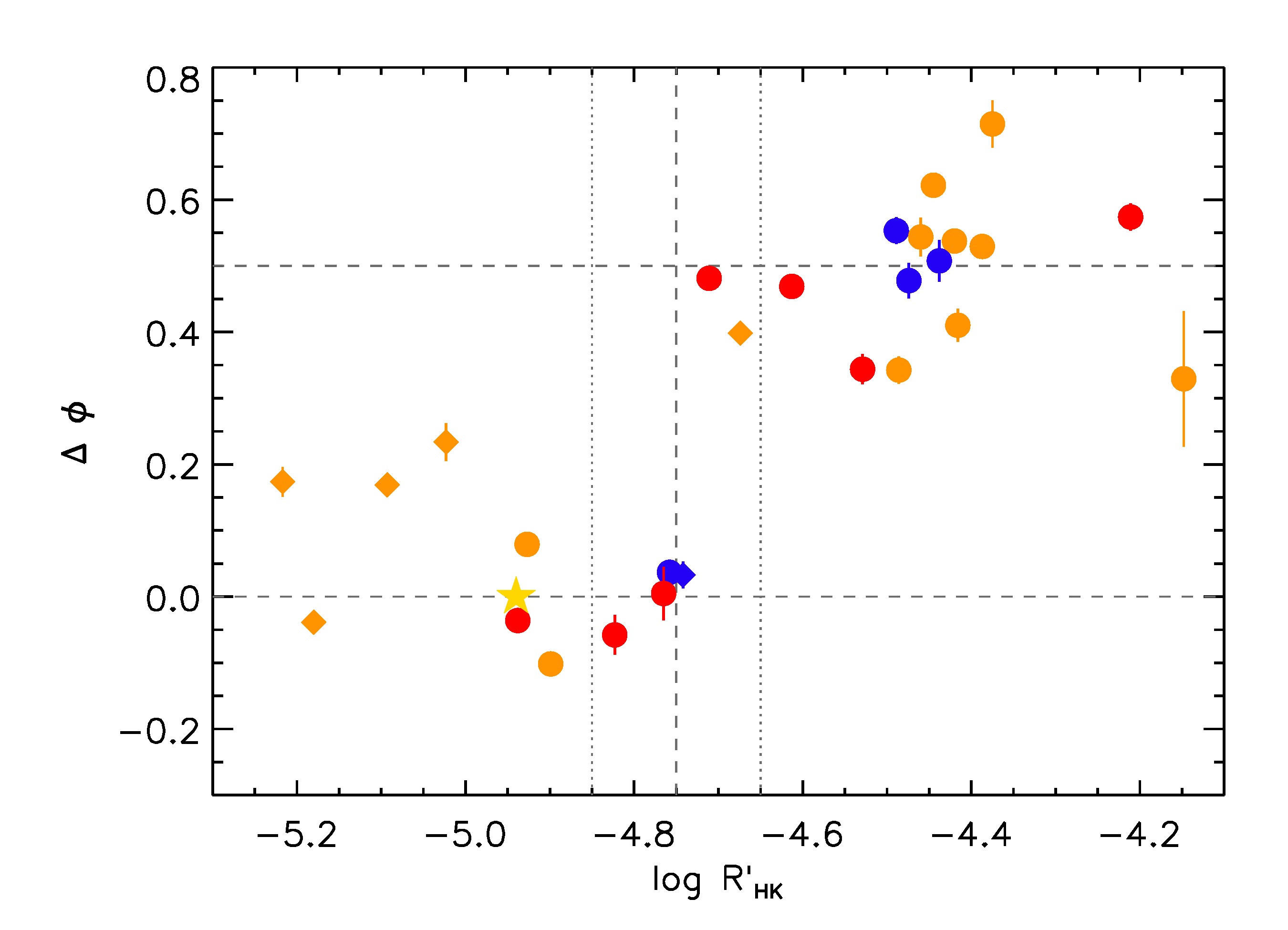}}
  \caption{Activity cycle phase difference $\Delta\phi$ against the magnetic 
  activity index $\logRHK$. Dots show dwarf stars (luminosity class V) and 
  diamonds show (sub-)giants (luminosity class III, IV, IV-V). The colors show 
  different spectral types: F~stars (blue), G~stars (orange), and K~stars (red).
  The vertical dashed line indicates the approximate location of the 
  Vaughan-Preston gap, with its boundaries indicated by the dotted lines, 
  adopted from Fig.~2 of \citet{Noyes1984}. The horizontal dashed lines at 
  $\Delta\phi=0.5$ and $\Delta\phi=0$ mark the region of spot and faculae 
  domination, respectively. The yellow star indicates the position of the Sun, 
  adopting the mean value $\log R'_{\rm HK, \,Sun} = -4.94$ from \citet{Egeland2017}.}
  \label{phase_diff_activity}
\end{figure}

The phase difference $\Delta\phi$ between the fits to the photometric and chromospheric 
time series reveals the dominant source of activity on the stellar surface, e.g., 
dark spots or bright faculae regions. A phase difference of $\Delta\phi \approx 0$ 
indicates that the star is brightest at activity maximum. This is the case in 
the Sun. The opposite case of minimum brightness during activity maximum, i.e. 
$\Delta\phi \approx 0.5$, demonstrates that the star is dominated by dark spots. 
In Fig.~\ref{phase_diff_activity} we show the phase difference $\Delta\phi$ 
against the mean stellar magnetic activity level $\logRHK$.

The maximum phase difference between two time series equals $|\Delta\phi|=0.5$. 
In Fig.~\ref{phase_diff_activity} we plot $\Delta\phi$ between $-0.25$ and 
$0.75$ to allow for a scatter around $\Delta\phi=0$ and $\Delta\phi=0.5$. We 
find two distinct groups of stars: one scattering around $\Delta\phi=0.5$ and the 
other scattering around $\Delta\phi=0$. The two groups are clearly separated 
by their mean activity level $\logRHK$. The transition from $\Delta\phi=0.5$ to 
$\Delta\phi=0$ occurs around $\logRHK \approx -4.75$, the location of the 
Vaughan-Preston gap (\citealt{VP1980,Noyes1984}, hereafter VP~gap). This clear
separation tells us that active stars (above the VP~gap) are spot-dominated
whereas inactive stars (below the VP~gap) are faculae-dominated.

\begin{figure}
  \resizebox{\hsize}{!}{\includegraphics{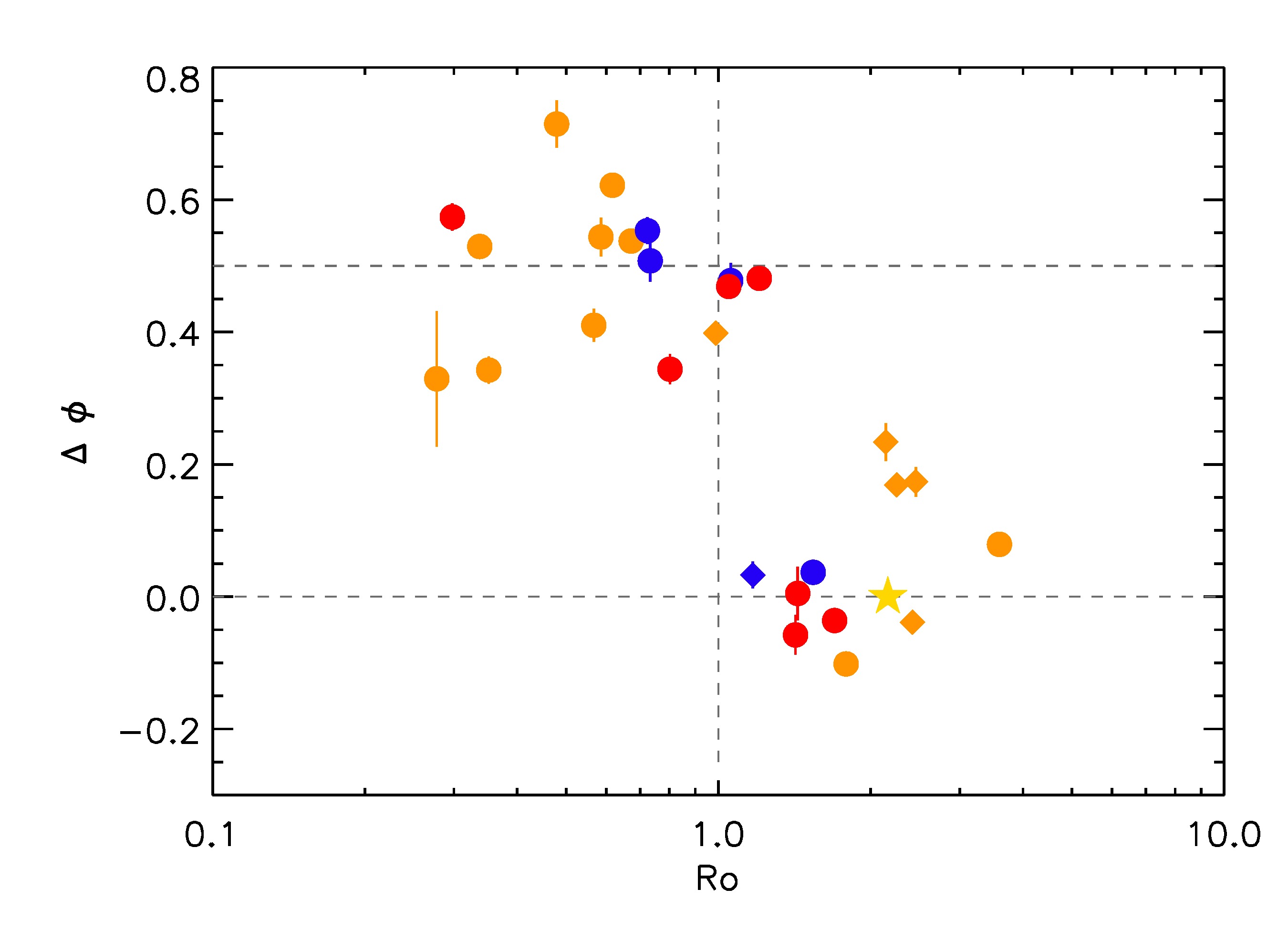}}
  \caption{Activity cycle phase difference $\Delta\phi$ against the Rossby number 
  $\rm{Ro} = \Prot/\tau$. Colors, symbols, and horizontal dashed lines have the same 
  meaning as in Fig.~\ref{phase_diff_activity}. The vertical dashed line indicates 
  the transition around $\rm{Ro}=1$. The yellow star indicates the position 
  of the Sun, $\rm{Ro_{Sun}=2.16}$, adopting the values $B-V_{\rm Sun} = 0.653$ 
  \citep{Ramirez2012}, and the synodic rotation period $P_{\rm rot, \,Sun} = 26.3$\,d.}
  \label{phase_diff_Rossby}
\end{figure}

As stars age they spin down because angular momentum is lost due to stellar 
winds. This spin-down is accompanied by a decrease of the magnetic activity
\citep{Skumanich1972}. Hence, we expect all stars to transition from spot-dominated 
to faculae-dominated surface activity, depending on their activity level and age. 
Taking advantage of the close connection between activity and rotation 
(Fig.~8 in \citealt{Noyes1984}), we show the phase difference versus
the Rossby number (instead of the activity level on the abscissa) in 
Fig.~\ref{phase_diff_Rossby}.
The Rossby number $\rm{Ro} = \Prot/\tau$ equals the rotation period divided by the 
convective turnover time $\tau$. The latter has been computed following 
\citet{Noyes1984}. As in Fig.~\ref{phase_diff_activity}, the separation into a 
spot-dominated and a faculae-dominated group is evident with the active stars
at $\rm{Ro} \lesssim 1$, and the inactive stars at $\rm{Ro} \gtrsim 1$. Below 
$\rm{Ro} = 1$ rotation is the driving source for surface activity; above 
$\rm{Ro} = 1$ stars rotate slower, and convection plays a more important role. 
The advantage of the Rossby number over the rotation periods is that the 
convective turnover time accounts for the mass dependence of the rotation period.

\begin{figure}
  \resizebox{\hsize}{!}{\includegraphics{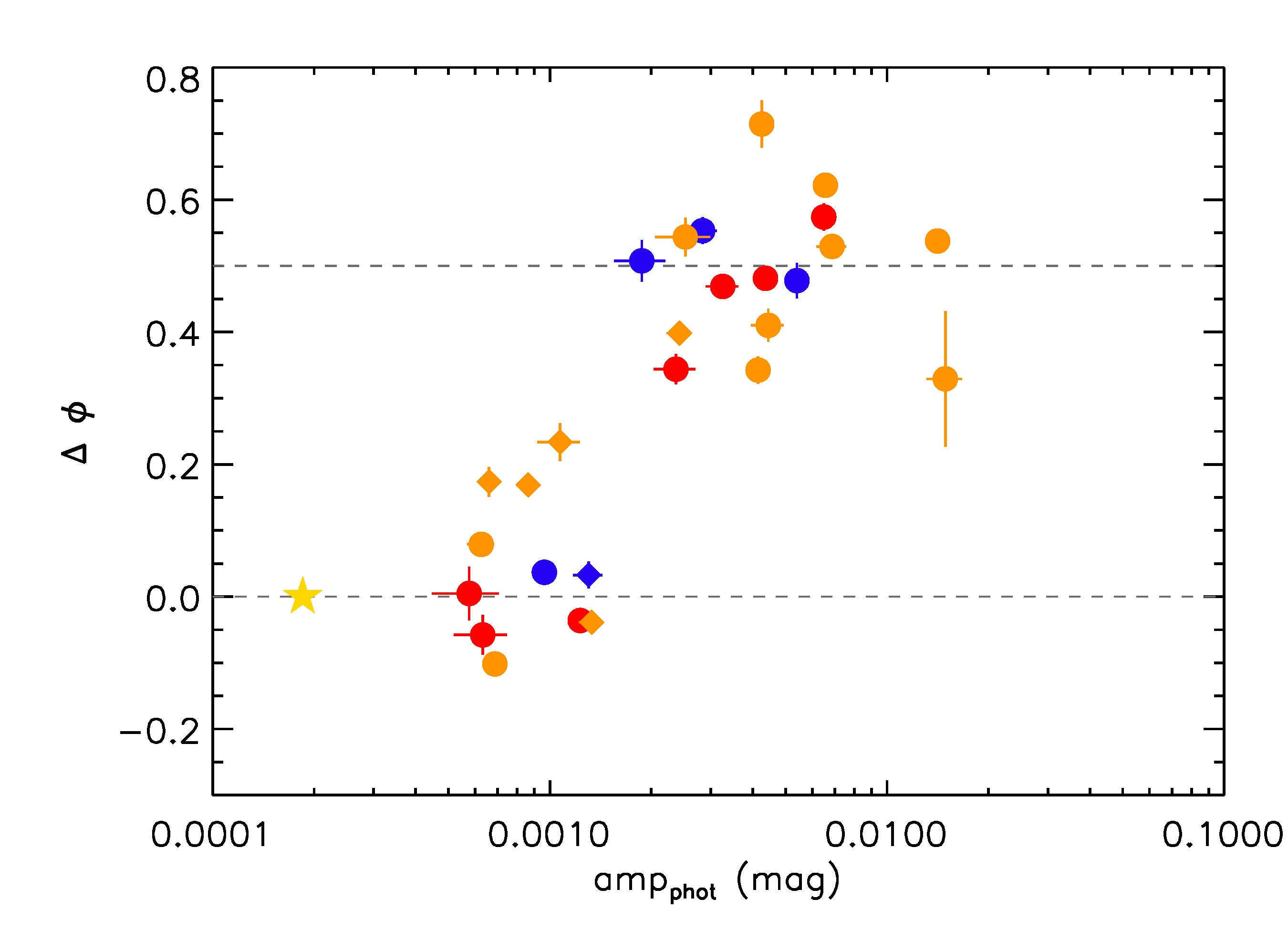}}
  \caption{Activity cycle phase difference $\Delta\phi$ against the photometric 
  sine amplitude. Colors and symbols have the same meaning as in 
  Fig.~\ref{phase_diff_activity}. 
  The photometric amplitude of the solar cycle (yellow star) was adopted 
  from \citet{Shapiro2016}. These authors calculated a total amplitude of 0.37\,mmag, 
  which results in a sine amplitude of $\rm amp_{phot,\,Sun}=0.185$\,mmag.}
  \label{phase_diff_amp_phot}
\end{figure}

The presence of two distinct groups in Fig.~\ref{phase_diff_amp_phot} provides 
further evidence for our interpretation that stars around $\Delta\phi=0$ are 
faculae-dominated and stars around $\Delta\phi=0.5$ are spot-dominated. The 
group of spot-dominated stars has on average six-times-larger photometric 
amplitudes than the faculae-dominated group. 
This result is consistent with the work of \citet{Shapiro2014}, who showed that 
with the increase of magnetic activity the spot contribution to the photometric 
variability increases faster than the faculae contribution. Consequently, less 
active stars with smaller photometric variability are faculae-dominated, while 
more active stars with larger photometric variability are spot-dominated. The 
total photometric amplitude of 0.37\,mmag for the solar cycle has been computed 
by \citet{Shapiro2016}. Such a small value was explained in \citet{Witzke2018} 
by the incident combination of solar fundamental parameters and the spectral 
location of the Str\"omgren b and y pass bands.
Furthermore, in Table~\ref{Pcyc_table} eight of the 27 stars with measured cycle 
periods exhibit a chromospheric activity cycle classified as excellent 
($\rm flag_{chrom}=3$). Concurrently, their photometric amplitudes are shallow. 
All these stars belong to the group around $\Delta\phi=0$, i.e., they are 
faculae-dominated. 

\section{Discussion}\label{discussion}
Taking advantage of contemporaneous photometric and chromospheric time series, 
we have determined activity cycle periods and the corresponding phase differences 
for 27 stars in our sample. 
For two stars in our sample, no activity cycle could be detected. HD\,124570
and HD\,143761 both are very inactive and show almost flat variability both in 
the photometric and the chromospheric time series. HD\,201092 shows an activity 
cycle of $\sim$10.8\,yr in the chromospheric time series, whereas no periodicity 
could be detected in the photometric time series. 

In Figs.~\ref{phase_diff_activity}-\ref{phase_diff_amp_phot} we showed that 
faculae-dominated stars are inactive slow rotators with small photometric 
amplitudes, compared to their active spot-dominated counterparts with shorter 
rotation periods and larger amplitudes. The transition from spot- to faculae-dominated 
activity occurs at an activity level of $\logRHK \approx -4.75$ (the VP~gap), 
and around a Rossby number $\rm{Ro} \approx 1$. 
This result is consistent with the work of \citet{Montet2017}. These authors 
searched for photometric activity cycles by analyzing \textit{Kepler} full-frame 
images (FFIs). They correlated the long-term brightness variations of the FFIs 
with photometric short-term variability measured by the magnetic activity proxy 
$S_{\rm ph}$ from the \textit{Kepler} light curves (see \citealt{Mathur2014}). 
These authors found that most stars with rotation periods shorter than 15--25~days 
show short-term photometric variability amplitudes that are anti-correlated with 
the seasonal brightness in the FFIs. This supports the interpretation that the 
activity cycles of fast rotating stars are spot-dominated. For stars with longer 
rotation periods, a correlation between short-term variability amplitudes and 
average brightness suggests that their activity cycles are faculae-dominated. 
In their sample of Sun-like stars with $0.6\lesssim B-V \lesssim 0.7$, the 
rotation period range between 15--25~days marks the transition from spot- to 
faculae-dominated activity. Using activity-rotation relations from 
\citet{Mamajek2008}, this period range coincides with the VP~gap, where we find 
the transition from spot- to faculae-dominated activity.
Using the same relations, this period range can be transferred to Rossby 
numbers: $\rm{Ro(\Prot=15\,d)}=1.25$ and $\rm{Ro(\Prot=25\,d)}=2.09$ using the 
mean color $B-V_{\rm mean}=0.65$. 

To further support the finding of a transition from spot- to 
faculae-dominated activity, we attempted a similar analysis to \citet{Montet2017} 
by correlating the following photometric time series: the first one consists of 
the seasonal mean brightness values, and the second one contains the variability 
range $\Rvar$ of each observing season (see \citealt{Basri2010,Basri2011}). This 
quantity measures the rotation-induced variability in each season similar to the 
quantity $S_{\rm ph}$ used by \citet{Montet2017}. We correlated these two time 
series to see whether the long- and short-term variability is in or out of phase. 
Our results are, however, inconclusive given the relative imprecision of the 
ground-based data.

Moreover, we find that the majority of faculae-dominated stars in our sample 
exhibits a chromospheric activity cycle with a strong sinusoidal shape 
($\rm flag_{chrom}=3$). All eight of these $\rm flag_{chrom}=3$ stars lie 
below the VP~gap. Activity-age relations from \citet{Mamajek2008} reveal ages 
greater than 2.7\,Gyr for these eight stars, suggesting that the transition from 
complex to sinusoidal or smooth cycles is complete at this age. This confirms earlier 
results by \citet{Olah2016} who analyzed activity cycles of Mount Wilson stars, 
finding that slowly rotating stars show smooth activity cycles, whereas more 
rapidly rotating stars exhibit more complex cycles, with a clear transition from 
complex to smooth cycles at ages between $\sim$2--3\,Gyr. 

For six stars in our sample, the magnetic field topology has been measured from 
Zeeman Doppler imaging \citep{See2016}. Their results suggest that spot-dominated 
stars show toroidal fields, whereas faculae-dominated stars exhibit mostly poloidal 
fields. A toroidal field has indeed been detected in HD\,20630 and HD\,131156A, 
for which we detect spot-dominance. HD\,10476 and HD\,201091 are faculae-dominated 
stars with smooth chromospheric activity cycle ($\rm flag_{chrom}=3$) showing 
strong poloidal fields and almost no toroidal fields. The spot-dominated star 
HD\,206860 does not fall in either group, showing 50\% of each field component, 
and HD\,120136 shows a slightly more toroidal than poloidal field, although our 
results suggest faculae-dominance for this star. Unfortunately, the number of 
stars with observed magnetic field topology in overlap with our sample is too 
small to either confirm or reject the scenario suggested by \citet{See2016}.

\subsection{Can the VP~gap explain the dearth of intermediate rotation periods?}
\citet{McQuillan2013} first discovered a dearth of rotation periods around 
$\sim$25~days in the rotation period distribution of the \textit{Kepler} M~dwarfs. 
It has been shown that this dearth extends to hotter stars covering a rotation period 
range of $\sim$15--25~days \citep{Reinhold2013, McQuillan2014, Reinhold2015}.
\citet{McQuillan2013} explained the detected bimodality of the rotation periods
by an overlap of two stellar populations of different age in the \textit{Kepler} 
field, with the younger stars rotating faster, on average, than the older stars.
\citet{Davenport2017} also discussed the bimodality seen in the \textit{Kepler} 
rotation period distribution. By using absolute G-band magnitudes from the Gaia 
mission, this author demonstrated that the \citet{McQuillan2014} sample might be 
strongly contaminated by subgiants. Because of their deeper convective envelopes, 
their rotation period distribution is significantly different from dwarf stars, 
which might explain the previous non-detection of the bimodality in these stars. 
Removing those evolved stars from the sample, \citet{Davenport2017} showed that 
the bimodality persists for stars hotter than 5000\,K. This author further 
finds that the fast and slow rotators in his sample also exhibit a different 
distribution of the total proper motion. Hence, the existence of two kinematically 
separate groups would favor the explanation of two epochs of stars formation in 
the \textit{Kepler} field. This explanation is further supported by the work of 
\citet{Davenport2018_arxiv}, showing that the bimodality also correlates with 
Galactic height, which is assumed to be related to stellar age.

\begin{figure*}
  \centering
  \includegraphics[width=17cm]{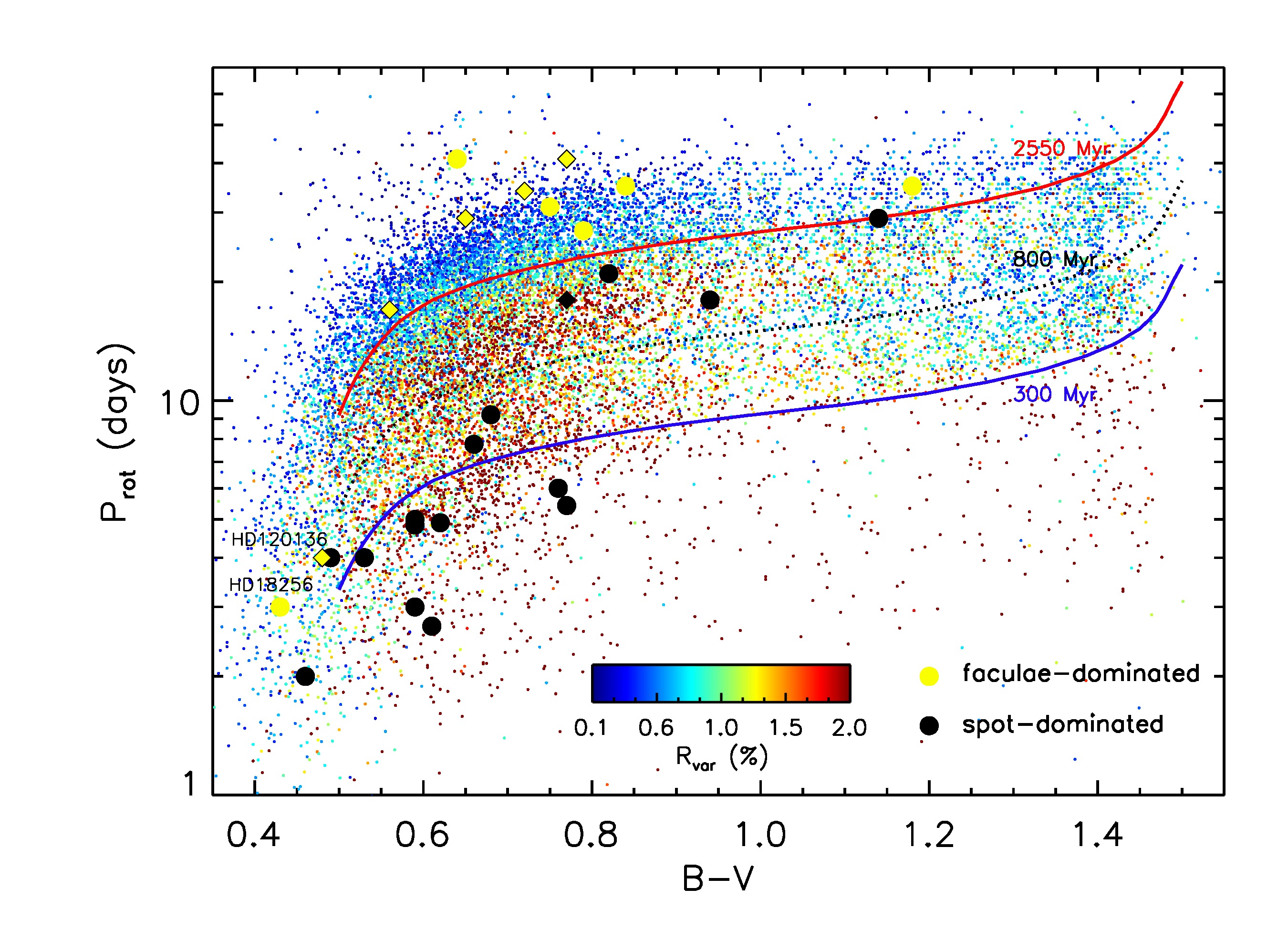}
  \caption{\textit{Kepler} rotation periods $\Prot$ against $B-V$ color taken from 
  \citet{Reinhold2015}. The data points are color-coded with the variability
  range $\Rvar$. Black and yellow symbols show spot- and faculae-dominated 
  stars, respectively, for the current sample. As before dots show dwarf stars 
  (luminosity class V) and diamond show (sub-)giants (luminosity class III, IV, 
  IV-V). The solid blue, dotted black, and solid red lines show 300\,Myr, 800\,Myr, 
  and 2550\,Myr isochrones, respectively, using gyrochronology relations from 
  \citet{Barnes2010}.}
  \label{kepler}
\end{figure*}

Over the last years it has been discussed whether the lack of 
intermediate active stars at the VP~gap can explain the dearth at intermediate 
rotation periods. To test this hypothesis, we revisit the rotation period 
distribution of the \textit{Kepler} sample measured by \citet{Reinhold2015} in 
Fig.~\ref{kepler}. The rotation periods are 
plotted against $B-V$ color, and the data points are color-coded with the 
variability range $\Rvar$ (we return to this quantity in Sect.~\ref{dearth}).
The dearth of rotation periods becomes visible at $\sim$15~days at $B-V\approx0.9$ 
and increases to $\sim$25~days at $B-V\approx 1.5$. We find that the 
\textit{Kepler} rotation periods are well represented by gyrochronology 
relations from \citet{Barnes2010}: the blue and red lines show the gyrochronology 
isochrones (gyrochrones) for ages of 300\,Myr and 2550\,Myr, respectively. The 
color-dependent dearth of rotation periods follows a gyrochrone with an age of 
$\sim$800\,Myr (dotted black line).

The age of 2550\,Myr has been selected because it refers to the activity level of 
$\logRHK \approx -4.75$ at the VP~gap. In Fig.~\ref{kepler} we over-plot the 
spot- and faculae-dominated stars of the current sample as black and yellow 
symbols, respectively. For colors $B-V>0.5$, where gyrochronology relations are well 
calibrated \citep{Barnes2007}, we find all spot-dominated stars below the 2550\,Myr 
isochrone, and all the faculae-dominated stars above this age. In the color range 
$B-V<0.5$ where gyrochronology is not well calibrated we use the same activity-age 
relations as before to infer the ages of the two F~stars HD\,18256 and HD\,120136. 
This analysis yields ages of $2.65$\,Gyr and $2.45$\,Gyr, respectively, consistent 
with the age of $2.55$\,Gyr associated with the activity level at the VP~gap. Hence, 
the expected age of stars at the VP~gap clearly separates the spot- and 
faculae-dominated stars in our sample. This result demonstrates that the observed 
dearth of stars at an age of $\sim$800\,Myr can not be explained in terms of the VP~gap.

\subsection{Alternate explanation for the dearth region}\label{dearth}
The \textit{Kepler} stars in Fig.~\ref{kepler} are color-coded by their 
variability range $\Rvar$, which describes the amplitude of the rotational 
variability. Interestingly, the dearth region around $\sim$800\,Myr is dominated 
by stars with small variability amplitudes (mostly blue points). Moreover, there 
seems to be a variability gradient towards the dearth region from either side, 
i.e., stars younger and older than $\sim$800\,Myr show larger variability. 
Based on the result that faculae are the dominant ingredient of the surface 
activity for the activity cycle after $2.55$\,Gyr, we expect that the faculae 
surface coverage may increase from an earlier age. To explain both the dearth of 
stars and the decreased variability, we propose the idea of a period determination 
bias, and suggest the following scenario: Fast rotating stars are able to 
generate strong magnetic fields, which in turn create 
large and long-lived spots. As the stars slow down, the magnetic field topology 
changes such that the magnetic field is not able to maintain large spots anymore, 
and instead generates smaller active regions surrounded by bright faculae. This 
process diminishes the variability amplitudes and continues until faculae become 
so prominent on the surface that bright and dark active regions partially cancel 
each other. The stars in the dearth region might have smaller and more homogeneously 
distributed spots, which simultaneously would decrease the variability, and render 
period determination more difficult. Hence, many stars with small variability 
amplitudes will not be detected due to certain amplitude constraints.

\begin{figure*}
  \centering
  \includegraphics[width=17cm]{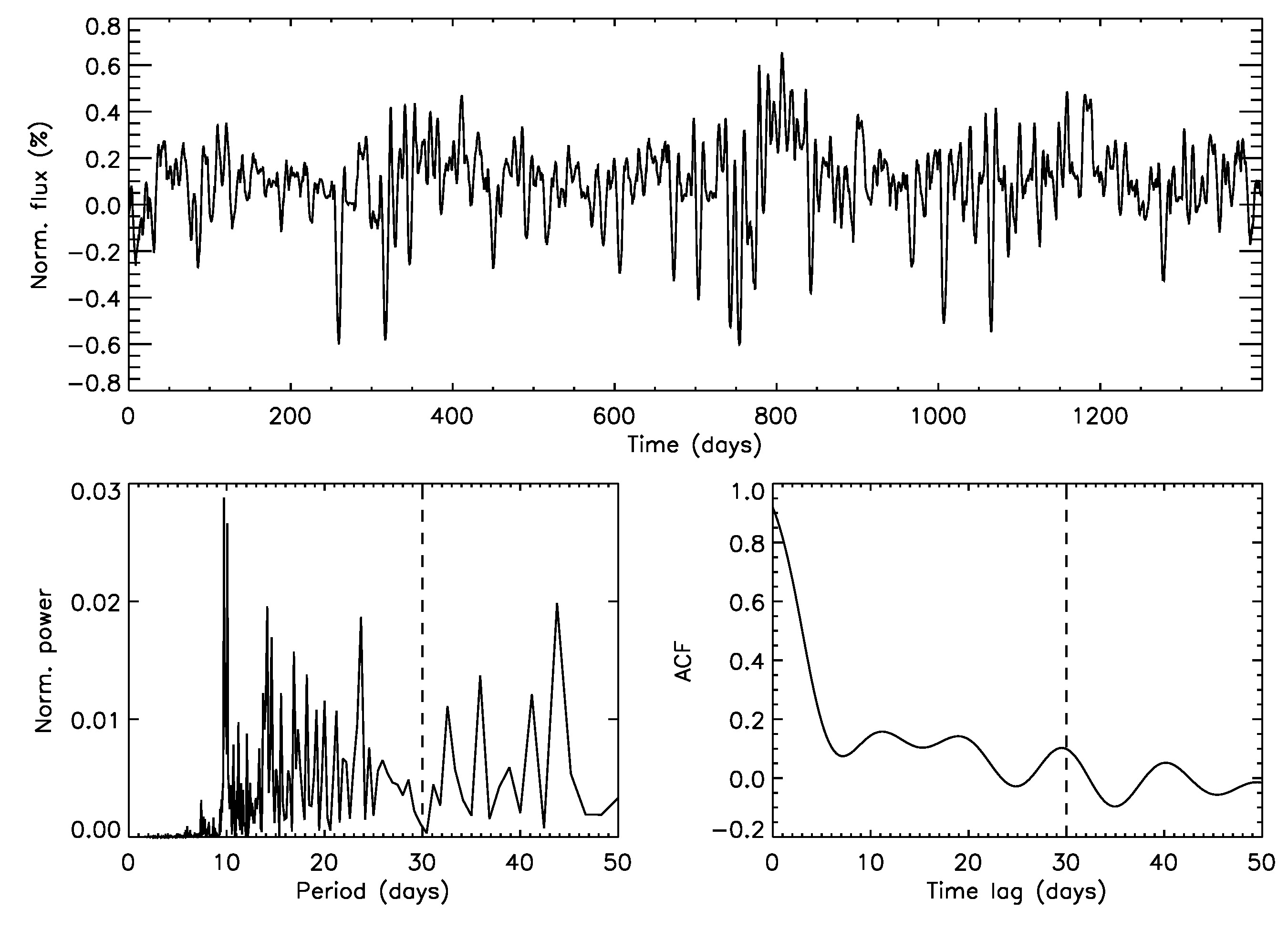}
  \caption{Top panel: Model light curve with a rotation period of 30~days viewed 
  at an inclination angle of 90 degrees.
  Lower left: Lomb-Scargle periodogram of the time 
  series. Lower right: auto-correlation function of the time series. The 
  vertical dashed line indicates the model rotation period.}
  \label{sim_lc}
\end{figure*}

To better illustrate the idea of spot and faculae cancellation, we applied the 
Spectral And Total Irradiance REconstruction (SATIRE, see \citealt{Fligge2000, 
Krivova2003}), which is the state-of-the-art model of solar brightness 
variations, for synthesizing light curves brought about by the mixture of 
facular and spot features. Such a model light curve is shown in Fig.~\ref{sim_lc}. 
We have considered a star with rotational period of 30~days at various inclination 
angles. The light curve was synthesized for 1400~days, which is comparable to 
the total duration of the \textit{Kepler} observations. The mean surface coverage of 
spots was chosen to be 0.1\% (i.e. the star is a bit more active than 
the Sun). We used a log-normal distribution of spot sizes from \citet{Baumann2005},
and assumed a linear decay of spot sizes with time (following 
\citealt{MartinezPillet1993}). We assumed that each spot is surrounded by a 
facular region, which is a reasonable assumption for modeling rotational stellar 
variability (see, e.g. \citealt{Shapiro2017}). The faculae-to-spot coverage ratio 
at the day of emergence was taken to be 3.4, and the lifetimes of faculae were 
considered to be three times longer than those of spots, which roughly corresponds 
to the solar case.

The lower left panel in Fig.~\ref{sim_lc} shows the Lomb-Scargle periodogram of 
the time series. The power was normalized to unity (i.e. perfect sinusoidal 
periodicity); none of the peaks is significant, following the threshold of a peak 
height of 0.1 as used by \citet{Reinhold2015}. The lower right panel shows the 
auto-correlation function (ACF) of the time series, which was used by 
\citet{McQuillan2014} to search for periodicity in the \textit{Kepler} sample. 
The peaks indicate a periodicity with multiples of 10~days, although 
with low power. The correct rotation period cannot be recovered either by the 
periodogram or by the ACF method. Thus, we conclude that 1) this star would have 
been considered as non-periodic using the periodogram approach, or 2) would have 
been considered as periodic using the ACF method, however measuring a wrong 
rotation period. This model light curve has a variability range of $\Rvar=0.56$\%;
if a significant period were detected, it would appear as blue dot in 
Fig.~\ref{kepler}.
We note that a similar spot and faculae cancellation happens in the solar case.  
\citet{Shapiro2017} showed that the 27-day peaks in the spot and facular component 
of the solar brightness variability (calculated for the period 1996--2015) cancel 
each other, so that the power spectrum of solar brightness variations is almost 
flat around the rotational period.

Fig.~\ref{kepler} further shows that the variability increases after an age of 
$\sim$800\,Myr, suggesting that spots again become the dominant source of 
activity. This scenario is supported by the presence of the four spot-dominated 
stars between the $\sim$800\,Myr and the $2.55$\,Gyr isochrone. It is worth 
mentioning that \citet{Lockwood2007} also detected spot dominance for these four 
stars. Once the stars slow down further and pass the VP~gap, faculae become the 
dominant source of surface activity on activity cycle time scales. This is 
supported by the large abundance of stars with small variability amplitudes 
above the transitionary age of $2.55$\,Gyr. Although the transition from spot to 
faculae domination has been inferred from variability on activity cycle time 
scales, this observation suggests that at least some rotation periods have 
been inferred from long-lived faculae rather than short-lived spots, as seen in 
the quiet Sun. Based on observations of solar activity this idea has already been 
proposed by \citet{Pettersen1992}. During solar activity maximum bright and dark 
features cancel each other, hampering a reliable period detection, whereas rotation 
can be inferred from faculae during activity minimum. To decide on the dominant 
surface feature spot-crossings during transits (e.g. \citealt{Morris2018b})
or Doppler images would be needed.

We emphasize that classifying a star to be spot- or faculae-dominated refers to 
the type of feature that causes it to appear overall fainter or brighter during 
the peaks of its activity cycle. On rotational time scales spots usually are the 
dominant source of variability seen in the light curve because they are more 
localized than faculae \citep{Shapiro2016}. Moreover, faculae-dominated stars 
older than $2.55$\,Gyr are still able to maintain spots, as is observed in the 
Sun. Applying the simultaneous sine fit to the individual observing seasons, and 
searching for the same rotation period instead of the same cycle period in both 
data sets, we indeed find spot dominance for most of the stars where a rotation 
period could be detected. That implies that a star can be spot-dominated on 
rotational time scales, whereas it can be faculae-dominated on activity cycle 
time scales.

\section{Conclusion}
In this paper we have determined activity cycle periods for 27 out of the 30 
stars in our sample. By computing the phase difference between the photometric 
and the chromospheric time series we have shown that stellar activity transitions 
from spot to faculae domination at an activity level of $\logRHK \approx -4.75$. 
This value equals the activity level at the Vaughan-Preston gap. Using 
activity-rotation-age relations this activity level is associated with a Rossby 
number $\rm{Ro} \approx 1$ and an age of $\sim$2.55\,Gyr. Furthermore, the 
transitional age of $\sim$2.55\,Gyr is consistent with measurements of a change 
of the activity cycle shape and the global magnetic field topology from the literature \citep{Olah2016,See2016}. These results provide observational constraints 
on the underlying dynamo. The connection between the dominant source of surface activity 
and the magnetic field topology sounds promising to either infer the type of surface 
activity from magnetic field measurements or vice versa. 

We further conclude that the lack of intermediate active stars at the 
Vaughan-Preston gap cannot explain the dearth of intermediate rotation periods 
discovered by the \textit{Kepler} telescope. Nevertheless, the results presented 
in this work strongly suggest that the importance of faculae contribution to the 
big picture of stellar activity has been underestimated so far. We propose a 
scenario to explain the dearth of rotation periods based on the cancellation of 
the spot and faculae contribution at rotation time scales. The proposed idea has 
been tested by simulating light curves with a solar faculae-to-spot ratio. We 
found that neither the periodogram nor the auto-correlation function were able 
to detect the correct rotation period. We plan to refine the light curve model, 
and extend it to stars of various spectral types in a forthcoming publication.

\begin{acknowledgements}
The research leading to the presented results has received funding from the 
European Research Council under the European Community's Seventh Framework 
Programme (/2007-2013) / ERC grant agreement no 338251 (StellarAges). A.I.S. 
acknowledges the support by the European Research Council (ERC) under the 
European Unions Horizon 2020 research and innovation program (grant agreement 
no. 715947). We thank S.~Messina and G.~Lockwood and collaborators for kindly 
sharing their data with us. 
We also like to thank T.~Metcalfe, H.~Schunker, and R.~H.~Cameron 
for valuable discussion. The HK\_Project\_v1995\_NSO data derived from the Mount 
Wilson Observatory HK Project, which was supported by both public and private 
funds through the Carnegie Observatories, the Mount Wilson Institute, and the 
Harvard-Smithsonian Center for Astrophysics starting in 1966 and continuing for 
over 36 years. These data are the result of the dedicated work of O.~Wilson, 
A.~Vaughan, G.~Preston, D.~Duncan, S.~Baliunas, and many others.
\end{acknowledgements}

\bibliographystyle{aa}
\bibliography{biblothek2}

\begin{thebibliography}{69}
\expandafter\ifx\csname natexlab\endcsname\relax\def\natexlab#1{#1}\fi

\bibitem[{{Ambikasaran} {et~al.}(2015){Ambikasaran}, {Foreman-Mackey},
  {Greengard}, {Hogg}, \& {O'Neil}}]{Ambikasaran2015}
{Ambikasaran}, S., {Foreman-Mackey}, D., {Greengard}, L., {Hogg}, D.~W., \&
  {O'Neil}, M. 2015, IEEE Transactions on Pattern Analysis and Machine
  Intelligence, 38 [\eprint[arXiv]{1403.6015}]

\bibitem[{{Angus} {et~al.}(2018){Angus}, {Morton}, {Aigrain}, {Foreman-Mackey},
  \& {Rajpaul}}]{Angus2018}
{Angus}, R., {Morton}, T., {Aigrain}, S., {Foreman-Mackey}, D., \& {Rajpaul},
  V. 2018, \mnras, 474, 2094

\bibitem[{{Balachandran}(1990)}]{Balachandran1990}
{Balachandran}, S. 1990, \apj, 354, 310

\bibitem[{{Baliunas} {et~al.}(1996){Baliunas}, {Sokoloff}, \&
  {Soon}}]{Baliunas1996}
{Baliunas}, S., {Sokoloff}, D., \& {Soon}, W. 1996, \apjl, 457, L99

\bibitem[{{Baliunas} {et~al.}(1995){Baliunas}, {Donahue}, {Soon}, {Horne},
  {Frazer}, {Woodard-Eklund}, {Bradford}, {Rao}, {Wilson}, {Zhang}, {Bennett},
  {Briggs}, {Carroll}, {Duncan}, {Figueroa}, {Lanning}, {Misch}, {Mueller},
  {Noyes}, {Poppe}, {Porter}, {Robinson}, {Russell}, {Shelton}, {Soyumer},
  {Vaughan}, \& {Whitney}}]{Baliunas1995}
{Baliunas}, S.~L., {Donahue}, R.~A., {Soon}, W.~H., {et~al.} 1995, \apj, 438,
  269

\bibitem[{{Barnes}(2007)}]{Barnes2007}
{Barnes}, S.~A. 2007, \apj, 669, 1167

\bibitem[{{Barnes}(2010)}]{Barnes2010}
{Barnes}, S.~A. 2010, \apj, 722, 222

\bibitem[{{Basri} {et~al.}(2011){Basri}, {Walkowicz}, {Batalha}, {Gilliland},
  {Jenkins}, {Borucki}, {Koch}, {Caldwell}, {Dupree}, {Latham}, {Marcy},
  {Meibom}, \& {Brown}}]{Basri2011}
{Basri}, G., {Walkowicz}, L.~M., {Batalha}, N., {et~al.} 2011, \aj, 141, 20

\bibitem[{{Basri} {et~al.}(2010){Basri}, {Walkowicz}, {Batalha}, {Gilliland},
  {Jenkins}, {Borucki}, {Koch}, {Caldwell}, {Dupree}, {Latham}, {Meibom},
  {Howell}, \& {Brown}}]{Basri2010}
{Basri}, G., {Walkowicz}, L.~M., {Batalha}, N., {et~al.} 2010, \apjl, 713, L155

\bibitem[{{Baumann} \& {Solanki}(2005)}]{Baumann2005}
{Baumann}, I. \& {Solanki}, S.~K. 2005, \aap, 443, 1061

\bibitem[{{Boeche} \& {Grebel}(2016)}]{Boeche2016}
{Boeche}, C. \& {Grebel}, E.~K. 2016, \aap, 587, A2

\bibitem[{{B{\"o}hm-Vitense}(2007)}]{BV2007}
{B{\"o}hm-Vitense}, E. 2007, \apj, 657, 486

\bibitem[{{Chen} {et~al.}(2000){Chen}, {Nissen}, {Zhao}, {Zhang}, \&
  {Benoni}}]{Chen2000}
{Chen}, Y.~Q., {Nissen}, P.~E., {Zhao}, G., {Zhang}, H.~W., \& {Benoni}, T.
  2000, \aaps, 141, 491

\bibitem[{{Davenport}(2017)}]{Davenport2017}
{Davenport}, J.~R.~A. 2017, \apj, 835, 16

\bibitem[{{Davenport} \& {Covey}(2018)}]{Davenport2018_arxiv}
{Davenport}, J.~R.~A. \& {Covey}, K.~R. 2018, ArXiv e-prints
  [\eprint[arXiv]{1807.09841}]

\bibitem[{{Duncan} {et~al.}(1991){Duncan}, {Vaughan}, {Wilson}, {Preston},
  {Frazer}, {Lanning}, {Misch}, {Mueller}, {Soyumer}, {Woodard}, {Baliunas},
  {Noyes}, {Hartmann}, {Porter}, {Zwaan}, {Middelkoop}, {Rutten}, \&
  {Mihalas}}]{Duncan1991}
{Duncan}, D.~K., {Vaughan}, A.~H., {Wilson}, O.~C., {et~al.} 1991, \apjs, 76,
  383

\bibitem[{{Egeland} {et~al.}(2017){Egeland}, {Soon}, {Baliunas}, {Hall},
  {Pevtsov}, \& {Bertello}}]{Egeland2017}
{Egeland}, R., {Soon}, W., {Baliunas}, S., {et~al.} 2017, \apj, 835, 25

\bibitem[{{Fligge} {et~al.}(2000){Fligge}, {Solanki}, \& {Unruh}}]{Fligge2000}
{Fligge}, M., {Solanki}, S.~K., \& {Unruh}, Y.~C. 2000, \aap, 353, 380

\bibitem[{{Foreman-Mackey} {et~al.}(2013){Foreman-Mackey}, {Hogg}, {Lang}, \&
  {Goodman}}]{Foreman-Mackey2013}
{Foreman-Mackey}, D., {Hogg}, D.~W., {Lang}, D., \& {Goodman}, J. 2013, \pasp,
  125, 306

\bibitem[{{Goodman} \& {Weare}(2010)}]{Goodman2010}
{Goodman}, J. \& {Weare}, J. 2010, Communications in Applied Mathematics and
  Computational Science, Vol.~5, No.~1, p.~65-80, 2010, 5, 65

\bibitem[{{K{\"o}nig} {et~al.}(2005){K{\"o}nig}, {Guenther}, {Woitas}, \&
  {Hatzes}}]{Koenig2005}
{K{\"o}nig}, B., {Guenther}, E.~W., {Woitas}, J., \& {Hatzes}, A.~P. 2005,
  \aap, 435, 215

\bibitem[{{Krivova} {et~al.}(2003){Krivova}, {Solanki}, {Fligge}, \&
  {Unruh}}]{Krivova2003}
{Krivova}, N.~A., {Solanki}, S.~K., {Fligge}, M., \& {Unruh}, Y.~C. 2003, \aap,
  399, L1

\bibitem[{{Lehtinen} {et~al.}(2016){Lehtinen}, {Jetsu}, {Hackman}, {Kajatkari},
  \& {Henry}}]{Lehtinen2016}
{Lehtinen}, J., {Jetsu}, L., {Hackman}, T., {Kajatkari}, P., \& {Henry}, G.~W.
  2016, \aap, 588, A38

\bibitem[{{Linsky} {et~al.}(1979){Linsky}, {McClintock}, {Robertson}, \&
  {Worden}}]{Linsky1979}
{Linsky}, J.~L., {McClintock}, W., {Robertson}, R.~M., \& {Worden}, S.~P. 1979,
  \apjs, 41, 47

\bibitem[{{Lockwood} {et~al.}(2007){Lockwood}, {Skiff}, {Henry}, {Henry},
  {Radick}, {Baliunas}, {Donahue}, \& {Soon}}]{Lockwood2007}
{Lockwood}, G.~W., {Skiff}, B.~A., {Henry}, G.~W., {et~al.} 2007, \apjs, 171,
  260

\bibitem[{{Lockwood} {et~al.}(1997){Lockwood}, {Skiff}, \&
  {Radick}}]{Lockwood1997}
{Lockwood}, G.~W., {Skiff}, B.~A., \& {Radick}, R.~R. 1997, \apj, 485, 789

\bibitem[{{Mamajek} \& {Hillenbrand}(2008)}]{Mamajek2008}
{Mamajek}, E.~E. \& {Hillenbrand}, L.~A. 2008, \apj, 687, 1264

\bibitem[{{Martinez Pillet} {et~al.}(1993){Martinez Pillet}, {Moreno-Insertis},
  \& {Vazquez}}]{MartinezPillet1993}
{Martinez Pillet}, V., {Moreno-Insertis}, F., \& {Vazquez}, M. 1993, \aap, 274,
  521

\bibitem[{{Mathur} {et~al.}(2014){Mathur}, {Garc{\'{\i}}a}, {Ballot},
  {Ceillier}, {Salabert}, {Metcalfe}, {R{\'e}gulo}, {Jim{\'e}nez}, \&
  {Bloemen}}]{Mathur2014}
{Mathur}, S., {Garc{\'{\i}}a}, R.~A., {Ballot}, J., {et~al.} 2014, \aap, 562,
  A124

\bibitem[{{McQuillan} {et~al.}(2013){McQuillan}, {Aigrain}, \&
  {Mazeh}}]{McQuillan2013}
{McQuillan}, A., {Aigrain}, S., \& {Mazeh}, T. 2013, \mnras, 432, 1203

\bibitem[{{McQuillan} {et~al.}(2014){McQuillan}, {Mazeh}, \&
  {Aigrain}}]{McQuillan2014}
{McQuillan}, A., {Mazeh}, T., \& {Aigrain}, S. 2014, \apjs, 211, 24

\bibitem[{{Messina} \& {Guinan}(2002)}]{Messina2002}
{Messina}, S. \& {Guinan}, E.~F. 2002, \aap, 393, 225

\bibitem[{{Messina} \& {Guinan}(2003)}]{Messina2003}
{Messina}, S. \& {Guinan}, E.~F. 2003, \aap, 409, 1017

\bibitem[{{Metcalfe} {et~al.}(2016){Metcalfe}, {Egeland}, \& {van
  Saders}}]{Metcalfe2016}
{Metcalfe}, T.~S., {Egeland}, R., \& {van Saders}, J. 2016, \apjl, 826, L2

\bibitem[{{Metcalfe} \& {van Saders}(2017)}]{Metcalfe2017}
{Metcalfe}, T.~S. \& {van Saders}, J. 2017, \solphys, 292, 126

\bibitem[{{Middelkoop}(1982)}]{Middelkoop1982}
{Middelkoop}, F. 1982, \aap, 107, 31

\bibitem[{{Mishenina} {et~al.}(2013){Mishenina}, {Pignatari}, {Korotin},
  {Soubiran}, {Charbonnel}, {Thielemann}, {Gorbaneva}, \&
  {Basak}}]{Mishenina2013}
{Mishenina}, T.~V., {Pignatari}, M., {Korotin}, S.~A., {et~al.} 2013, \aap,
  552, A128

\bibitem[{{Montet} {et~al.}(2017){Montet}, {Tovar}, \&
  {Foreman-Mackey}}]{Montet2017}
{Montet}, B.~T., {Tovar}, G., \& {Foreman-Mackey}, D. 2017, \apj, 851, 116

\bibitem[{{Morris} {et~al.}(2018){Morris}, {Agol}, {Hebb}, \&
  {Hawley}}]{Morris2018b}
{Morris}, B.~M., {Agol}, E., {Hebb}, L., \& {Hawley}, S.~L. 2018, \aj, 156, 91

\bibitem[{{Niedzielski} {et~al.}(2016){Niedzielski}, {Deka-Szymankiewicz},
  {Adamczyk}, {Adam{\'o}w}, {Nowak}, \& {Wolszczan}}]{Niedzielski2016}
{Niedzielski}, A., {Deka-Szymankiewicz}, B., {Adamczyk}, M., {et~al.} 2016,
  \aap, 585, A73

\bibitem[{{Noyes} {et~al.}(1984){Noyes}, {Hartmann}, {Baliunas}, {Duncan}, \&
  {Vaughan}}]{Noyes1984}
{Noyes}, R.~W., {Hartmann}, L.~W., {Baliunas}, S.~L., {Duncan}, D.~K., \&
  {Vaughan}, A.~H. 1984, \apj, 279, 763

\bibitem[{{Ol{\'a}h} {et~al.}(2016){Ol{\'a}h}, {K{\H o}v{\'a}ri}, {Petrovay},
  {Soon}, {Baliunas}, {Koll{\'a}th}, \& {Vida}}]{Olah2016}
{Ol{\'a}h}, K., {K{\H o}v{\'a}ri}, Z., {Petrovay}, K., {et~al.} 2016, \aap,
  590, A133

\bibitem[{{Olspert} {et~al.}(2017){Olspert}, {Lehtinen}, {K{\"a}pyl{\"a}},
  {Pelt}, \& {Grigorievskiy}}]{Olspert2017_II_arxiv}
{Olspert}, N., {Lehtinen}, J.~J., {K{\"a}pyl{\"a}}, M.~J., {Pelt}, J., \&
  {Grigorievskiy}, A. 2017, ArXiv e-prints [\eprint[arXiv]{1712.08240}]

\bibitem[{{Pettersen} {et~al.}(1992){Pettersen}, {Hawley}, \&
  {Fisher}}]{Pettersen1992}
{Pettersen}, B.~R., {Hawley}, S.~L., \& {Fisher}, G.~H. 1992, \solphys, 142,
  197

\bibitem[{{Preminger} {et~al.}(2011){Preminger}, {Chapman}, \&
  {Cookson}}]{Preminger2011}
{Preminger}, D.~G., {Chapman}, G.~A., \& {Cookson}, A.~M. 2011, \apjl, 739, L45

\bibitem[{{Prugniel} {et~al.}(2011){Prugniel}, {Vauglin}, \&
  {Koleva}}]{Prugniel2011}
{Prugniel}, P., {Vauglin}, I., \& {Koleva}, M. 2011, \aap, 531, A165

\bibitem[{{Radick} {et~al.}(2018){Radick}, {Lockwood}, {Henry}, {Hall}, \&
  {Pevtsov}}]{Radick2018}
{Radick}, R.~R., {Lockwood}, G.~W., {Henry}, G.~W., {Hall}, J.~C., \&
  {Pevtsov}, A.~A. 2018, \apj, 855, 75

\bibitem[{{Radick} {et~al.}(1998){Radick}, {Lockwood}, {Skiff}, \&
  {Baliunas}}]{Radick1998}
{Radick}, R.~R., {Lockwood}, G.~W., {Skiff}, B.~A., \& {Baliunas}, S.~L. 1998,
  \apjs, 118, 239

\bibitem[{{Ram{\'{\i}}rez} {et~al.}(2013){Ram{\'{\i}}rez}, {Allende Prieto}, \&
  {Lambert}}]{Ramirez2013}
{Ram{\'{\i}}rez}, I., {Allende Prieto}, C., \& {Lambert}, D.~L. 2013, \apj,
  764, 78

\bibitem[{{Ram{\'{\i}}rez} {et~al.}(2012){Ram{\'{\i}}rez}, {Michel}, {Sefako},
  {Tucci Maia}, {Schuster}, {van Wyk}, {Mel{\'e}ndez}, {Casagrande}, \&
  {Castilho}}]{Ramirez2012}
{Ram{\'{\i}}rez}, I., {Michel}, R., {Sefako}, R., {et~al.} 2012, \apj, 752, 5

\bibitem[{Rasmussen \& Williams(2005)}]{Rasmussen2005}
Rasmussen, C.~E. \& Williams, C. K.~I. 2005, Gaussian Processes for Machine
  Learning (Adaptive Computation and Machine Learning) (The MIT Press)

\bibitem[{{Reinhold} {et~al.}(2017){Reinhold}, {Cameron}, \&
  {Gizon}}]{Reinhold2017}
{Reinhold}, T., {Cameron}, R.~H., \& {Gizon}, L. 2017, \aap, 603, A52

\bibitem[{{Reinhold} \& {Gizon}(2015)}]{Reinhold2015}
{Reinhold}, T. \& {Gizon}, L. 2015, \aap, 583, A65

\bibitem[{{Reinhold} {et~al.}(2013){Reinhold}, {Reiners}, \&
  {Basri}}]{Reinhold2013}
{Reinhold}, T., {Reiners}, A., \& {Basri}, G. 2013, \aap, 560, A4

\bibitem[{{Rutten}(1984)}]{Rutten1984}
{Rutten}, R.~G.~M. 1984, \aap, 130, 353

\bibitem[{{Saar} \& {Brandenburg}(1999)}]{SB1999}
{Saar}, S.~H. \& {Brandenburg}, A. 1999, \apj, 524, 295

\bibitem[{{Saikia} {et~al.}(2018){Saikia}, {Marvin}, {Jeffers}, {Reiners},
  {Cameron}, {Marsden}, {Petit}, {Warnecke}, \& {Yadav}}]{Saikia2018}
{Saikia}, S.~B., {Marvin}, C.~J., {Jeffers}, S.~V., {et~al.} 2018, \aap, 616,
  A108

\bibitem[{{See} {et~al.}(2016){See}, {Jardine}, {Vidotto}, {Donati}, {Boro
  Saikia}, {Bouvier}, {Fares}, {Folsom}, {Gregory}, {Hussain}, {Jeffers},
  {Marsden}, {Morin}, {Moutou}, {do Nascimento}, {Petit}, \& {Waite}}]{See2016}
{See}, V., {Jardine}, M., {Vidotto}, A.~A., {et~al.} 2016, \mnras, 462, 4442

\bibitem[{{Shapiro} {et~al.}(2017){Shapiro}, {Solanki}, {Krivova}, {Cameron},
  {Yeo}, \& {Schmutz}}]{Shapiro2017}
{Shapiro}, A.~I., {Solanki}, S.~K., {Krivova}, N.~A., {et~al.} 2017, Nature
  Astronomy, 1, 612

\bibitem[{{Shapiro} {et~al.}(2014){Shapiro}, {Solanki}, {Krivova}, {Schmutz},
  {Ball}, {Knaack}, {Rozanov}, \& {Unruh}}]{Shapiro2014}
{Shapiro}, A.~I., {Solanki}, S.~K., {Krivova}, N.~A., {et~al.} 2014, \aap, 569,
  A38

\bibitem[{{Shapiro} {et~al.}(2016){Shapiro}, {Solanki}, {Krivova}, {Yeo}, \&
  {Schmutz}}]{Shapiro2016}
{Shapiro}, A.~I., {Solanki}, S.~K., {Krivova}, N.~A., {Yeo}, K.~L., \&
  {Schmutz}, W.~K. 2016, \aap, 589, A46

\bibitem[{{Skumanich}(1972)}]{Skumanich1972}
{Skumanich}, A. 1972, \apj, 171, 565

\bibitem[{{Takeda}(2007)}]{Takeda2007}
{Takeda}, Y. 2007, \pasj, 59, 335

\bibitem[{{Vaughan} \& {Preston}(1980)}]{VP1980}
{Vaughan}, A.~H. \& {Preston}, G.~W. 1980, \pasp, 92, 385

\bibitem[{{Vaughan} {et~al.}(1978){Vaughan}, {Preston}, \&
  {Wilson}}]{Vaughan1978}
{Vaughan}, A.~H., {Preston}, G.~W., \& {Wilson}, O.~C. 1978, \pasp, 90, 267

\bibitem[{{Wilson}(1968)}]{Wilson1968}
{Wilson}, O.~C. 1968, \apj, 153, 221

\bibitem[{{Wilson}(1978)}]{Wilson1978}
{Wilson}, O.~C. 1978, \apj, 226, 379

\bibitem[{{Witzke} {et~al.}(2018){Witzke}, {Shapiro}, {Solanki}, {Krivova}, \&
  {Schmutz}}]{Witzke2018}
{Witzke}, V., {Shapiro}, A.~I., {Solanki}, S.~K., {Krivova}, N.~A., \&
  {Schmutz}, W. 2018, ArXiv e-prints [\eprint[arXiv]{1809.04360}]

\bibitem[{{Zechmeister} \& {K{\"u}rster}(2009)}]{Zechmeister2009}
{Zechmeister}, M. \& {K{\"u}rster}, M. 2009, \aap, 496, 577

\end{thebibliography}
\begin{appendix}
  \section{Photometric data}\label{app_A}
  \begin{table}[ht]
    \small
    \caption{HD\,1835 photometric observing seasons (in JD-2440000).}
    \label{HD1835_phot}
    \centering
    \begin{tabular}{ccccc}
\hline\hline
JD range & $\rm N_{obs}$ & $\rm JD_{mean}$ & $\rm \Delta (b+y)/2_{mean}$ & $\sigma_{by}$ \\
\hline
5949.83--6078.77 & 30 & 6034.32 & -0.0064 & 0.0088 \\
6287.86--6433.77 & 53 & 6348.44 & 0.0059 & 0.0102 \\
6642.86--6778.77 & 38 & 6714.40 & -0.0039 & 0.0088 \\
7037.85--7137.78 & 27 & 7084.78 & 0.0048 & 0.0064 \\
7419.84--7528.77 & 34 & 7472.45 & 0.0167 & 0.0064 \\
7770.84--7836.80 & 32 & 7800.20 & 0.0062 & 0.0120 \\
8125.85--8230.78 & 38 & 8170.29 & 0.0134 & 0.0099 \\
8514.84--8598.78 & 22 & 8544.36 & -0.0075 & 0.0067 \\
8863.85--8992.77 & 38 & 8913.08 & -0.0088 & 0.0084 \\
9215.85--9353.60 & 37 & 9294.45 & -0.0057 & 0.0184 \\
9547.00--9706.60 & 57 & 9647.00 & -0.0111 & 0.0119 \\
9974.83--10087.60 & 47 & 10009.30 & 0.0035 & 0.0073 \\
10307.86--10458.60 & 37 & 10390.79 & 0.0020 & 0.0075 \\
10712.83--10799.60 & 38 & 10742.73 & 0.0078 & 0.0054 \\
11072.83--11197.60 & 60 & 11127.88 & -0.0087 & 0.0078 \\
11450.80--11549.60 & 55 & 11490.14 & -0.0109 & 0.0101 \\
11809.80--11928.60 & 30 & 11862.86 & 0.0017 & 0.0074 \\
12193.70--12283.60 & 19 & 12239.49 & 0.0018 & 0.0092 \\
12533.80--12658.60 & 36 & 12597.38 & 0.0149 & 0.0070 \\
\hline
\end{tabular}
  \end{table}
  
  \begin{table}[ht]
    \small
    \caption{HD\,10476 photometric observing seasons (in JD-2440000).}
    \label{HD10476_phot}
    \centering
    \begin{tabular}{ccccc}
\hline\hline
JD range & $\rm N_{obs}$ & $\rm JD_{mean}$ & $\rm \Delta (b+y)/2_{mean}$ & $\sigma_{by}$ \\
\hline
5949.83--6069.77 & 32 & 6016.92 & 0.0032 & 0.0014 \\
6294.86--6425.79 & 30 & 6365.41 & 0.0032 & 0.0013 \\
6658.87--6808.77 & 28 & 6713.82 & -0.0008 & 0.0015 \\
7039.85--7158.77 & 30 & 7094.75 & -0.0002 & 0.0012 \\
7419.84--7535.77 & 32 & 7469.50 & -0.0004 & 0.0015 \\
7771.85--7853.80 & 40 & 7816.02 & -0.0027 & 0.0017 \\
8125.85--8230.79 & 36 & 8175.15 & -0.0025 & 0.0011 \\
8514.84--8598.78 & 18 & 8545.38 & -0.0021 & 0.0009 \\
8863.86--8929.80 & 28 & 8891.69 & -0.0017 & 0.0009 \\
9215.86--9389.60 & 58 & 9309.29 & -0.0002 & 0.0013 \\
9595.86--9751.60 & 46 & 9647.90 & -0.0000 & 0.0014 \\
9976.83--10123.60 & 53 & 10025.85 & 0.0000 & 0.0013 \\
10348.83--10494.60 & 47 & 10396.72 & 0.0009 & 0.0013 \\
10715.84--10856.60 & 43 & 10765.05 & 0.0007 & 0.0017 \\
11072.84--11218.60 & 69 & 11128.76 & 0.0011 & 0.0016 \\
11435.00--11577.70 & 77 & 11495.25 & -0.0002 & 0.0017 \\
11807.00--11962.60 & 40 & 11867.40 & -0.0005 & 0.0018 \\
12196.80--12328.60 & 27 & 12261.18 & -0.0002 & 0.0011 \\
12533.90--12673.70 & 40 & 12606.90 & 0.0004 & 0.0017 \\
\hline
\end{tabular}
  \end{table}
  
  \begin{table}[ht]
    \small
    \caption{HD\,13421 photometric observing seasons (in JD-2440000).}
    \label{HD13421_phot}
    \centering
    \begin{tabular}{ccccc}
\hline\hline
JD range & $\rm N_{obs}$ & $\rm JD_{mean}$ & $\rm \Delta (b+y)/2_{mean}$ & $\sigma_{by}$ \\
\hline
5956.84--6066.78 & 24 & 6010.06 & 0.0019 & 0.0023 \\
6307.86--6451.77 & 18 & 6382.03 & 0.0018 & 0.0018 \\
6679.86--6809.77 & 29 & 6732.89 & 0.0006 & 0.0020 \\
7039.86--7173.77 & 17 & 7104.99 & -0.0007 & 0.0017 \\
7419.85--7546.77 & 22 & 7479.63 & -0.0004 & 0.0015 \\
7793.84--7884.78 & 26 & 7840.58 & -0.0012 & 0.0022 \\
8147.85--8218.80 & 15 & 8186.22 & -0.0005 & 0.0013 \\
8532.84--8598.79 & 12 & 8556.66 & 0.0003 & 0.0013 \\
8867.86--8977.78 & 14 & 8910.97 & 0.0001 & 0.0014 \\
9248.80--9389.60 & 40 & 9324.31 & -0.0003 & 0.0012 \\
9595.87--9751.60 & 46 & 9657.35 & -0.0001 & 0.0014 \\
9977.84--10130.60 & 59 & 10039.30 & -0.0005 & 0.0013 \\
10348.84--10486.60 & 48 & 10397.22 & -0.0001 & 0.0013 \\
10715.84--10861.60 & 48 & 10775.12 & 0.0009 & 0.0012 \\
11085.90--11223.60 & 70 & 11146.90 & -0.0007 & 0.0016 \\
11435.00--11595.60 & 75 & 11507.73 & 0.0009 & 0.0012 \\
11807.90--11959.60 & 46 & 11876.82 & 0.0007 & 0.0013 \\
12179.90--12314.60 & 23 & 12255.73 & -0.0020 & 0.0010 \\
12533.90--12692.60 & 27 & 12612.88 & -0.0023 & 0.0013 \\
\hline
\end{tabular}
  \end{table}

  \begin{table}[ht]
    \small
    \caption{HD\,18256 photometric observing seasons (in JD-2440000).}
    \label{HD18256_phot}
    \centering
    \begin{tabular}{ccccc}
\hline\hline
JD range & $\rm N_{obs}$ & $\rm JD_{mean}$ & $\rm \Delta (b+y)/2_{mean}$ & $\sigma_{by}$ \\
\hline
5956.84--6083.78 & 18 & 6016.37 & 0.0021 & 0.0019 \\
6307.86--6460.78 & 19 & 6390.71 & 0.0019 & 0.0013 \\
6680.86--6809.78 & 15 & 6740.82 & 0.0001 & 0.0016 \\
7064.85--7173.78 & 15 & 7118.28 & -0.0005 & 0.0014 \\
7524.80--7556.77 & 10 & 7537.19 & -0.0003 & 0.0005 \\
7824.82--7884.79 & 20 & 7848.81 & -0.0015 & 0.0024 \\
8125.86--8219.80 & 16 & 8176.33 & -0.0003 & 0.0011 \\
8533.83--8598.80 & 8 & 8551.83 & -0.0030 & 0.0015 \\
8863.86--8989.78 & 20 & 8912.83 & -0.0003 & 0.0011 \\
9239.90--9399.60 & 80 & 9318.09 & 0.0004 & 0.0013 \\
9611.86--9769.60 & 64 & 9669.54 & -0.0001 & 0.0013 \\
9980.85--10122.60 & 69 & 10029.47 & -0.0006 & 0.0014 \\
10348.85--10511.60 & 46 & 10395.42 & -0.0018 & 0.0018 \\
10719.00--10861.70 & 38 & 10766.19 & -0.0019 & 0.0015 \\
11086.90--11222.70 & 54 & 11144.28 & 0.0017 & 0.0010 \\
11447.80--11582.70 & 66 & 11505.95 & -0.0012 & 0.0015 \\
11808.00--11956.70 & 37 & 11876.75 & 0.0007 & 0.0012 \\
12200.90--12342.60 & 26 & 12261.84 & 0.0029 & 0.0013 \\
12533.90--12692.70 & 39 & 12604.36 & 0.0014 & 0.0014 \\
\hline
\end{tabular}
  \end{table}
  
  \begin{table}[ht]
    \small
    \caption{HD\,20630 photometric observing seasons (in JD-2440000).}
    \label{HD20630_phot}
    \centering
    \begin{tabular}{ccccc}
\hline\hline
JD range & $\rm N_{obs}$ & $\rm JD_{mean}$ & $\rm V_{mean}$ & $\sigma_V$ \\
\hline
7908.21--8301.29 & 29 & 8166.21 & 4.8458 & 0.0121 \\
8435.31--9031.42 & 41 & 8606.20 & 4.8405 & 0.0123 \\
9239.92--9393.66 & 87 & 9309.28 & 4.8182 & 0.0079 \\
9624.92--9773.60 & 60 & 9676.74 & 4.8209 & 0.0058 \\
9986.90--10123.64 & 50 & 10036.29 & 4.8429 & 0.0065 \\
10391.93--10482.70 & 32 & 10436.63 & 4.8459 & 0.0132 \\
10718.95--10836.71 & 27 & 10764.94 & 4.8399 & 0.0065 \\
11102.94--11240.61 & 33 & 11160.70 & 4.8258 & 0.0068 \\
\hline
\end{tabular}
  \end{table}
  
  \begin{table}[ht]
    \small
    \caption{HD\,25998 photometric observing seasons (in JD-2440000).}
    \label{HD25998_phot}
    \centering
    \begin{tabular}{ccccc}
\hline\hline
JD range & $\rm N_{obs}$ & $\rm JD_{mean}$ & $\rm \Delta (b+y)/2_{mean}$ & $\sigma_{by}$ \\
\hline
5956.86--6087.78 & 16 & 6043.69 & 0.0013 & 0.0057 \\
6331.86--6433.80 & 21 & 6381.40 & 0.0018 & 0.0027 \\
6679.86--6828.80 & 23 & 6736.87 & -0.0033 & 0.0040 \\
7069.85--7173.78 & 10 & 7126.42 & -0.0030 & 0.0039 \\
7524.81--7530.81 & 6 & 7527.47 & 0.0037 & 0.0033 \\
7837.84--7933.79 & 12 & 7882.31 & -0.0001 & 0.0053 \\
8206.82--8276.80 & 12 & 8225.14 & -0.0004 & 0.0038 \\
8534.84--8632.78 & 8 & 8576.06 & -0.0022 & 0.0079 \\
8889.85--9000.80 & 8 & 8949.32 & 0.0018 & 0.0029 \\
9326.81--9385.79 & 16 & 9350.55 & 0.0019 & 0.0038 \\
9611.86--9687.80 & 24 & 9650.83 & 0.0003 & 0.0048 \\
9991.85--10098.79 & 26 & 10043.58 & -0.0038 & 0.0054 \\
10348.86--10395.83 & 24 & 10365.59 & 0.0053 & 0.0032 \\
\hline
\end{tabular}
  \end{table}
  
  \begin{table}[ht]
    \small
    \caption{HD\,35296 photometric observing seasons (in JD-2440000).}
    \label{HD35296_phot}
    \centering
    \begin{tabular}{ccccc}
\hline\hline
JD range & $\rm N_{obs}$ & $\rm JD_{mean}$ & $\rm \Delta (b+y)/2_{mean}$ & $\sigma_{by}$ \\
\hline
5761.79--5779.78 & 10 & 5770.38 & 0.0006 & 0.0041 \\
5956.86--6150.79 & 30 & 6089.81 & 0.0037 & 0.0072 \\
6331.87--6513.78 & 63 & 6421.10 & -0.0013 & 0.0068 \\
6705.88--6882.78 & 32 & 6777.12 & -0.0071 & 0.0057 \\
7104.85--7234.78 & 33 & 7179.26 & 0.0029 & 0.0083 \\
7523.82--7607.78 & 24 & 7563.55 & 0.0042 & 0.0055 \\
7844.84--7973.78 & 26 & 7918.34 & -0.0020 & 0.0076 \\
8218.84--8313.78 & 14 & 8262.52 & -0.0014 & 0.0054 \\
8598.83--8698.78 & 20 & 8645.00 & 0.0025 & 0.0050 \\
8928.85--9055.78 & 16 & 8985.32 & 0.0021 & 0.0081 \\
9326.82--9396.80 & 19 & 9357.18 & -0.0022 & 0.0067 \\
9634.86--9688.83 & 24 & 9663.68 & 0.0041 & 0.0040 \\
10008.85--10102.80 & 24 & 10051.41 & 0.0030 & 0.0059 \\
10367.86--10519.78 & 24 & 10470.97 & -0.0030 & 0.0055 \\
10884.78--10892.78 & 6 & 10889.11 & -0.0120 & 0.0061 \\
11212.78--11248.78 & 4 & 11230.78 & 0.0005 & 0.0010 \\
11525.81--11579.79 & 8 & 11556.30 & 0.0039 & 0.0011 \\
\hline
\end{tabular}
  \end{table}
  
  \begin{table}[ht]
    \small
    \caption{HD\,39587 photometric observing seasons (in JD-2440000).}
    \label{HD39587_phot}
    \centering
    \begin{tabular}{ccccc}
\hline\hline
JD range & $\rm N_{obs}$ & $\rm JD_{mean}$ & $\rm \Delta (b+y)/2_{mean}$ & $\sigma_{by}$ \\
\hline
5761.79--5779.78 & 10 & 5770.38 & 0.0140 & 0.0019 \\
5956.86--6150.79 & 30 & 6089.81 & 0.0065 & 0.0068 \\
6331.87--6513.78 & 63 & 6421.10 & 0.0004 & 0.0057 \\
6705.88--6882.78 & 32 & 6777.12 & -0.0006 & 0.0059 \\
7104.85--7234.78 & 33 & 7179.26 & -0.0032 & 0.0062 \\
7523.82--7607.78 & 24 & 7563.55 & -0.0016 & 0.0054 \\
7844.84--7973.78 & 26 & 7918.34 & -0.0072 & 0.0062 \\
8218.84--8313.78 & 12 & 8266.81 & 0.0004 & 0.0015 \\
8598.83--8698.78 & 20 & 8645.00 & -0.0051 & 0.0059 \\
8928.85--9055.78 & 16 & 8985.32 & -0.0071 & 0.0094 \\
9326.82--9396.80 & 19 & 9357.18 & 0.0045 & 0.0042 \\
9634.86--9688.83 & 24 & 9663.68 & 0.0048 & 0.0060 \\
10008.85--10102.80 & 24 & 10051.41 & 0.0075 & 0.0062 \\
10367.86--10519.78 & 24 & 10470.97 & -0.0009 & 0.0082 \\
10884.78--10892.78 & 6 & 10889.11 & -0.0062 & 0.0043 \\
11183.80--11248.78 & 6 & 11215.12 & -0.0057 & 0.0062 \\
11525.81--11579.79 & 8 & 11556.30 & -0.0087 & 0.0030 \\
\hline
\end{tabular}
  \end{table}
  
  \begin{table}[ht]
    \small
    \caption{HD\,72905 photometric observing seasons (in JD-2440000).}
    \label{HD72905_phot}
    \centering
    \begin{tabular}{ccccc}
\hline\hline
JD range & $\rm N_{obs}$ & $\rm JD_{mean}$ & $\rm V_{mean}$ & $\sigma_V$ \\
\hline
7625.66--7650.69 & 15 & 7638.21 & 5.6785 & 0.0066 \\
7815.03--8024.19 & 57 & 7917.05 & 5.6734 & 0.0087 \\
8133.73--8492.62 & 116 & 8302.56 & 5.6755 & 0.0103 \\
8560.34--8765.22 & 60 & 8677.77 & 5.6809 & 0.0085 \\
8972.11--9062.71 & 22 & 9027.58 & 5.6691 & 0.0074 \\
9284.98--9482.65 & 122 & 9368.75 & 5.6472 & 0.0089 \\
9645.00--9808.77 & 94 & 9725.89 & 5.6529 & 0.0094 \\
10029.95--10214.65 & 94 & 10131.35 & 5.6471 & 0.0067 \\
10405.99--10569.68 & 81 & 10484.01 & 5.6519 & 0.0064 \\
10736.02--10949.65 & 102 & 10848.40 & 5.6351 & 0.0049 \\
11118.97--11317.66 & 82 & 11225.23 & 5.6509 & 0.0065 \\
11546.81--11634.68 & 31 & 11584.92 & 5.6535 & 0.0069 \\
\hline
\end{tabular}
  \end{table}
  
  \begin{table}[ht]
    \small
    \caption{HD\,75332 photometric observing seasons (in JD-2440000).}
    \label{HD75332_phot}
    \centering
    \begin{tabular}{ccccc}
\hline\hline
JD range & $\rm N_{obs}$ & $\rm JD_{mean}$ & $\rm \Delta (b+y)/2_{mean}$ & $\sigma_{by}$ \\
\hline
6363.44--6517.79 & 34 & 6433.53 & 0.0033 & 0.0048 \\
6741.87--6906.79 & 29 & 6848.85 & -0.0150 & 0.0105 \\
7149.84--7239.79 & 32 & 7208.00 & -0.0078 & 0.0058 \\
7525.84--7627.79 & 29 & 7575.99 & 0.0008 & 0.0037 \\
7917.82--7992.79 & 20 & 7949.91 & 0.0020 & 0.0044 \\
8276.83--8324.81 & 12 & 8297.32 & -0.0013 & 0.0026 \\
8632.83--8719.78 & 22 & 8661.64 & 0.0027 & 0.0027 \\
8928.86--9124.70 & 20 & 9049.63 & 0.0080 & 0.0044 \\
9287.00--9475.70 & 94 & 9377.07 & 0.0024 & 0.0045 \\
9663.00--9832.70 & 70 & 9748.82 & 0.0071 & 0.0032 \\
10030.00--10218.70 & 47 & 10125.57 & 0.0008 & 0.0051 \\
10404.90--10577.70 & 47 & 10484.15 & -0.0064 & 0.0046 \\
10737.00--10956.70 & 53 & 10859.31 & -0.0023 & 0.0044 \\
11093.00--11314.70 & 42 & 11208.17 & -0.0060 & 0.0060 \\
11471.00--11668.70 & 47 & 11574.77 & -0.0007 & 0.0040 \\
11833.00--12046.70 & 55 & 11963.18 & -0.0040 & 0.0074 \\
12216.00--12422.70 & 53 & 12317.58 & 0.0062 & 0.0034 \\
12567.00--12785.70 & 65 & 12684.54 & 0.0025 & 0.0057 \\
\hline
\end{tabular}
  \end{table}
  
  \begin{table}[ht]
    \small
    \caption{HD\,81809 photometric observing seasons (in JD-2440000).}
    \label{HD81809_phot}
    \centering
    \begin{tabular}{ccccc}
\hline\hline
JD range & $\rm N_{obs}$ & $\rm JD_{mean}$ & $\rm \Delta (b+y)/2_{mean}$ & $\sigma_{by}$ \\
\hline
5763.79--5815.79 & 21 & 5789.41 & -0.0000 & 0.0024 \\
6065.85--6159.79 & 26 & 6123.74 & 0.0011 & 0.0024 \\
6404.44--6512.81 & 14 & 6465.66 & 0.0004 & 0.0017 \\
6808.84--6908.78 & 28 & 6865.67 & -0.0037 & 0.0025 \\
7198.83--7251.80 & 33 & 7220.84 & -0.0009 & 0.0019 \\
7536.85--7627.79 & 21 & 7578.06 & -0.0005 & 0.0020 \\
7920.83--8008.79 & 20 & 7959.01 & -0.0008 & 0.0024 \\
8276.84--8378.79 & 14 & 8322.10 & -0.0013 & 0.0015 \\
8637.84--8733.79 & 22 & 8675.00 & -0.0000 & 0.0027 \\
9021.83--9121.60 & 17 & 9076.07 & -0.0013 & 0.0030 \\
9295.00--9482.60 & 97 & 9386.77 & 0.0008 & 0.0020 \\
9662.00--9840.70 & 98 & 9749.32 & 0.0002 & 0.0019 \\
10030.00--10213.60 & 64 & 10122.37 & -0.0007 & 0.0016 \\
10406.00--10552.70 & 56 & 10490.81 & 0.0009 & 0.0015 \\
10761.00--10944.70 & 49 & 10863.14 & -0.0003 & 0.0015 \\
11124.00--11315.70 & 39 & 11225.24 & -0.0009 & 0.0025 \\
11479.00--11669.70 & 54 & 11572.73 & 0.0017 & 0.0014 \\
11865.00--12036.70 & 44 & 11957.70 & 0.0006 & 0.0018 \\
12245.90--12403.70 & 40 & 12305.69 & 0.0008 & 0.0018 \\
12574.00--12771.70 & 50 & 12666.29 & -0.0006 & 0.0019 \\
\hline
\end{tabular}
  \end{table}
  
  \begin{table}[ht]
    \small
    \caption{HD\,82443 photometric observing seasons (in JD-2440000).}
    \label{HD82443_phot}
    \centering
    \begin{tabular}{ccccc}
\hline\hline
JD range & $\rm N_{obs}$ & $\rm JD_{mean}$ & $\rm V_{mean}$ & $\sigma_V$ \\
\hline
7570.82--7673.66 & 63 & 7617.97 & 7.0558 & 0.0243 \\
7976.36--8065.31 & 35 & 8031.52 & 7.0455 & 0.0165 \\
8193.90--8393.43 & 32 & 8312.40 & 7.0633 & 0.0243 \\
8540.72--8761.57 & 53 & 8667.59 & 7.0848 & 0.0274 \\
8906.01--9137.65 & 110 & 9030.65 & 7.0516 & 0.0117 \\
9284.96--9511.65 & 208 & 9416.95 & 7.0478 & 0.0128 \\
9644.97--9840.69 & 126 & 9741.10 & 7.0667 & 0.0234 \\
9996.02--10216.71 & 145 & 10104.08 & 7.0639 & 0.0247 \\
10392.92--10575.69 & 240 & 10497.37 & 7.0496 & 0.0214 \\
10728.01--10976.65 & 160 & 10857.47 & 7.0641 & 0.0184 \\
11094.00--11303.70 & 142 & 11203.53 & 7.0583 & 0.0139 \\
11546.91--11612.79 & 17 & 11568.05 & 7.0412 & 0.0145 \\
\hline
\end{tabular}
  \end{table}
  
  \begin{table}[ht]
    \small
    \caption{HD\,82885 photometric observing seasons (in JD-2440000).}
    \label{HD82885_phot}
    \centering
    \begin{tabular}{ccccc}
\hline\hline
JD range & $\rm N_{obs}$ & $\rm JD_{mean}$ & $\rm \Delta (b+y)/2_{mean}$ & $\sigma_{by}$ \\
\hline
5763.80--5815.79 & 22 & 5792.34 & 0.0061 & 0.0063 \\
6065.86--6197.79 & 54 & 6138.66 & -0.0039 & 0.0127 \\
6373.88--6571.79 & 55 & 6487.74 & 0.0019 & 0.0053 \\
6741.44--6921.79 & 41 & 6865.83 & 0.0031 & 0.0034 \\
7199.84--7259.79 & 38 & 7218.82 & 0.0088 & 0.0026 \\
7536.84--7665.79 & 34 & 7597.17 & 0.0074 & 0.0040 \\
7917.83--8023.79 & 32 & 7973.43 & 0.0043 & 0.0043 \\
8276.84--8382.79 & 24 & 8338.81 & 0.0055 & 0.0035 \\
8632.84--8733.79 & 22 & 8670.46 & 0.0006 & 0.0027 \\
9021.82--9139.60 & 35 & 9090.83 & 0.0062 & 0.0037 \\
9298.00--9486.70 & 119 & 9397.43 & -0.0020 & 0.0060 \\
9675.00--9870.60 & 100 & 9757.83 & -0.0024 & 0.0060 \\
10030.00--10223.70 & 95 & 10134.76 & 0.0024 & 0.0041 \\
10404.90--10600.60 & 81 & 10490.79 & -0.0012 & 0.0054 \\
10737.00--10959.70 & 75 & 10869.01 & -0.0037 & 0.0093 \\
11094.00--11325.70 & 60 & 11229.36 & 0.0056 & 0.0035 \\
11471.00--11690.70 & 84 & 11593.82 & 0.0000 & 0.0067 \\
11833.00--12060.70 & 55 & 11962.38 & -0.0001 & 0.0049 \\
12216.00--12425.70 & 65 & 12330.08 & -0.0121 & 0.0068 \\
12567.00--12786.70 & 79 & 12687.88 & -0.0018 & 0.0058 \\
\hline
\end{tabular}
  \end{table}
  
  \begin{table}[ht]
    \small
    \caption{HD\,103095 photometric observing seasons (in JD-2440000).}
    \label{HD103095_phot}
    \centering
    \begin{tabular}{ccccc}
\hline\hline
JD range & $\rm N_{obs}$ & $\rm JD_{mean}$ & $\rm \Delta (b+y)/2_{mean}$ & $\sigma_{by}$ \\
\hline
5763.82--5838.79 & 20 & 5796.60 & 0.0007 & 0.0014 \\
6065.87--6165.80 & 28 & 6120.69 & -0.0009 & 0.0011 \\
6421.86--6540.79 & 21 & 6497.30 & 0.0011 & 0.0014 \\
6809.86--6920.80 & 20 & 6873.52 & 0.0014 & 0.0025 \\
7199.85--7247.81 & 34 & 7219.66 & 0.0013 & 0.0016 \\
7543.86--7667.80 & 26 & 7624.81 & 0.0013 & 0.0013 \\
7967.82--8032.79 & 26 & 8006.34 & 0.0017 & 0.0017 \\
8276.85--8390.79 & 28 & 8343.82 & 0.0016 & 0.0019 \\
8637.86--8756.79 & 16 & 8719.94 & 0.0007 & 0.0020 \\
9021.84--9164.70 & 62 & 9115.47 & -0.0014 & 0.0020 \\
9317.00--9528.70 & 125 & 9426.35 & -0.0019 & 0.0014 \\
9689.00--9891.70 & 114 & 9791.35 & -0.0012 & 0.0014 \\
10052.00--10251.70 & 98 & 10145.47 & -0.0009 & 0.0013 \\
10405.00--10633.70 & 97 & 10522.94 & -0.0005 & 0.0014 \\
10762.00--10996.70 & 104 & 10903.46 & 0.0008 & 0.0015 \\
11149.00--11354.70 & 76 & 11264.31 & 0.0012 & 0.0021 \\
11504.00--11709.70 & 95 & 11621.09 & 0.0011 & 0.0016 \\
11862.00--12076.70 & 77 & 11998.97 & 0.0009 & 0.0013 \\
12246.00--12446.70 & 75 & 12348.87 & 0.0002 & 0.0017 \\
12601.00--12813.70 & 76 & 12709.92 & 0.0001 & 0.0015 \\
\hline
\end{tabular}
  \end{table}
  
  \begin{table}[ht]
    \small
    \caption{HD\,115383 photometric observing seasons (in JD-2440000).}
    \label{HD115383_phot}
    \centering
    \begin{tabular}{ccccc}
\hline\hline
JD range & $\rm N_{obs}$ & $\rm JD_{mean}$ & $\rm \Delta (b+y)/2_{mean}$ & $\sigma_{by}$ \\
\hline
5781.83--5840.80 & 11 & 5802.55 & -0.0095 & 0.0014 \\
6104.86--6226.79 & 12 & 6157.33 & 0.0016 & 0.0046 \\
6487.85--6571.81 & 21 & 6521.60 & 0.0059 & 0.0027 \\
6881.84--6965.79 & 9 & 6908.71 & 0.0008 & 0.0023 \\
7200.86--7320.80 & 28 & 7263.90 & -0.0049 & 0.0041 \\
7602.84--7697.79 & 18 & 7649.81 & 0.0062 & 0.0039 \\
8008.80--8061.80 & 18 & 8042.58 & 0.0030 & 0.0044 \\
8301.85--8405.80 & 22 & 8368.36 & -0.0045 & 0.0025 \\
8732.82--8785.79 & 10 & 8757.80 & -0.0064 & 0.0027 \\
9052.84--9164.70 & 62 & 9127.20 & 0.0003 & 0.0056 \\
9339.00--9546.70 & 75 & 9460.55 & -0.0020 & 0.0041 \\
9718.00--9908.70 & 93 & 9819.91 & 0.0054 & 0.0035 \\
10071.00--10249.70 & 79 & 10175.46 & 0.0001 & 0.0032 \\
10434.00--10635.70 & 89 & 10540.95 & -0.0010 & 0.0032 \\
10795.00--10993.70 & 109 & 10909.06 & -0.0065 & 0.0045 \\
11181.00--11344.70 & 52 & 11287.39 & -0.0048 & 0.0033 \\
11530.00--11710.79 & 67 & 11631.35 & -0.0004 & 0.0048 \\
11888.00--12075.70 & 33 & 11998.83 & 0.0026 & 0.0051 \\
12263.10--12462.70 & 68 & 12356.41 & 0.0081 & 0.0037 \\
12616.00--12829.70 & 65 & 12725.22 & 0.0021 & 0.0039 \\
\hline
\end{tabular}
  \end{table}
  
  \begin{table}[ht]
    \small
    \caption{HD\,115404 photometric observing seasons (in JD-2440000).}
    \label{HD115404_phot}
    \centering
    \begin{tabular}{ccccc}
\hline\hline
JD range & $\rm N_{obs}$ & $\rm JD_{mean}$ & $\rm \Delta (b+y)/2_{mean}$ & $\sigma_{by}$ \\
\hline
5781.84--5850.80 & 18 & 5824.93 & -0.0167 & 0.0052 \\
6104.86--6235.79 & 39 & 6190.82 & -0.0081 & 0.0062 \\
6487.84--6571.81 & 25 & 6518.20 & -0.0021 & 0.0086 \\
6881.84--6969.79 & 11 & 6913.64 & -0.0074 & 0.0083 \\
7215.85--7320.80 & 22 & 7269.27 & -0.0082 & 0.0031 \\
7602.85--7705.80 & 18 & 7652.48 & -0.0072 & 0.0077 \\
8008.81--8066.80 & 20 & 8045.00 & -0.0075 & 0.0077 \\
8328.85--8405.80 & 21 & 8374.15 & -0.0082 & 0.0079 \\
8716.83--8786.79 & 9 & 8754.14 & -0.0018 & 0.0015 \\
9066.85--9164.70 & 64 & 9126.91 & -0.0071 & 0.0053 \\
9335.00--9540.70 & 99 & 9440.05 & -0.0071 & 0.0104 \\
9725.00--9908.70 & 75 & 9809.45 & 0.0021 & 0.0049 \\
10063.00--10265.70 & 64 & 10172.32 & 0.0042 & 0.0047 \\
10419.00--10635.70 & 76 & 10535.51 & 0.0083 & 0.0080 \\
10788.00--10993.70 & 75 & 10899.02 & 0.0006 & 0.0068 \\
11149.00--11360.70 & 60 & 11275.34 & 0.0028 & 0.0065 \\
11531.00--11709.80 & 61 & 11638.53 & 0.0022 & 0.0066 \\
11888.00--12076.70 & 62 & 12008.56 & 0.0024 & 0.0062 \\
12246.00--12447.70 & 54 & 12350.97 & 0.0079 & 0.0053 \\
12618.00--12813.70 & 65 & 12722.58 & 0.0097 & 0.0052 \\
\hline
\end{tabular}
  \end{table}
  
  \begin{table}[ht]
    \small
    \caption{HD\,120136 photometric observing seasons (in JD-2440000).}
    \label{HD120136_phot}
    \centering
    \begin{tabular}{ccccc}
\hline\hline
JD range & $\rm N_{obs}$ & $\rm JD_{mean}$ & $\rm \Delta (b+y)/2_{mean}$ & $\sigma_{by}$ \\
\hline
5777.83--5865.80 & 24 & 5831.81 & 0.0054 & 0.0023 \\
6104.87--6238.79 & 16 & 6181.95 & 0.0045 & 0.0016 \\
6487.85--6595.80 & 15 & 6547.09 & -0.0030 & 0.0041 \\
6883.84--6970.79 & 16 & 6933.75 & -0.0045 & 0.0016 \\
7200.44--7320.81 & 26 & 7267.51 & -0.0024 & 0.0022 \\
7602.85--7704.80 & 18 & 7663.37 & -0.0035 & 0.0014 \\
8019.81--8064.80 & 18 & 8045.91 & -0.0042 & 0.0017 \\
8356.83--8425.80 & 16 & 8387.19 & -0.0036 & 0.0015 \\
8733.83--8801.80 & 11 & 8773.98 & -0.0029 & 0.0010 \\
9079.84--9164.70 & 61 & 9129.00 & -0.0008 & 0.0026 \\
9345.00--9547.70 & 83 & 9460.59 & -0.0002 & 0.0013 \\
9725.00--9890.70 & 81 & 9824.10 & 0.0010 & 0.0019 \\
10085.00--10241.80 & 76 & 10177.83 & 0.0005 & 0.0013 \\
10436.00--10635.70 & 254 & 10553.47 & -0.0004 & 0.0015 \\
10795.00--10993.70 & 83 & 10913.88 & 0.0015 & 0.0017 \\
11182.00--11342.80 & 57 & 11290.73 & 0.0015 & 0.0014 \\
11531.00--11729.70 & 63 & 11640.16 & 0.0015 & 0.0016 \\
11894.00--12076.80 & 43 & 12012.53 & 0.0025 & 0.0017 \\
12268.00--12448.70 & 41 & 12356.12 & -0.0011 & 0.0016 \\
12650.00--12829.70 & 45 & 12736.26 & -0.0027 & 0.0020 \\
\hline
\end{tabular}
  \end{table}
  
  \begin{table}[ht]
    \small
    \caption{HD\,124570 photometric observing seasons (in JD-2440000).}
    \label{HD124570_phot}
    \centering
    \begin{tabular}{ccccc}
\hline\hline
JD range & $\rm N_{obs}$ & $\rm JD_{mean}$ & $\rm \Delta (b+y)/2_{mean}$ & $\sigma_{by}$ \\
\hline
5777.84--5865.80 & 16 & 5815.45 & -0.0037 & 0.0017 \\
6145.84--6226.80 & 19 & 6190.40 & -0.0032 & 0.0009 \\
6487.86--6602.80 & 29 & 6550.58 & -0.0017 & 0.0015 \\
6883.85--6970.79 & 14 & 6931.74 & -0.0014 & 0.0021 \\
7204.87--7349.80 & 18 & 7271.83 & -0.0016 & 0.0011 \\
7630.82--7709.80 & 20 & 7675.71 & -0.0016 & 0.0011 \\
8008.82--8065.80 & 18 & 8044.36 & -0.0018 & 0.0020 \\
8369.83--8405.80 & 16 & 8388.69 & -0.0013 & 0.0016 \\
8755.82--8800.79 & 10 & 8782.60 & -0.0003 & 0.0013 \\
9094.80--9164.70 & 27 & 9133.34 & -0.0003 & 0.0010 \\
9354.00--9519.70 & 30 & 9465.50 & -0.0006 & 0.0010 \\
9747.00--9890.70 & 42 & 9830.80 & 0.0005 & 0.0012 \\
10122.00--10249.70 & 28 & 10184.23 & 0.0005 & 0.0012 \\
10435.00--10635.70 & 53 & 10540.62 & 0.0009 & 0.0017 \\
10798.00--10994.70 & 61 & 10909.38 & 0.0011 & 0.0017 \\
11178.00--11342.80 & 49 & 11288.36 & 0.0014 & 0.0013 \\
11531.00--11710.80 & 61 & 11640.99 & 0.0024 & 0.0015 \\
11895.00--12084.80 & 54 & 12019.94 & 0.0009 & 0.0013 \\
12263.00--12462.70 & 72 & 12364.38 & -0.0005 & 0.0009 \\
12629.00--12829.70 & 91 & 12735.89 & -0.0006 & 0.0010 \\
\hline
\end{tabular}
  \end{table}
   
  \begin{table}[ht]
    \small
    \caption{HD\,129333 photometric observing seasons (in JD-2440000).}
    \label{HD129333_phot}
    \centering
    \begin{tabular}{ccccc}
\hline\hline
JD range & $\rm N_{obs}$ & $\rm JD_{mean}$ & $\rm \Delta (b+y)/2_{mean}$ & $\sigma_{by}$ \\
\hline
5781.86--5874.80 & 14 & 5843.96 & 0.0728 & 0.0146 \\
6104.47--6252.79 & 54 & 6211.03 & 0.0722 & 0.0204 \\
6488.03--6640.79 & 52 & 6555.81 & 0.0519 & 0.0136 \\
6829.87--6983.80 & 26 & 6937.55 & 0.0143 & 0.0161 \\
7199.86--7348.80 & 36 & 7256.06 & 0.0223 & 0.0137 \\
7580.84--7725.79 & 44 & 7668.11 & 0.0179 & 0.0164 \\
8008.82--8066.80 & 40 & 8039.91 & 0.0214 & 0.0168 \\
8369.84--8439.80 & 35 & 8399.53 & 0.0067 & 0.0252 \\
8732.83--8806.79 & 26 & 8775.70 & 0.0214 & 0.0142 \\
9030.86--9163.70 & 72 & 9127.20 & 0.0077 & 0.0234 \\
9357.00--9548.79 & 84 & 9468.00 & 0.0274 & 0.0232 \\
9725.00--9908.70 & 95 & 9828.22 & 0.0032 & 0.0326 \\
10087.00--10260.70 & 93 & 10186.45 & -0.0026 & 0.0288 \\
10480.00--10639.79 & 46 & 10567.82 & -0.0089 & 0.0440 \\
10800.00--10996.70 & 59 & 10919.59 & -0.0279 & 0.0196 \\
11176.00--11342.80 & 55 & 11276.69 & -0.0491 & 0.0272 \\
11557.00--11709.80 & 54 & 11646.17 & -0.0561 & 0.0288 \\
11962.00--12074.80 & 44 & 12027.82 & -0.0416 & 0.0379 \\
12281.00--12447.80 & 33 & 12378.15 & -0.0682 & 0.0216 \\
12636.00--12811.80 & 46 & 12734.45 & -0.0476 & 0.0244 \\
\hline
\end{tabular}
  \end{table}
  
  \begin{table}[ht]
    \small
    \caption{HD\,131156A photometric observing seasons (in JD-2440000).}
    \label{HD131156A_phot}
    \centering
    \begin{tabular}{ccccc}
\hline\hline
JD range & $\rm N_{obs}$ & $\rm JD_{mean}$ & $\rm \Delta (b+y)/2_{mean}$ & $\sigma_{by}$ \\
\hline
5777.84--5874.80 & 28 & 5840.46 & -0.0023 & 0.0065 \\
6105.18--6251.80 & 43 & 6205.60 & -0.0013 & 0.0058 \\
6487.86--6602.80 & 33 & 6554.10 & 0.0016 & 0.0074 \\
6882.87--6980.80 & 35 & 6943.67 & 0.0047 & 0.0075 \\
7232.86--7349.80 & 16 & 7288.83 & -0.0160 & 0.0032 \\
7627.84--7708.79 & 33 & 7676.06 & 0.0053 & 0.0055 \\
8008.83--8066.80 & 32 & 8045.06 & 0.0033 & 0.0100 \\
8381.84--8416.81 & 14 & 8394.11 & 0.0102 & 0.0022 \\
8732.84--8806.80 & 25 & 8782.37 & 0.0011 & 0.0057 \\
9107.83--9139.80 & 23 & 9125.69 & 0.0120 & 0.0086 \\
9443.86--9546.79 & 24 & 9498.40 & -0.0062 & 0.0056 \\
9800.85--9892.80 & 24 & 9848.24 & -0.0154 & 0.0063 \\
10161.86--10233.80 & 24 & 10194.42 & -0.0136 & 0.0060 \\
10571.82--10629.79 & 26 & 10618.34 & 0.0106 & 0.0085 \\
\hline
\end{tabular}
  \end{table}
  
  \begin{table}[ht]
    \small
    \caption{HD\,143761 photometric observing seasons (in JD-2440000).}
    \label{HD143761_phot}
    \centering
    \begin{tabular}{ccccc}
\hline\hline
JD range & $\rm N_{obs}$ & $\rm JD_{mean}$ & $\rm \Delta (b+y)/2_{mean}$ & $\sigma_{by}$ \\
\hline
5783.86--5867.81 & 22 & 5839.83 & -0.0035 & 0.0018 \\
6145.86--6223.82 & 18 & 6194.72 & -0.0024 & 0.0012 \\
6487.87--6602.81 & 18 & 6553.48 & -0.0022 & 0.0010 \\
6885.86--6976.80 & 19 & 6942.82 & -0.0019 & 0.0011 \\
7284.84--7320.82 & 8 & 7301.58 & -0.0019 & 0.0017 \\
7644.84--7723.79 & 34 & 7693.22 & -0.0009 & 0.0015 \\
8016.83--8065.81 & 32 & 8050.00 & -0.0010 & 0.0014 \\
8381.84--8382.84 & 4 & 8382.34 & -0.0003 & 0.0015 \\
8732.84--8803.80 & 22 & 8778.45 & -0.0003 & 0.0015 \\
9094.85--9163.82 & 72 & 9129.31 & -0.0006 & 0.0012 \\
9372.05--9547.76 & 80 & 9476.56 & -0.0005 & 0.0010 \\
9750.02--9923.80 & 97 & 9845.46 & -0.0004 & 0.0013 \\
10109.04--10265.76 & 76 & 10190.36 & -0.0003 & 0.0012 \\
10466.01--10631.83 & 83 & 10578.93 & 0.0006 & 0.0014 \\
10823.03--11000.79 & 163 & 10948.59 & 0.0004 & 0.0018 \\
11184.04--11362.79 & 67 & 11309.71 & 0.0014 & 0.0017 \\
11557.03--11711.80 & 62 & 11650.55 & 0.0020 & 0.0013 \\
11941.97--12076.84 & 47 & 12028.41 & 0.0018 & 0.0018 \\
\hline
\end{tabular}
  \end{table}
  
  \begin{table}[ht]
    \small
    \caption{HD\,149661 photometric observing seasons (in JD-2440000).}
    \label{HD149661_phot}
    \centering
    \begin{tabular}{ccccc}
\hline\hline
JD range & $\rm N_{obs}$ & $\rm JD_{mean}$ & $\rm \Delta (b+y)/2_{mean}$ & $\sigma_{by}$ \\
\hline
5781.87--5874.81 & 18 & 5842.27 & -0.0101 & 0.0058 \\
6145.87--6251.80 & 26 & 6224.43 & -0.0070 & 0.0063 \\
6567.84--6644.79 & 20 & 6612.61 & -0.0013 & 0.0053 \\
6909.86--6982.80 & 20 & 6966.41 & -0.0106 & 0.0039 \\
7644.85--7725.80 & 21 & 7697.05 & -0.0089 & 0.0071 \\
8016.84--8066.81 & 26 & 8047.98 & -0.0060 & 0.0053 \\
8399.83--8474.79 & 13 & 8430.43 & -0.0036 & 0.0042 \\
8756.84--8825.79 & 27 & 8796.00 & -0.0013 & 0.0029 \\
9096.90--9164.80 & 52 & 9138.90 & 0.0043 & 0.0067 \\
9399.00--9547.70 & 65 & 9487.34 & 0.0015 & 0.0063 \\
9774.00--9924.80 & 82 & 9864.26 & 0.0015 & 0.0069 \\
10122.00--10260.70 & 68 & 10205.23 & 0.0018 & 0.0052 \\
10490.00--10639.80 & 51 & 10591.16 & 0.0035 & 0.0048 \\
10847.00--10996.70 & 41 & 10929.11 & 0.0088 & 0.0029 \\
11224.00--11342.90 & 42 & 11299.85 & -0.0016 & 0.0060 \\
11591.00--11707.80 & 32 & 11652.68 & -0.0088 & 0.0030 \\
11960.00--12076.90 & 36 & 12030.87 & -0.0038 & 0.0050 \\
12400.90--12539.60 & 22 & 12433.23 & 0.0081 & 0.0036 \\
12706.00--12811.90 & 34 & 12766.44 & 0.0089 & 0.0032 \\
\hline
\end{tabular}
  \end{table}
  
  \begin{table}[ht]
    \small
    \caption{HD\,158614 photometric observing seasons (in JD-2440000).}
    \label{HD158614_phot}
    \centering
    \begin{tabular}{ccccc}
\hline\hline
JD range & $\rm N_{obs}$ & $\rm JD_{mean}$ & $\rm \Delta (b+y)/2_{mean}$ & $\sigma_{by}$ \\
\hline
5808.86--5874.82 & 14 & 5851.69 & 0.0039 & 0.0019 \\
6197.84--6245.81 & 14 & 6229.11 & 0.0041 & 0.0016 \\
6567.84--6625.80 & 12 & 6593.16 & -0.0016 & 0.0025 \\
6909.86--6983.81 & 18 & 6966.04 & -0.0006 & 0.0017 \\
7310.84--7384.79 & 6 & 7359.81 & 0.0003 & 0.0017 \\
7702.82--7727.80 & 12 & 7714.31 & -0.0007 & 0.0027 \\
8016.84--8112.79 & 20 & 8052.63 & -0.0026 & 0.0008 \\
8410.83--8474.80 & 12 & 8437.48 & -0.0018 & 0.0014 \\
8756.84--8825.80 & 28 & 8793.04 & -0.0006 & 0.0019 \\
9130.84--9157.82 & 20 & 9145.33 & -0.0007 & 0.0019 \\
9483.84--9530.81 & 24 & 9509.33 & -0.0001 & 0.0019 \\
9862.84--9924.80 & 29 & 9896.25 & -0.0000 & 0.0017 \\
10193.86--10259.81 & 24 & 10237.49 & 0.0008 & 0.0019 \\
10611.82--10639.80 & 18 & 10625.26 & -0.0000 & 0.0018 \\
10949.84--11011.80 & 20 & 10987.01 & 0.0004 & 0.0019 \\
11361.80--11362.80 & 4 & 11362.30 & 0.0010 & 0.0013 \\
\hline
\end{tabular}
  \end{table}
  
  \begin{table}[ht]
    \small
    \caption{HD\,161239 photometric observing seasons (in JD-2440000).}
    \label{HD161239_phot}
    \centering
    \begin{tabular}{ccccc}
\hline\hline
JD range & $\rm N_{obs}$ & $\rm JD_{mean}$ & $\rm \Delta (b+y)/2_{mean}$ & $\sigma_{by}$ \\
\hline
5808.86--5958.78 & 22 & 5885.05 & 0.0015 & 0.0046 \\
6193.84--6278.80 & 30 & 6229.50 & 0.0012 & 0.0020 \\
6517.86--6692.79 & 21 & 6613.30 & -0.0017 & 0.0020 \\
6908.85--6983.82 & 12 & 6959.49 & -0.0004 & 0.0015 \\
7289.84--7428.78 & 12 & 7376.47 & 0.0002 & 0.0012 \\
7667.84--7793.78 & 14 & 7755.80 & 0.0027 & 0.0019 \\
8028.84--8126.79 & 15 & 8062.43 & 0.0005 & 0.0026 \\
8410.84--8437.84 & 4 & 8424.34 & -0.0025 & 0.0025 \\
8776.84--8827.82 & 18 & 8799.49 & -0.0012 & 0.0025 \\
9116.85--9170.83 & 22 & 9152.47 & -0.0015 & 0.0016 \\
9473.85--9529.84 & 24 & 9507.34 & 0.0006 & 0.0014 \\
9812.86--9924.81 & 41 & 9880.83 & -0.0009 & 0.0026 \\
10167.87--10255.83 & 24 & 10225.17 & 0.0011 & 0.0033 \\
10617.83--10639.81 & 16 & 10624.69 & -0.0033 & 0.0018 \\
10972.84--11045.79 & 20 & 11001.21 & 0.0027 & 0.0052 \\
11331.82--11425.78 & 7 & 11358.38 & -0.0030 & 0.0007 \\
\hline
\end{tabular}
  \end{table}
  
  \begin{table}[ht]
    \small
    \caption{HD\,182572 photometric observing seasons (in JD-2440000).}
    \label{HD182572_phot}
    \centering
    \begin{tabular}{ccccc}
\hline\hline
JD range & $\rm N_{obs}$ & $\rm JD_{mean}$ & $\rm \Delta (b+y)/2_{mean}$ & $\sigma_{by}$ \\
\hline
5862.84--5996.77 & 28 & 5938.52 & 0.0021 & 0.0020 \\
6219.85--6326.78 & 21 & 6250.92 & 0.0011 & 0.0014 \\
6592.85--6719.77 & 22 & 6635.60 & 0.0014 & 0.0017 \\
6957.84--7072.78 & 25 & 7005.94 & 0.0019 & 0.0019 \\
7310.86--7425.79 & 12 & 7381.48 & -0.0004 & 0.0013 \\
7741.81--7800.78 & 22 & 7777.16 & 0.0006 & 0.0013 \\
8066.84--8175.77 & 20 & 8131.90 & -0.0002 & 0.0012 \\
8824.82--8873.80 & 8 & 8848.06 & 0.0014 & 0.0018 \\
9095.00--9301.60 & 95 & 9174.20 & 0.0016 & 0.0014 \\
9443.00--9654.60 & 71 & 9536.42 & 0.0000 & 0.0015 \\
9806.00--10017.60 & 60 & 9911.08 & 0.0000 & 0.0013 \\
10173.00--10365.77 & 49 & 10266.27 & 0.0005 & 0.0018 \\
10528.00--10777.60 & 72 & 10676.21 & 0.0000 & 0.0014 \\
10897.00--11138.60 & 92 & 11021.03 & -0.0007 & 0.0014 \\
11262.00--11507.60 & 67 & 11400.17 & -0.0015 & 0.0015 \\
11621.00--11874.60 & 50 & 11720.52 & -0.0023 & 0.0016 \\
11986.00--12076.90 & 31 & 12044.77 & -0.0012 & 0.0014 \\
12179.60--12228.60 & 7 & 12215.03 & -0.0000 & 0.0008 \\
12401.00--12576.60 & 48 & 12467.52 & -0.0008 & 0.0012 \\
12737.00--12813.90 & 25 & 12770.86 & 0.0004 & 0.0014 \\
\hline
\end{tabular}
  \end{table}
  
  \begin{table}[ht]
    \small
    \caption{HD\,185144 photometric observing seasons (in JD-2440000).}
    \label{HD185144_phot}
    \centering
    \begin{tabular}{ccccc}
\hline\hline
JD range & $\rm N_{obs}$ & $\rm JD_{mean}$ & $\rm \Delta (b+y)/2_{mean}$ & $\sigma_{by}$ \\
\hline
5862.85--5993.77 & 29 & 5939.05 & 0.0000 & 0.0016 \\
6225.85--6331.78 & 18 & 6268.04 & 0.0004 & 0.0015 \\
6596.84--6718.78 & 14 & 6642.96 & -0.0044 & 0.0012 \\
6963.84--7076.77 & 22 & 7016.17 & -0.0029 & 0.0027 \\
7383.81--7438.77 & 10 & 7417.98 & -0.0009 & 0.0024 \\
7727.83--7793.78 & 17 & 7775.85 & -0.0029 & 0.0023 \\
8100.82--8174.77 & 20 & 8137.40 & -0.0008 & 0.0016 \\
8517.78--8525.79 & 6 & 8522.78 & 0.0002 & 0.0014 \\
8792.84--8874.79 & 12 & 8831.32 & -0.0017 & 0.0015 \\
9102.00--9291.60 & 64 & 9178.87 & -0.0004 & 0.0019 \\
9471.00--9689.60 & 68 & 9571.66 & 0.0001 & 0.0017 \\
9814.00--10054.60 & 58 & 9917.75 & 0.0009 & 0.0013 \\
10204.00--10266.00 & 24 & 10222.96 & 0.0023 & 0.0016 \\
10614.90--10781.60 & 36 & 10690.41 & 0.0004 & 0.0018 \\
10893.00--11123.60 & 36 & 11047.45 & 0.0020 & 0.0017 \\
11275.00--11496.60 & 33 & 11441.12 & 0.0003 & 0.0013 \\
11629.00--11867.60 & 24 & 11750.03 & -0.0010 & 0.0016 \\
12037.90--12067.90 & 6 & 12060.57 & 0.0016 & 0.0011 \\
12413.00--12590.60 & 39 & 12489.92 & 0.0001 & 0.0019 \\
12741.00--12815.90 & 13 & 12788.49 & 0.0019 & 0.0018 \\
\hline
\end{tabular}
  \end{table}
  
  \begin{table}[ht]
    \small
    \caption{HD\,190007 photometric observing seasons (in JD-2440000).}
    \label{HD190007_phot}
    \centering
    \begin{tabular}{ccccc}
\hline\hline
JD range & $\rm N_{obs}$ & $\rm JD_{mean}$ & $\rm \Delta (b+y)/2_{mean}$ & $\sigma_{by}$ \\
\hline
5862.86--6015.77 & 36 & 5958.41 & 0.0027 & 0.0076 \\
6219.86--6376.77 & 37 & 6270.12 & -0.0005 & 0.0038 \\
6588.85--6732.77 & 48 & 6641.82 & -0.0074 & 0.0059 \\
6957.85--7086.77 & 58 & 7021.71 & 0.0029 & 0.0083 \\
7349.84--7456.77 & 40 & 7415.05 & 0.0044 & 0.0045 \\
7741.82--7793.79 & 36 & 7772.75 & 0.0074 & 0.0036 \\
8094.83--8190.77 & 40 & 8138.60 & 0.0010 & 0.0039 \\
8474.82--8537.78 & 16 & 8518.92 & 0.0075 & 0.0046 \\
8824.83--8892.80 & 20 & 8859.01 & 0.0027 & 0.0042 \\
9095.00--9318.60 & 59 & 9192.31 & 0.0042 & 0.0040 \\
9466.00--9662.60 & 62 & 9567.10 & -0.0078 & 0.0092 \\
9832.00--10040.60 & 63 & 9951.08 & 0.0013 & 0.0055 \\
10198.00--10366.77 & 42 & 10282.24 & -0.0104 & 0.0101 \\
10563.00--10768.60 & 42 & 10692.74 & -0.0010 & 0.0038 \\
10937.00--11134.60 & 42 & 11056.81 & 0.0020 & 0.0050 \\
11296.00--11354.90 & 8 & 11327.19 & 0.0040 & 0.0021 \\
11448.79--11499.60 & 24 & 11462.16 & 0.0032 & 0.0022 \\
11678.90--11710.90 & 9 & 11694.27 & 0.0030 & 0.0035 \\
12061.90--12077.00 & 8 & 12068.94 & -0.0058 & 0.0028 \\
12214.60--12228.60 & 5 & 12220.00 & 0.0012 & 0.0008 \\
12392.00--12461.90 & 19 & 12426.67 & -0.0032 & 0.0048 \\
12537.70--12600.60 & 20 & 12560.46 & 0.0008 & 0.0029 \\
12748.00--12812.90 & 16 & 12784.92 & -0.0027 & 0.0026 \\
\hline
\end{tabular}
  \end{table}
  
  \begin{table}[ht]
    \small
    \caption{HD\,201091 photometric observing seasons (in JD-2440000).}
    \label{HD201091_phot}
    \centering
    \begin{tabular}{ccccc}
\hline\hline
JD range & $\rm N_{obs}$ & $\rm JD_{mean}$ & $\rm \Delta (b+y)/2_{mean}$ & $\sigma_{by}$ \\
\hline
5862.86--6015.77 & 38 & 5958.62 & 0.0023 & 0.0026 \\
6234.85--6376.77 & 34 & 6302.88 & 0.0041 & 0.0020 \\
6590.85--6728.78 & 28 & 6658.64 & -0.0008 & 0.0018 \\
6965.84--7086.79 & 39 & 7039.35 & 0.0001 & 0.0027 \\
7349.84--7456.78 & 38 & 7411.18 & 0.0023 & 0.0026 \\
7722.84--7792.80 & 34 & 7767.81 & 0.0010 & 0.0021 \\
8094.84--8202.79 & 38 & 8134.95 & -0.0003 & 0.0017 \\
8438.84--8464.84 & 3 & 8447.51 & 0.0005 & 0.0000 \\
8792.84--8888.81 & 28 & 8849.54 & -0.0019 & 0.0017 \\
9178.84--9255.80 & 12 & 9213.32 & -0.0022 & 0.0031 \\
9528.84--9611.79 & 27 & 9564.46 & -0.0006 & 0.0036 \\
9922.83--10003.78 & 24 & 9976.88 & 0.0013 & 0.0024 \\
10301.83--10441.60 & 39 & 10363.30 & 0.0011 & 0.0017 \\
10613.90--10800.60 & 48 & 10720.87 & -0.0003 & 0.0024 \\
10914.00--11172.60 & 40 & 11085.72 & -0.0053 & 0.0025 \\
11449.80--11535.60 & 27 & 11482.61 & -0.0017 & 0.0018 \\
11807.90--11908.60 & 16 & 11869.75 & -0.0003 & 0.0016 \\
12228.70--12275.60 & 9 & 12263.17 & -0.0010 & 0.0024 \\
12369.00--12448.90 & 28 & 12430.55 & 0.0002 & 0.0011 \\
12537.80--12634.60 & 45 & 12583.88 & 0.0000 & 0.0024 \\
12733.00--12816.00 & 28 & 12792.65 & 0.0002 & 0.0028 \\
\hline
\end{tabular}
  \end{table}
  
  \begin{table}[ht]
    \small
    \caption{HD\,201092 photometric observing seasons (in JD-2440000).}
    \label{HD201092_phot}
    \centering
    \begin{tabular}{ccccc}
\hline\hline
JD range & $\rm N_{obs}$ & $\rm JD_{mean}$ & $\rm \Delta (b+y)/2_{mean}$ & $\sigma_{by}$ \\
\hline
5862.86--6015.77 & 39 & 5958.45 & 0.0020 & 0.0022 \\
6234.85--6376.77 & 33 & 6302.48 & 0.0036 & 0.0020 \\
6590.85--6728.78 & 28 & 6658.64 & 0.0032 & 0.0016 \\
6965.84--7086.79 & 39 & 7039.35 & 0.0021 & 0.0016 \\
7349.84--7456.78 & 38 & 7411.18 & 0.0037 & 0.0026 \\
7722.84--7792.80 & 34 & 7767.81 & -0.0002 & 0.0027 \\
8094.84--8202.79 & 38 & 8134.95 & -0.0023 & 0.0020 \\
8438.84--8464.84 & 4 & 8451.84 & -0.0021 & 0.0017 \\
8792.84--8888.81 & 28 & 8849.54 & -0.0021 & 0.0025 \\
9178.84--9255.80 & 12 & 9213.32 & -0.0046 & 0.0024 \\
9528.84--9611.79 & 27 & 9564.46 & -0.0024 & 0.0025 \\
9922.83--10003.78 & 24 & 9976.88 & -0.0025 & 0.0018 \\
10301.83--10373.77 & 28 & 10341.73 & -0.0030 & 0.0021 \\
10712.80--10754.77 & 20 & 10726.50 & -0.0015 & 0.0019 \\
11045.81--11094.79 & 14 & 11079.94 & -0.0041 & 0.0024 \\
11449.80--11500.77 & 19 & 11477.63 & 0.0007 & 0.0019 \\
\hline
\end{tabular}
  \end{table}
  
  \begin{table}[ht]
    \small
    \caption{HD\,206860 photometric observing seasons (in JD-2440000).}
    \label{HD206860_phot}
    \centering
    \begin{tabular}{ccccc}
\hline\hline
JD range & $\rm N_{obs}$ & $\rm JD_{mean}$ & $\rm V_{mean}$ & $\sigma_V$ \\
\hline
7901.57--7901.75 & 3 & 7901.66 & 5.9817 & 0.0040 \\
8017.71--8066.95 & 9 & 8056.78 & 5.9810 & 0.0105 \\
8202.57--8249.68 & 8 & 8231.97 & 5.9822 & 0.0046 \\
8363.59--8412.84 & 12 & 8382.81 & 5.9782 & 0.0068 \\
8550.01--8565.75 & 8 & 8557.88 & 5.9654 & 0.0046 \\
8758.04--8806.57 & 16 & 8783.24 & 5.9782 & 0.0109 \\
9126.97--9163.92 & 22 & 9147.16 & 5.9724 & 0.0062 \\
9239.73--9337.57 & 43 & 9272.48 & 5.9729 & 0.0058 \\
9491.97--9547.95 & 18 & 9525.71 & 5.9769 & 0.0044 \\
9633.70--9706.57 & 36 & 9663.89 & 5.9753 & 0.0066 \\
9862.96--9908.84 & 33 & 9881.87 & 5.9786 & 0.0101 \\
9986.70--10071.57 & 33 & 10022.10 & 5.9804 & 0.0112 \\
10392.71--10443.58 & 20 & 10422.48 & 5.9752 & 0.0116 \\
10620.97--10642.95 & 3 & 10628.96 & 5.9820 & 0.0026 \\
10714.82--10806.59 & 33 & 10755.31 & 5.9725 & 0.0075 \\
11010.53--11024.48 & 12 & 11015.37 & 5.9701 & 0.0136 \\
11077.74--11179.57 & 51 & 11124.51 & 5.9731 & 0.0066 \\
\hline
\end{tabular}
  \end{table}
  
  \section{Mount Wilson data}\label{app_B}
  \begin{table}[ht]
    \small
    \caption{HD\,1835 S-index observing seasons (in JD-2440000).}
    \label{HD1835_chrom}
    \centering
    \begin{tabular}{ccccc}
\hline\hline
JD range & $\rm N_{obs}$ & $\rm JD_{mean}$ & $\rm S_{mean}$ & $\sigma_S$ \\
\hline
4423.93--4608.60 & 244 & 4487.05 & 0.3457 & 0.0126 \\
4803.99--4920.69 & 177 & 4869.14 & 0.3319 & 0.0155 \\
5198.93--5309.63 & 68 & 5236.36 & 0.3423 & 0.0182 \\
5542.98--5725.61 & 130 & 5632.87 & 0.3475 & 0.0170 \\
5905.97--6069.64 & 160 & 5983.71 & 0.3408 & 0.0180 \\
6283.96--6427.70 & 139 & 6353.10 & 0.3257 & 0.0122 \\
6654.99--6808.61 & 165 & 6710.21 & 0.3313 & 0.0126 \\
7027.91--7183.60 & 159 & 7089.48 & 0.3161 & 0.0132 \\
7388.95--7541.60 & 161 & 7453.66 & 0.3002 & 0.0116 \\
7766.92--7878.67 & 75 & 7826.33 & 0.3341 & 0.0180 \\
8108.98--8244.62 & 75 & 8183.79 & 0.3367 & 0.0148 \\
8489.97--8649.60 & 45 & 8602.55 & 0.3643 & 0.0138 \\
8869.91--8988.64 & 117 & 8918.89 & 0.3636 & 0.0179 \\
9227.95--9344.62 & 94 & 9287.36 & 0.3515 & 0.0270 \\
9581.90--9736.61 & 183 & 9647.64 & 0.3444 & 0.0176 \\
9951.93--10096.62 & 108 & 10013.95 & 0.3534 & 0.0144 \\
10311.99--10458.61 & 117 & 10376.43 & 0.3586 & 0.0182 \\
10676.95--10809.62 & 72 & 10719.09 & 0.3473 & 0.0139 \\
11047.93--11186.62 & 102 & 11120.61 & 0.3869 & 0.0160 \\
11406.91--11557.60 & 86 & 11487.55 & 0.3744 & 0.0196 \\
11778.89--11900.64 & 33 & 11841.12 & 0.3693 & 0.0186 \\
12121.94--12197.79 & 42 & 12163.00 & 0.3655 & 0.0203 \\
\hline
\end{tabular}
  \end{table}
  
  \begin{table}[ht]
    \small
    \caption{HD\,10476 S-index observing seasons (in JD-2440000).}
    \label{HD10476_chrom}
    \centering
    \begin{tabular}{ccccc}
\hline\hline
JD range & $\rm N_{obs}$ & $\rm JD_{mean}$ & $\rm S_{mean}$ & $\sigma_S$ \\
\hline
3347.80--3481.80 & 9 & 3414.80 & 0.2067 & 0.0098 \\
3742.80--3892.80 & 7 & 3795.23 & 0.1963 & 0.0018 \\
4172.80--4226.80 & 5 & 4203.40 & 0.1841 & 0.0035 \\
4412.00--4566.76 & 190 & 4477.34 & 0.1770 & 0.0046 \\
4828.01--4900.91 & 30 & 4866.93 & 0.1731 & 0.0045 \\
5197.97--5345.67 & 107 & 5252.16 & 0.1765 & 0.0047 \\
5544.97--5743.62 & 138 & 5658.10 & 0.1969 & 0.0071 \\
5883.99--6070.69 & 54 & 5961.79 & 0.2165 & 0.0102 \\
6286.00--6470.63 & 114 & 6383.20 & 0.2093 & 0.0082 \\
6672.99--6817.62 & 93 & 6731.20 & 0.2108 & 0.0099 \\
7024.93--7198.65 & 111 & 7098.47 & 0.1942 & 0.0059 \\
7406.00--7559.62 & 78 & 7484.01 & 0.1788 & 0.0044 \\
7766.00--7912.61 & 75 & 7832.82 & 0.1824 & 0.0102 \\
8122.98--8256.66 & 57 & 8181.18 & 0.1734 & 0.0047 \\
8597.70--8677.63 & 42 & 8635.03 & 0.1750 & 0.0047 \\
8873.88--9015.62 & 78 & 8936.08 & 0.1742 & 0.0042 \\
9251.93--9368.63 & 99 & 9310.46 & 0.1777 & 0.0032 \\
9581.94--9755.62 & 132 & 9660.45 & 0.1846 & 0.0056 \\
9957.00--10107.61 & 87 & 10032.03 & 0.1878 & 0.0077 \\
10315.95--10482.61 & 101 & 10395.07 & 0.2005 & 0.0096 \\
10700.92--10841.61 & 53 & 10755.48 & 0.1875 & 0.0052 \\
11073.92--11211.60 & 78 & 11140.87 & 0.1878 & 0.0085 \\
11457.92--11557.66 & 51 & 11513.83 & 0.1829 & 0.0081 \\
11779.94--11907.62 & 33 & 11855.31 & 0.1740 & 0.0072 \\
12141.99--12190.86 & 24 & 12163.53 & 0.1691 & 0.0059 \\
\hline
\end{tabular}
  \end{table}
  
  \begin{table}[ht]
    \small
    \caption{HD\,13421 S-index observing seasons (in JD-2440000).}
    \label{HD13421_chrom}
    \centering
    \begin{tabular}{ccccc}
\hline\hline
JD range & $\rm N_{obs}$ & $\rm JD_{mean}$ & $\rm S_{mean}$ & $\sigma_S$ \\
\hline
4423.97--4566.79 & 184 & 4479.53 & 0.1287 & 0.0029 \\
4828.00--4986.65 & 52 & 4906.68 & 0.1305 & 0.0034 \\
5193.00--5375.60 & 159 & 5280.31 & 0.1285 & 0.0040 \\
5544.98--5743.64 & 194 & 5653.71 & 0.1283 & 0.0042 \\
5918.96--6119.65 & 271 & 6009.33 & 0.1288 & 0.0045 \\
6291.98--6472.63 & 243 & 6392.43 & 0.1285 & 0.0030 \\
6673.94--6845.61 & 244 & 6751.18 & 0.1281 & 0.0038 \\
7024.97--7213.61 & 332 & 7118.74 & 0.1296 & 0.0034 \\
7416.94--7579.61 & 258 & 7489.30 & 0.1288 & 0.0042 \\
7762.01--7942.61 & 282 & 7856.87 & 0.1292 & 0.0030 \\
8109.01--8301.61 & 199 & 8217.11 & 0.1310 & 0.0032 \\
8576.80--8657.62 & 63 & 8631.66 & 0.1321 & 0.0035 \\
8873.89--9020.62 & 185 & 8942.78 & 0.1291 & 0.0029 \\
9250.92--9375.62 & 185 & 9313.10 & 0.1275 & 0.0025 \\
9600.93--9769.61 & 235 & 9685.95 & 0.1271 & 0.0035 \\
9951.99--10127.60 & 178 & 10034.67 & 0.1275 & 0.0033 \\
10315.00--10485.61 & 189 & 10395.99 & 0.1285 & 0.0037 \\
10704.93--10844.61 & 93 & 10763.09 & 0.1297 & 0.0043 \\
11067.93--11213.62 & 165 & 11141.12 & 0.1302 & 0.0046 \\
11441.91--11571.61 & 131 & 11512.28 & 0.1291 & 0.0039 \\
11779.95--11913.64 & 79 & 11851.57 & 0.1288 & 0.0041 \\
12138.96--12205.83 & 63 & 12165.73 & 0.1249 & 0.0045 \\
\hline
\end{tabular}
  \end{table}
  
  \begin{table}[ht]
    \small
    \caption{HD\,18256 S-index observing seasons (in JD-2440000).}
    \label{HD18256_chrom}
    \centering
    \begin{tabular}{ccccc}
\hline\hline
JD range & $\rm N_{obs}$ & $\rm JD_{mean}$ & $\rm S_{mean}$ & $\sigma_S$ \\
\hline
4849.99--5026.64 & 24 & 4888.04 & 0.1819 & 0.0040 \\
5211.97--5346.64 & 54 & 5264.49 & 0.1755 & 0.0045 \\
5569.02--5741.65 & 123 & 5670.71 & 0.1798 & 0.0056 \\
5918.94--6121.63 & 134 & 6011.01 & 0.1906 & 0.0061 \\
6304.00--6486.62 & 219 & 6398.11 & 0.1886 & 0.0041 \\
6673.99--6831.65 & 168 & 6740.01 & 0.1790 & 0.0048 \\
7025.00--7215.61 & 209 & 7118.95 & 0.1798 & 0.0052 \\
7420.97--7564.64 & 126 & 7484.25 & 0.1749 & 0.0048 \\
7803.99--7942.62 & 75 & 7870.86 & 0.1759 & 0.0038 \\
8152.99--8300.64 & 69 & 8233.79 & 0.1808 & 0.0053 \\
8645.65--8657.60 & 14 & 8652.63 & 0.1697 & 0.0034 \\
8886.00--9019.66 & 120 & 8948.63 & 0.1823 & 0.0040 \\
9264.97--9369.66 & 39 & 9317.41 & 0.1837 & 0.0054 \\
9615.89--9766.61 & 37 & 9680.20 & 0.1764 & 0.0044 \\
9978.97--10127.61 & 36 & 10038.65 & 0.1690 & 0.0044 \\
10316.95--10482.65 & 45 & 10402.28 & 0.1642 & 0.0035 \\
10709.94--10844.64 & 33 & 10771.54 & 0.1640 & 0.0057 \\
11070.96--11212.66 & 36 & 11144.71 & 0.1851 & 0.0064 \\
11442.92--11572.61 & 36 & 11510.28 & 0.1667 & 0.0043 \\
11803.96--11913.66 & 15 & 11879.35 & 0.2012 & 0.0254 \\
12146.00--12181.89 & 12 & 12162.19 & 0.1790 & 0.0056 \\
\hline
\end{tabular}
  \end{table}
  
  \begin{table}[ht]
    \small
    \caption{HD\,20630 S-index observing seasons (in JD-2440000).}
    \label{HD20630_chrom}
    \centering
    \begin{tabular}{ccccc}
\hline\hline
JD range & $\rm N_{obs}$ & $\rm JD_{mean}$ & $\rm S_{mean}$ & $\sigma_S$ \\
\hline
4172.80--4260.80 & 12 & 4207.38 & 0.3661 & 0.0128 \\
4426.98--4659.64 & 253 & 4511.54 & 0.3519 & 0.0112 \\
4848.00--4900.96 & 102 & 4872.72 & 0.3362 & 0.0105 \\
5199.01--5375.61 & 69 & 5260.97 & 0.3362 & 0.0083 \\
5563.99--5756.61 & 63 & 5700.05 & 0.3514 & 0.0125 \\
5918.98--6125.61 & 177 & 6022.36 & 0.3561 & 0.0133 \\
6301.97--6487.67 & 177 & 6388.84 & 0.3416 & 0.0108 \\
6678.97--6843.64 & 148 & 6743.18 & 0.3402 & 0.0125 \\
7024.99--7216.63 & 189 & 7108.04 & 0.3290 & 0.0139 \\
7420.98--7574.65 & 135 & 7484.70 & 0.3454 & 0.0095 \\
7804.02--7942.63 & 79 & 7861.40 & 0.3698 & 0.0114 \\
8166.00--8311.62 & 84 & 8248.89 & 0.3641 & 0.0144 \\
8597.75--8694.63 & 66 & 8642.69 & 0.3555 & 0.0141 \\
8873.91--9051.64 & 105 & 8946.18 & 0.3370 & 0.0142 \\
9265.92--9399.65 & 84 & 9330.95 & 0.3273 & 0.0123 \\
9615.91--9771.62 & 126 & 9691.19 & 0.3309 & 0.0156 \\
\hline
\end{tabular}
  \end{table}
  
  \begin{table}[ht]
    \small
    \caption{HD\,25998 S-index observing seasons (in JD-2440000).}
    \label{HD25998_chrom}
    \centering
    \begin{tabular}{ccccc}
\hline\hline
JD range & $\rm N_{obs}$ & $\rm JD_{mean}$ & $\rm S_{mean}$ & $\sigma_S$ \\
\hline
4446.99--4658.66 & 235 & 4514.18 & 0.2903 & 0.0084 \\
4869.03--4886.91 & 15 & 4877.14 & 0.2877 & 0.0068 \\
5212.00--5349.69 & 29 & 5268.76 & 0.2805 & 0.0083 \\
5576.03--5772.65 & 126 & 5680.79 & 0.2771 & 0.0089 \\
5945.01--6121.65 & 102 & 6028.96 & 0.2858 & 0.0092 \\
6329.98--6496.63 & 174 & 6412.01 & 0.2826 & 0.0066 \\
6703.03--6845.68 & 116 & 6757.27 & 0.2825 & 0.0076 \\
7047.01--7240.64 & 157 & 7151.76 & 0.2746 & 0.0069 \\
7444.84--7574.68 & 123 & 7500.30 & 0.2685 & 0.0060 \\
7805.00--7942.66 & 63 & 7867.74 & 0.2735 & 0.0058 \\
8121.98--8312.66 & 75 & 8256.29 & 0.2755 & 0.0075 \\
8639.68--8696.64 & 45 & 8662.27 & 0.2744 & 0.0058 \\
8908.90--9019.69 & 87 & 8959.05 & 0.2570 & 0.0057 \\
9300.84--9397.67 & 18 & 9341.59 & 0.2705 & 0.0056 \\
9642.92--9759.68 & 27 & 9697.92 & 0.2746 & 0.0053 \\
9983.00--10137.62 & 30 & 10050.24 & 0.2784 & 0.0132 \\
10342.95--10500.62 & 30 & 10411.43 & 0.2658 & 0.0070 \\
10766.89--10844.66 & 18 & 10810.41 & 0.2629 & 0.0075 \\
11067.98--11222.66 & 27 & 11145.61 & 0.2632 & 0.0077 \\
11441.98--11571.67 & 15 & 11499.85 & 0.2538 & 0.0076 \\
11805.01--11943.63 & 24 & 11884.92 & 0.2632 & 0.0151 \\
12153.00--12205.89 & 9 & 12176.28 & 0.2673 & 0.0118 \\
\hline
\end{tabular}
  \end{table}
  
  \begin{table}[ht]
    \small
    \caption{HD\,35296 S-index observing seasons (in JD-2440000).}
    \label{HD35296_chrom}
    \centering
    \begin{tabular}{ccccc}
\hline\hline
JD range & $\rm N_{obs}$ & $\rm JD_{mean}$ & $\rm S_{mean}$ & $\sigma_S$ \\
\hline
4545.96--4681.64 & 69 & 4609.95 & 0.3147 & 0.0085 \\
4871.03--5026.70 & 33 & 4913.23 & 0.3182 & 0.0074 \\
5247.97--5432.65 & 96 & 5335.69 & 0.3105 & 0.0092 \\
5591.03--5794.69 & 203 & 5727.50 & 0.3096 & 0.0102 \\
5958.99--6122.70 & 105 & 6045.11 & 0.3082 & 0.0088 \\
6330.01--6516.67 & 207 & 6427.72 & 0.3165 & 0.0089 \\
6713.03--6877.63 & 141 & 6777.85 & 0.3117 & 0.0082 \\
7058.02--7242.67 & 156 & 7158.63 & 0.3007 & 0.0076 \\
7450.98--7592.67 & 105 & 7515.54 & 0.2933 & 0.0072 \\
7832.99--7957.68 & 87 & 7889.91 & 0.2938 & 0.0068 \\
8135.99--8346.63 & 78 & 8271.90 & 0.2948 & 0.0075 \\
8576.91--8722.63 & 69 & 8668.41 & 0.2869 & 0.0081 \\
8929.00--9069.64 & 81 & 9000.97 & 0.2845 & 0.0114 \\
9296.95--9433.66 & 121 & 9359.44 & 0.2920 & 0.0070 \\
9647.92--9784.67 & 156 & 9715.56 & 0.2894 & 0.0101 \\
10011.99--10160.65 & 114 & 10089.97 & 0.2922 & 0.0115 \\
10345.97--10522.65 & 132 & 10445.48 & 0.2850 & 0.0093 \\
10761.93--10884.66 & 96 & 10826.98 & 0.2846 & 0.0131 \\
11106.93--11250.64 & 121 & 11169.40 & 0.3019 & 0.0121 \\
11471.96--11623.63 & 111 & 11531.38 & 0.2866 & 0.0119 \\
11837.94--11940.70 & 48 & 11897.94 & 0.2982 & 0.0159 \\
\hline
\end{tabular}
  \end{table}
  
  \begin{table}[ht]
    \small
    \caption{HD\,39587 S-index observing seasons (in JD-2440000).}
    \label{HD39587_chrom}
    \centering
    \begin{tabular}{ccccc}
\hline\hline
JD range & $\rm N_{obs}$ & $\rm JD_{mean}$ & $\rm S_{mean}$ & $\sigma_S$ \\
\hline
4545.97--4681.65 & 72 & 4610.79 & 0.3101 & 0.0089 \\
4874.98--5026.69 & 66 & 4947.15 & 0.3165 & 0.0115 \\
5248.00--5431.66 & 90 & 5332.46 & 0.3203 & 0.0103 \\
5592.02--5802.66 & 840 & 5702.05 & 0.3085 & 0.0055 \\
5978.01--6156.67 & 141 & 6051.21 & 0.3162 & 0.0105 \\
6330.01--6525.63 & 893 & 6448.73 & 0.3183 & 0.0068 \\
6713.02--6877.64 & 132 & 6779.05 & 0.3125 & 0.0131 \\
7065.01--7263.69 & 201 & 7152.61 & 0.3131 & 0.0118 \\
7421.03--7621.64 & 153 & 7511.13 & 0.3007 & 0.0094 \\
7826.96--7957.69 & 81 & 7877.63 & 0.3018 & 0.0077 \\
8209.96--8347.64 & 90 & 8265.18 & 0.2978 & 0.0085 \\
8597.85--8732.65 & 102 & 8664.44 & 0.3087 & 0.0066 \\
8874.93--9069.64 & 84 & 8977.03 & 0.2989 & 0.0119 \\
9299.95--9433.67 & 111 & 9356.35 & 0.2927 & 0.0098 \\
9660.94--9784.68 & 138 & 9716.19 & 0.2936 & 0.0084 \\
10012.00--10160.66 & 104 & 10097.80 & 0.2860 & 0.0174 \\
10347.00--10537.62 & 151 & 10467.07 & 0.3022 & 0.0110 \\
10761.98--10884.67 & 90 & 10825.55 & 0.2892 & 0.0103 \\
11117.97--11257.62 & 114 & 11186.37 & 0.3251 & 0.0154 \\
11471.98--11623.63 & 105 & 11536.90 & 0.3183 & 0.0134 \\
11862.89--11988.62 & 48 & 11910.68 & 0.3209 & 0.0170 \\
\hline
\end{tabular}
  \end{table}
  
  \begin{table}[ht]
    \small
    \caption{HD\,72905 S-index observing seasons (in JD-2440000).}
    \label{HD72905_chrom}
    \centering
    \begin{tabular}{ccccc}
\hline\hline
JD range & $\rm N_{obs}$ & $\rm JD_{mean}$ & $\rm S_{mean}$ & $\sigma_S$ \\
\hline
5634.06--5812.69 & 114 & 5739.52 & 0.3524 & 0.0102 \\
6004.00--6189.69 & 90 & 6129.99 & 0.3583 & 0.0118 \\
6375.03--6553.71 & 116 & 6472.71 & 0.3644 & 0.0103 \\
6740.02--6906.64 & 93 & 6830.12 & 0.3577 & 0.0127 \\
7107.99--7295.71 & 118 & 7199.89 & 0.3584 & 0.0126 \\
7467.02--7631.66 & 90 & 7560.80 & 0.3544 & 0.0162 \\
7863.00--7975.76 & 26 & 7898.64 & 0.3671 & 0.0093 \\
8167.99--8361.68 & 36 & 8283.42 & 0.3855 & 0.0188 \\
8656.81--8732.69 & 12 & 8709.71 & 0.3693 & 0.0136 \\
9015.91--9102.67 & 27 & 9079.39 & 0.3500 & 0.0095 \\
9326.96--9464.67 & 46 & 9406.74 & 0.3659 & 0.0128 \\
9759.85--9793.74 & 9 & 9775.13 & 0.3781 & 0.0061 \\
\hline
\end{tabular}
  \end{table}
  
  \begin{table}[ht]
    \small
    \caption{HD\,75332 S-index observing seasons (in JD-2440000).}
    \label{HD75332_chrom}
    \centering
    \begin{tabular}{ccccc}
\hline\hline
JD range & $\rm N_{obs}$ & $\rm JD_{mean}$ & $\rm S_{mean}$ & $\sigma_S$ \\
\hline
3236.80--3238.80 & 3 & 3237.80 & 0.2845 & 0.0077 \\
3476.80--3542.80 & 4 & 3494.05 & 0.2688 & 0.0208 \\
3891.80--3984.80 & 4 & 3917.05 & 0.2761 & 0.0031 \\
4169.80--4277.80 & 15 & 4215.20 & 0.2771 & 0.0054 \\
4545.97--4710.65 & 90 & 4651.94 & 0.2511 & 0.0082 \\
4875.00--5082.67 & 39 & 5014.33 & 0.2517 & 0.0091 \\
5308.97--5440.64 & 120 & 5373.58 & 0.2720 & 0.0084 \\
5634.02--5853.68 & 309 & 5758.25 & 0.2735 & 0.0090 \\
6010.02--6193.72 & 180 & 6141.00 & 0.2776 & 0.0090 \\
6383.03--6553.75 & 227 & 6477.82 & 0.2724 & 0.0074 \\
6748.04--6921.66 & 157 & 6842.01 & 0.2928 & 0.0111 \\
7110.01--7297.66 & 226 & 7195.71 & 0.2825 & 0.0070 \\
7472.02--7631.67 & 138 & 7566.47 & 0.2720 & 0.0086 \\
7832.04--7981.72 & 99 & 7907.84 & 0.2800 & 0.0106 \\
8186.97--8375.66 & 123 & 8293.28 & 0.2976 & 0.0094 \\
8646.78--8741.66 & 72 & 8695.46 & 0.2843 & 0.0082 \\
8977.00--9118.66 & 125 & 9056.03 & 0.2736 & 0.0085 \\
9313.01--9463.72 & 24 & 9388.10 & 0.2798 & 0.0079 \\
9688.98--9795.76 & 18 & 9740.05 & 0.2777 & 0.0065 \\
10085.99--10179.69 & 18 & 10132.98 & 0.2737 & 0.0083 \\
10434.93--10564.67 & 24 & 10495.44 & 0.2861 & 0.0102 \\
10805.99--10932.64 & 21 & 10870.67 & 0.2749 & 0.0091 \\
11165.96--11304.65 & 63 & 11257.08 & 0.2946 & 0.0126 \\
11525.94--11646.70 & 72 & 11592.43 & 0.2837 & 0.0122 \\
11898.96--12001.67 & 42 & 11942.68 & 0.2813 & 0.0112 \\
\hline
\end{tabular}
  \end{table}
  
  \begin{table}[ht]
    \caption{HD\,81809 S-index observing seasons (in JD-2440000).}
    \label{HD81809_chrom}
    \centering
    \begin{tabular}{ccccc}
\hline\hline
JD range & $\rm N_{obs}$ & $\rm JD_{mean}$ & $\rm S_{mean}$ & $\sigma_S$ \\
\hline
3236.80--3257.80 & 5 & 3245.60 & 0.1778 & 0.0072 \\
3891.80--3984.80 & 5 & 3930.40 & 0.1698 & 0.0033 \\
4191.80--4277.80 & 6 & 4235.97 & 0.1612 & 0.0049 \\
4546.02--4710.68 & 105 & 4654.37 & 0.1628 & 0.0058 \\
4940.02--5082.69 & 36 & 5009.36 & 0.1783 & 0.0044 \\
5298.04--5440.69 & 138 & 5366.08 & 0.1817 & 0.0075 \\
5634.04--5813.73 & 60 & 5729.59 & 0.1841 & 0.0064 \\
6071.02--6189.66 & 78 & 6141.55 & 0.1780 & 0.0042 \\
6410.05--6570.67 & 120 & 6489.82 & 0.1722 & 0.0042 \\
6767.07--6922.66 & 86 & 6843.56 & 0.1669 & 0.0061 \\
7125.05--7306.67 & 99 & 7212.95 & 0.1626 & 0.0047 \\
7495.99--7639.69 & 76 & 7569.84 & 0.1613 & 0.0051 \\
7866.03--7981.74 & 39 & 7923.90 & 0.1684 & 0.0065 \\
8243.99--8375.68 & 48 & 8304.11 & 0.1803 & 0.0052 \\
8598.02--8760.67 & 108 & 8678.56 & 0.1809 & 0.0051 \\
9011.96--9106.67 & 24 & 9073.02 & 0.1775 & 0.0044 \\
9335.96--9486.67 & 141 & 9411.89 & 0.1709 & 0.0046 \\
9705.00--9819.74 & 81 & 9760.91 & 0.1629 & 0.0050 \\
10084.98--10207.68 & 102 & 10145.65 & 0.1614 & 0.0051 \\
10434.98--10570.66 & 114 & 10504.72 & 0.1675 & 0.0082 \\
10814.94--10934.66 & 66 & 10888.41 & 0.1660 & 0.0055 \\
11161.97--11305.66 & 87 & 11243.93 & 0.1766 & 0.0065 \\
11525.96--11660.65 & 72 & 11595.98 & 0.1795 & 0.0064 \\
11891.98--11996.72 & 42 & 11939.36 & 0.1850 & 0.0088 \\
\hline
\end{tabular}
  \end{table}
  
  \begin{table}[ht]
    \small
    \caption{HD\,82443 S-index observing seasons (in JD-2440000).}
    \label{HD82443_chrom}
    \centering
    \begin{tabular}{ccccc}
\hline\hline
JD range & $\rm N_{obs}$ & $\rm JD_{mean}$ & $\rm S_{mean}$ & $\sigma_S$ \\
\hline
6558.69--6575.67 & 21 & 6566.82 & 0.5911 & 0.0256 \\
6768.05--6921.68 & 111 & 6848.00 & 0.6372 & 0.0394 \\
7133.05--7308.66 & 205 & 7218.38 & 0.5939 & 0.0222 \\
7496.02--7648.69 & 123 & 7573.08 & 0.5799 & 0.0294 \\
7863.04--7981.76 & 84 & 7911.85 & 0.6939 & 0.0308 \\
8240.99--8376.71 & 72 & 8303.00 & 0.6599 & 0.0221 \\
8651.92--8771.70 & 75 & 8720.71 & 0.6850 & 0.0291 \\
8991.98--9130.67 & 93 & 9076.42 & 0.6593 & 0.0210 \\
9335.97--9487.67 & 160 & 9413.54 & 0.6423 & 0.0279 \\
9714.00--9819.76 & 84 & 9770.78 & 0.7078 & 0.0571 \\
\hline
\end{tabular}
  \end{table}
  
  \begin{table}[ht]
    \small
    \caption{HD\,82885 S-index observing seasons (in JD-2440000).}
    \label{HD82885_chrom}
    \centering
    \begin{tabular}{ccccc}
\hline\hline
JD range & $\rm N_{obs}$ & $\rm JD_{mean}$ & $\rm S_{mean}$ & $\sigma_S$ \\
\hline
4546.03--4775.67 & 199 & 4702.87 & 0.2825 & 0.0207 \\
4875.02--5121.69 & 155 & 5018.97 & 0.2776 & 0.0135 \\
5298.04--5440.65 & 150 & 5363.79 & 0.2877 & 0.0111 \\
5634.03--5853.70 & 123 & 5755.63 & 0.2847 & 0.0125 \\
6010.04--6211.72 & 87 & 6146.29 & 0.2981 & 0.0155 \\
6417.96--6577.67 & 141 & 6498.35 & 0.2942 & 0.0135 \\
6768.06--6934.67 & 96 & 6854.33 & 0.2720 & 0.0134 \\
7142.05--7307.69 & 108 & 7221.67 & 0.2596 & 0.0138 \\
7496.04--7648.70 & 84 & 7572.12 & 0.2451 & 0.0108 \\
7866.04--7981.76 & 51 & 7924.25 & 0.2446 & 0.0088 \\
8193.04--8375.70 & 51 & 8296.42 & 0.2351 & 0.0082 \\
8598.06--8771.70 & 111 & 8686.65 & 0.2797 & 0.0130 \\
8992.00--9126.69 & 60 & 9080.38 & 0.2438 & 0.0075 \\
9359.94--9485.69 & 59 & 9429.23 & 0.2767 & 0.0171 \\
9753.90--9818.77 & 27 & 9783.72 & 0.2682 & 0.0119 \\
10090.94--10219.67 & 89 & 10155.17 & 0.2606 & 0.0133 \\
10434.99--10570.67 & 96 & 10504.54 & 0.2619 & 0.0103 \\
10814.96--10934.67 & 57 & 10886.00 & 0.2642 & 0.0152 \\
11197.91--11308.67 & 69 & 11255.40 & 0.2376 & 0.0115 \\
11527.96--11682.66 & 66 & 11620.71 & 0.2628 & 0.0168 \\
11891.96--12025.64 & 42 & 11948.48 & 0.2770 & 0.0097 \\
\hline
\end{tabular}
  \end{table}
  
  \begin{table}[ht]
    \small
    \caption{HD\,103095 S-index observing seasons (in JD-2440000).}
    \label{HD103095_chrom}
    \centering
    \input{appendix/HD103095_season_table_chrom.tex}
  \end{table}
  
  \begin{table}[ht]
    \small
    \caption{HD\,115383 S-index observing seasons (in JD-2440000).}
    \label{HD115383_chrom}
    \centering
    \input{appendix/HD115383_season_table_chrom.tex}
  \end{table}
  
  \begin{table}[ht]
    \small
    \caption{HD\,115404 S-index observing seasons (in JD-2440000).}
    \label{HD115404_chrom}
    \centering
    \input{appendix/HD115404_season_table_chrom.tex}
  \end{table}
  
  \begin{table}[ht]
    \small
    \caption{HD\,120136 S-index observing seasons (in JD-2440000).}
    \label{HD120136_chrom}
    \centering
    \input{appendix/HD120136_season_table_chrom.tex}
  \end{table}
  
  \begin{table}[ht]
    \small
    \caption{HD\,124570 S-index observing seasons (in JD-2440000).}
    \label{HD124570_chrom}
    \centering
    \input{appendix/HD124570_season_table_chrom.tex}
  \end{table}
  
  \begin{table}[ht]
    \small
    \caption{HD\,129333 S-index observing seasons (in JD-2440000).}
    \label{HD129333_chrom}
    \centering
    \input{appendix/HD129333_season_table_chrom.tex}
  \end{table}
  
  \begin{table}[ht]
    \small
    \caption{HD\,131156A S-index observing seasons (in JD-2440000).}
    \label{HD131156A_chrom}
    \centering
    \input{appendix/HD131156A_season_table_chrom.tex}
  \end{table}
  
  \begin{table}[ht]
    \small
    \caption{HD\,143761 S-index observing seasons (in JD-2440000).}
    \label{HD143761_chrom}
    \centering
    \input{appendix/HD143761_season_table_chrom.tex}
  \end{table}
  
  \begin{table}[ht]
    \small
    \caption{HD\,149661 S-index observing seasons (in JD-2440000).}
    \label{HD149661_chrom}
    \centering
    \input{appendix/HD149661_season_table_chrom.tex}
  \end{table}
  
  \begin{table}[ht]
    \small
    \caption{HD\,158614 S-index observing seasons (in JD-2440000).}
    \label{HD158614_chrom}
    \centering
    \input{appendix/HD158614_season_table_chrom.tex}
  \end{table}
  
  \begin{table}[ht]
    \small
    \caption{HD\,161239 S-index observing seasons (in JD-2440000).}
    \label{HD161239_chrom}
    \centering
    \input{appendix/HD161239_season_table_chrom.tex}
  \end{table}
  
  \begin{table}[ht]
    \small
    \caption{HD\,182572 S-index observing seasons (in JD-2440000).}
    \label{HD182572_chrom}
    \centering
    \input{appendix/HD182572_season_table_chrom.tex}
  \end{table}
  
  \begin{table}[ht]
    \small
    \caption{HD\,185144 S-index observing seasons (in JD-2440000).}
    \label{HD185144_chrom}
    \centering
    \input{appendix/HD185144_season_table_chrom.tex}
  \end{table}
  
  \begin{table}[ht]
    \small
    \caption{HD\,190007 S-index observing seasons (in JD-2440000).}
    \label{HD190007_chrom}
    \centering
    \input{appendix/HD190007_season_table_chrom.tex}
  \end{table}
  
  \begin{table}[ht]
    \small
    \caption{HD\,201091 S-index observing seasons (in JD-2440000).}
    \label{HD201091_chrom}
    \centering
    \input{appendix/HD201091_season_table_chrom.tex}
  \end{table}
  
  \begin{table}[ht]
    \small
    \caption{HD\,201092 S-index observing seasons (in JD-2440000).}
    \label{HD201092_chrom}
    \centering
    \input{appendix/HD201092_season_table_chrom.tex}
  \end{table}
  
  \begin{table}[ht]
    \small
    \caption{HD\,206860 S-index observing seasons (in JD-2440000).}
    \label{HD206860_chrom}
    \centering
    \input{appendix/HD206860_season_table_chrom.tex}
  \end{table}
  \section{Example light curves}\label{app_C}
  \begin{figure*}
    \centering
    \includegraphics[width=17cm]{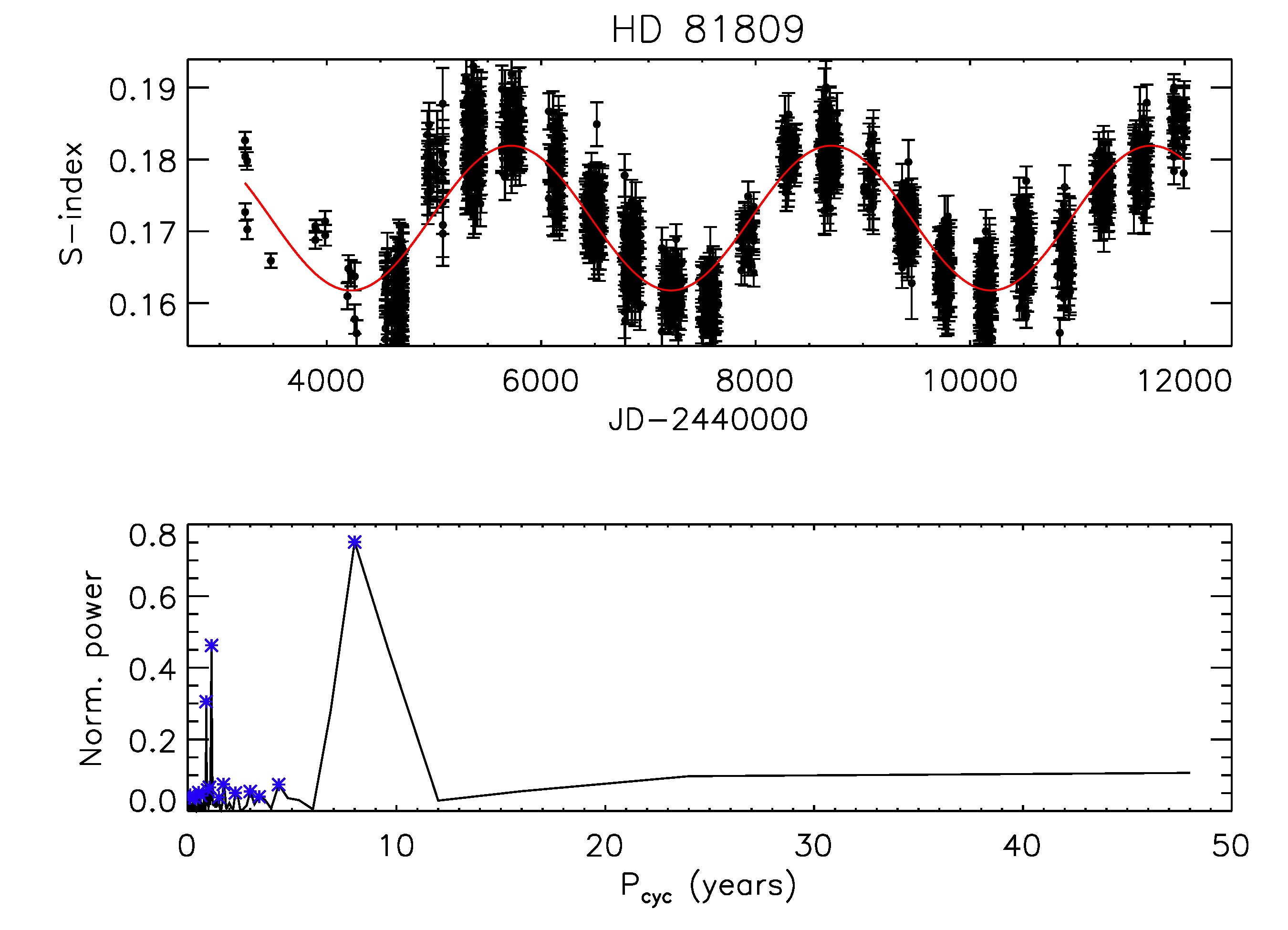}
    \caption{Top panel: Chromospheric time series of the star HD\,81809 flagged 
    as excellent periodicity ($\rm flag_{chrom} = 3$). The red curve shows the 
    best sine fit to the data. Bottom panel: Lomb-Scargle periodogram of the 
    time series.}
    \label{example_flag3}
  \end{figure*}
  
  \begin{figure*}
    \centering
    \includegraphics[width=17cm]{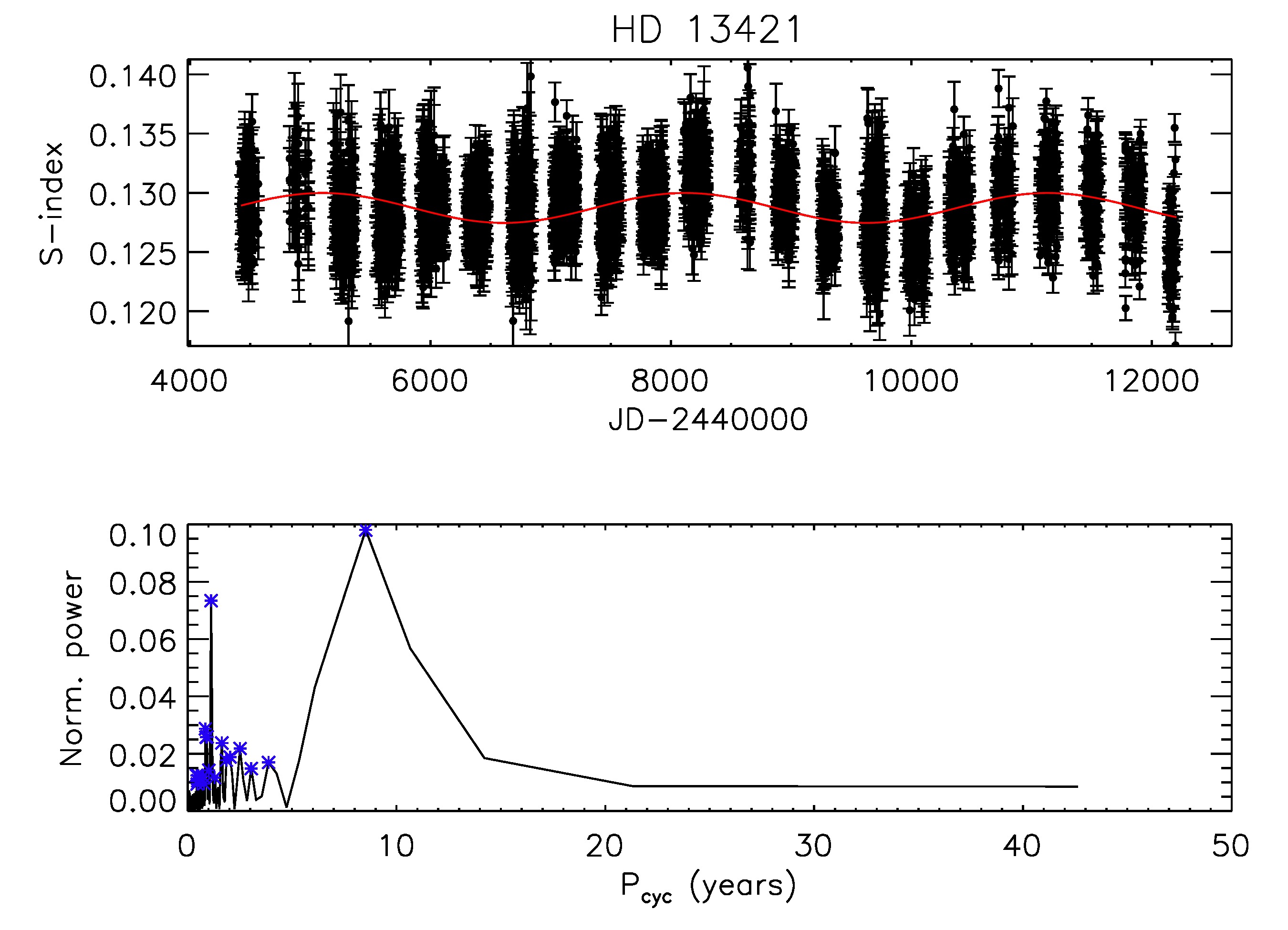}
    \caption{Top panel: Chromospheric time series of the star HD\,13421 flagged 
    as weak periodicity ($\rm flag_{chrom} = 1$). The red curve shows the 
    best sine fit to the data. Bottom panel: Lomb-Scargle periodogram of the 
    time series.}
    \label{example_flag1}
  \end{figure*}
\end{appendix}

\end{document}